\documentclass[a4paper,11pt]{article}

\usepackage[dvips]{graphicx}
\usepackage{amsfonts, color, float,  amsmath}
\usepackage{amssymb}\usepackage{subfigure}
\usepackage{url}
\usepackage{array}
\usepackage{psfrag}

\usepackage{slashed}
\usepackage[svgnames]{xcolor} 

\usepackage[small,bf,sf]{caption} 
\usepackage{microtype}
\usepackage{cite}
\usepackage{authblk}
\usepackage{datetime}
\usepackage[bottom]{footmisc} 
\usepackage{booktabs}  

\allowdisplaybreaks

\def\beq {\begin{equation}}
\def\eeq {\end{equation}}
\def\bea {\begin{eqnarray}}
\def\eea {\end{eqnarray}}

\AtBeginDocument{}

\title{\LARGE {\bf \sffamily \boldmath QCD corrections to $B \to \pi$ form factors
from light-cone sum rules}}

\author[a]{Yu-Ming Wang}
\author[b]{Yue-Long Shen}

\affil[a]{\sl Physik Department T31, Technische Universit\"at M\"unchen,
James-Franck-Stra\ss e~1, D-85748 Garching, Germany}

\affil[b]{\sl College of Information Science and Engineering,
Ocean University of China, Qingdao, Shandong 266100, P.R. China}

\date{\today}

\begin{document}
\begin{flushright}
{\small  TUM-HEP-995/15}
\end{flushright}

\vspace*{-0.7cm}
\begingroup
\let\newpage\relax
\maketitle
\endgroup
\date{}

\begin{abstract}
\noindent
We compute  perturbative corrections to $B \to \pi$ form factors from QCD light-cone sum rules
with $B$-meson distribution amplitudes. Applying the method of regions we demonstrate  factorization of
the vacuum-to-$B$-meson correlation function defined with an interpolating current for pion,
at one-loop level, explicitly in the heavy quark limit. The short-distance functions in the factorization
formulae of the correlation function involves both hard  and hard-collinear scales; and these functions
can be further factorized into hard coefficients by integrating out the hard fluctuations and  jet
functions encoding the hard-collinear information. Resummation of large logarithms in the short-distance
functions is then achieved via the standard renormalization-group approach. We further show  that
structures of the factorization formulae for $f_{B \pi}^{+}(q^2)$ and $f_{B \pi}^{0}(q^2)$
at large hadronic recoil from QCD light-cone sum rules match that derived in QCD factorization.
In particular,  we perform an exploratory phenomenological analysis
of  $B \to \pi$ form factors, paying attention to various sources of perturbative and systematic uncertainties, and extract
$|V_{ub}|= \left(3.05^{+0.54}_{-0.38} |_{\rm th.} \pm 0.09 |_{\rm exp.}\right)  \times 10^{-3}$ with the inverse
moment of the $B$-meson distribution amplitude $\phi_B^{+}(\omega)$ determined by reproducing  $f_{B \pi}^{+}(q^2=0)$
obtained from the light-cone sum rules with $\pi$ distribution amplitudes.
Furthermore, we present the invariant-mass distributions
of the lepton pair for $B \to \pi \ell \nu_{\ell}$ ($\ell= \mu \,, \tau$) in the whole kinematic region.
Finally, we discuss non-valence Fock state contributions to the $B \to \pi$ form factors
$f_{B \pi}^{+}(q^2)$ and $f_{B \pi}^{0}(q^2)$ in brief.

\end{abstract}

\thispagestyle{empty}


\section{Introduction}

Making every  endeavor to achieve precision determinations of heavy-to-light transition form factors
is of utmost importance to, on the one hand, test the CKM sector of the Standard Model,
and on the other side  to sharpen our knowledge towards  diverse facets  of the theory of
strong interaction  (QCD).  We are continually surprised by complexities
and subtleties of factorization properties and heavy quark expansions of even the simplest
$B \to \pi$ form factors in the context of both QCD factorization and  QCD sum rules
on the light-cone (LCSR), not to mention more sophisticated $B \to \rho \,, K^{\ast}$
form factors with an unstable particle in the final states.
The purposes of this paper are to pursue an  endeavor to understand factorization structures of
$B \to \pi$ form factors  from the LCSR with  $B$-meson distribution amplitudes (DAs) at ${\cal O}(\alpha_s)$
in QCD  \cite{Khodjamirian:2005ea,Khodjamirian:2006st}  (see also \cite{DeFazio:2005dx,DeFazio:2007hw}
for an alternative  formulation in the framework of soft-collinear effective theory (SCET));
and to provide a complementary approach to anatomize the topical $|V_{ub}|$ tension arising from
the mismatch in exclusive and inclusive determinations.

Constructions of the LCSR with $B$-meson DAs are accomplished by introducing the $B$-meson-to-vacuum
correlation function, demonstrating factorization of the considered correlator in the proper kinematic regime,
and applying the parton-hadron duality ansatz in the light-meson channel.
It is evident that proof of QCD factorization for the correlation function defined with an on-shell $B$-meson state
at next-to-leading order (NLO) constitutes a primary task in such program,
in addition to further refinements of the duality relation.
Inspecting the tree-level contribution to the correlation function shows that three different
momentum modes with the scaling behaviors
\begin{eqnarray}
P_{\mu} \equiv (n \cdot P \,, \bar n \cdot P \,, P_{\perp} ) \,, & \qquad  &
P_{h \,, \,  \mu} \sim {\cal O}(1,1,1) \,,  \nonumber \\
P_{hc \,, \, \mu} \sim {\cal O}(1,\lambda,\lambda^{1/2}) \,, & \qquad  &
P_{s \,, \, \mu} \sim {\cal O}(\lambda,\lambda,\lambda)\,,
\end{eqnarray}
appear in the problem under consideration, where $n_{\mu}$ and $\bar n_{\mu}$ are  light-cone vectors,
satisfying $n^2=\bar n^2=0$ and $n \cdot \bar n=2$, and are chosen such that the four-momentum of
the fast-moving pion  state has a large component $n \cdot p$ of order $m_b$.
$P_{h \,, \,  \mu}$, $P_{hc \,, \, \mu}$ and $P_{s \,, \, \mu}$
corresponding to the four-momentum  of the  external $b$-quark,  of the interpolating current of pion
and  of the light-spectator quark, will be called hard, hard-collinear and soft modes hereafter.
The transfer momentum $q_{\mu}$ of the weak current $\bar u \, \Gamma \, b$ can
correspond to either a hard mode or a hard-collinear mode dependent on the kinematic region;
a unified description for the purpose of demonstrating factorization of the correlation function
at NLO can be achieved by focusing on the kinematic variable $n \cdot p$.
The heavy-quark expansion parameter $\lambda$ scales as $\Lambda/m_b$
where $\Lambda$ is a hadronic  scale of order $\Lambda_{\rm QCD}$.
It is well known that computing multi-scale amplitudes at loop level can be  facilitated
by applying the method of regions \cite{Beneke:1997zp} in dimensional regularization,
which has been extensively used for evaluating multi-loop integrals in heavy quarkonium decays,
top-quark pair production near threshold, Higgs production at hadron colliders and last but not least
$B$-meson decays.  More importantly, we also benefit from a separation of dynamics at distinct energy scales
allowing for resummation of large logarithms in the resulting matching coefficients and non-perturbative
distribution functions with the standard renormalization-group (RG) approach in the momentum space.
It is then our favored strategy to establish a factorization formula of the considered correlation function
at leading power in $\Lambda/m_b$ and  at  ${\cal O}(\alpha_s)$ using the method of regions.

The fundamental non-perturbative inputs entering LCSR discussed in this paper  are the $B$-meson
DAs defined by hadronic matrix elements of HQET string operators, which also serve as essential
ingredients for the theoretical description of many other exclusive $B$-meson decays, e.g.,
the radiative leptonic $B \to \gamma \, \ell \, \nu_{\ell}$ decays.
It will be shown that the constructed $B$-meson LCSR for $B \to \pi$ form factors are not only sensitive
to the inverse moment of $\phi_{B}^{+}(\omega,\mu)$, i.e., $\lambda_B(\mu)$,
but also dependent heavily on  small $\omega$ behaviors of the $B$-meson DAs (see also \cite{DeFazio:2007hw}).
We are therefore provided with a golden opportunity to probe  more actuate images of the $B$ meson in terms
of the elementary constituents (quarks and gluons), anticipating precision measurements of differential
$q^2$ distributions of $B \to \pi \ell \, \nu_{l}$ at high luminosity experiments and alternative (refined)
determinations of $|V_{ub}|$ exclusively (for instance, the leptonic $B \to \tau \nu_{\tau}$ decay).
We should also mention that understanding renormalization  properties of the $B$-meson DAs and
perturbative QCD constraints of $\phi_{B}^{\pm}(\omega,\mu)$ at  high $\omega$ are also of conceptual
interests for many reasons.

As diverse techniques for computing $B \to \pi$ form factors have been developed so far
and theory predictions are continuously refined with yet higher precision,
several comments on the state-of-art of QCD calculations might be meaningful.

\begin{itemize}

\item {The up-to-date calculations  of $B \to \pi$ form factors from the LCSR  with pion DAs  are
restricted to NLO corrections to twist-2 and twist-3 terms \cite{Ball:2001fp,Ball:2004ye,Duplancic:2008ix}
where asymptotic expressions of twist-3 DAs were taken to demonstrate factorization of the
relevant correlation functions without bothering about mixing of the two- and three-particle
DAs under renormalization. In addition, next-to-next-to-leading-order (NNLO)
perturbative corrections to the twist-2 part
induced by the running QCD coupling  were fulfilled recently in \cite{Bharucha:2012wy}.
Such computations should however be taken cum grano salis, because the large-$\beta_0$
approximation generally overestimates the complete perturbative corrections strongly.
Further improvements of the pion LCSR, including  complete NLO calculations
of the twist-3 terms beyond the asymptotic limit and detailed analysis of the sub-leading power corrections from  twist-5
and 6 parts, are highly  desirable.}

\item {The industries of investigating heavy-to-light $B$-meson form factors
in QCD factorization were initiated in \cite{Beneke:2000wa}
where ${\cal O}(\alpha_s)$ corrections were found to be dominated by the spectator-scattering terms
suffering  sizeable uncertainty from $\lambda_B(\mu)$.
Perturbative corrections to hard matching coefficients were carried out at one loop
\cite{Bauer:2000yr,Beneke:2004rc} for $A$-type currents and \cite{Beneke:2004rc,Becher:2004kk} for
$B$-type currents, and at two loops \cite{Bonciani:2008wf,Asatrian:2008uk,Beneke:2008ei,Bell:2008ws,Bell:2010mg}
for $A$-type currents. The jet functions from integrating out dynamics of the hard-collinear fluctuation were accomplished
at one-loop level \cite{Hill:2004if,Becher:2004kk,Beneke:2005gs}.
One should however keep in mind that hadronic matrix elements of $A$-type currents
(up to perturbatively calculable contributions dependent on the factorization schemes)
cannot be further factorized  in ${\rm SCET(c,s)}$ \cite{Beneke:2003pa} and
must be taken as fundamental inputs from   other approaches.
}

\item {Yet another approach to compute $B \to \pi$ form factors is based upon transverse-momentum-dependent
(TMD) QCD factorization for hard processes developed from the theory of on-shell Sudakov form factor \cite{Collins:1989bt}
and the asymptotic behavior of elastic  hadron-hadron scattering at high energy \cite{Botts:1989kf} with the
underlying physical principle that the elastic scattering of an isolated parton suffers a strong suppression
at high energy  from radiative QCD corrections. Recently, computations of $B \to \pi$ form factors with  TMD factorization
approach have been pushed to ${\cal O}(\alpha_s)$ for twist-2 \cite{Li:2012nk,Li:2012md}
and twist-3 \cite{Cheng:2014fwa} contributions of pion DAs.
However, one needs to be aware of the fact that TMD factorization of hard exclusive processes becomes
extraordinarily  delicate  due to complex infrared subtractions beyond the leading order in $\alpha_s$ \cite{Li:2014xda}
and a complete understanding of TMD factorization for exclusive processes with large momentum transfer
has not been achieved to date  on the conceptual side. }

\end{itemize}

The remainder of this paper is structured as follows.
In section \ref{Recapitulation of the LCSR method} we briefly review the method of the LCSR with
$B$-meson DAs by illustrating the tree-level calculation of $B \to \pi$ form factors.
To facilitate proof of QCD factorization for the considered correlation function at NLO
we recapitalize basic of the diagrammatical factorization  approach at tree level as
an instructive example. We then generalize  factorization proof of the correction
function to the one-loop order in section \ref{section: NLO factorization}
by showing a complete cancelation of soft contributions
to the one-loop QCD diagrams and infrared subtractions determined by convolutions
of the one-loop partonic DAs of the $B$-meson and  the tree-level hard-scattering kernel,
at leading power in $\Lambda/m_b$.
Hard functions and jet functions entering  factorization formulae of the correlation function
are simultaneously obtained by computing the relevant one-loop integrals with the method of regions.
Next-to-leading-logarithmic (NLL) resummation of the hard coefficient functions is performed by virtue of the RG
approach and a detailed comparison of the obtained perturbative matching coefficients  with
the equivalent expressions computed in SCET is also presented in section \ref{section: NLO factorization}.
We further derive NLL resummation improved LCSR for $B \to \pi$ form factors in section
\ref{section: NLL resummation improved sum rules}, which constitute the main results of this paper.
Phenomenological applications of the new sum rules are explored in section \ref{section: numerical analysis},
including determinations of the $q^2$ shapes of $B \to \pi$ form factors, extractions of the CKM matrix element
$|V_{ub}|$ and predictions of the normalized $q^2$ distributions in $B \to \pi \ell \nu_{\ell}$.
In section \ref{section: three-particle DAs} we turn to discuss the impact of three-particle $B$-meson
DAs on $B \to \pi$ form factors, which is still the missing ingredient of our calculations.
The concluding discussions are presented in section  \ref{section: summaries}.
Appendix \ref{loop integrals} contains some useful expressions of one-loop integrals after expanding
integrands with the method of regions. Spectral representations of the  convolution integrals
for constructing the LCSR  with  $B$-meson DAs and two-point QCD sum rules for the decay constants
of the $B$-meson and the pion are collected in the Appendices \ref{appendix B} and \ref{Appendix C}.

\section{Recapitulation of the LCSR method}
\label{Recapitulation of the LCSR method}

We construct  LCSR of the form factors $f_{B \pi}^{+}(q^2)$ and $f_{B \pi}^{0}(q^2)$ with the correlation function
\begin{eqnarray}
\Pi_{\mu}(n \cdot p,\bar n \cdot p)&=& \int d^4x ~e^{i p\cdot x}
\langle 0 |T\left\{\bar{d}(x) \not n \, \gamma_5 \, u(x),
\bar{u}(0) \, \gamma_\mu  \, b(0) \right\}|\bar B(p+q) \rangle
\nonumber\\
&=&\Pi(n \cdot p,\bar n \cdot p) \,  n_\mu +\widetilde{\Pi}(n \cdot p,\bar n \cdot p) \, \bar n_\mu\,,
\label{correlator: definition}
\end{eqnarray}
defined with a pion interpolating current carrying a four-momentum $p_{\mu}$ and a weak $b \to u$ transition current.
We work in the rest frame of the $B$-meson with the velocity vector satisfying $n \cdot v =\bar n \cdot v=1$
and $v_{\perp}=0$. For definiteness, we adopt the following conventions
\begin{eqnarray}
n \cdot p \simeq \frac{m_B^2+m_{\pi}^2-q^2}{m_B}= 2 E_{\pi} \,,
\qquad \bar n \cdot p \sim \cal{O} ({\rm \Lambda_{QCD}}) \,.
\end{eqnarray}
The correlation function $\Pi_{\mu}(n \cdot p,\bar n \cdot p)$ can be computed from
light-cone operator-product-expansion (OPE) at $\bar n \cdot p <0$.
\begin{figure}
\begin{center}
\includegraphics[width=0.3\columnwidth]{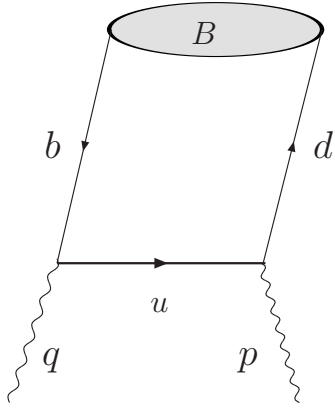}
\vspace*{0.1cm}
\caption{Diagrammatical representation of the correlation function
$\Pi_{\mu}(n \cdot p,\bar n \cdot p)$ at tree level. }
\label{fig:tree diagram}
\end{center}
\end{figure}
Evaluating the diagram in Fig. \ref{fig:tree diagram} yields
\begin{eqnarray}
\widetilde{\Pi}(n \cdot p,\bar n \cdot p) &=& \tilde f_B(\mu) \, m_B \,
\int_0^{\infty} d \omega^{\prime} \,
\frac{\phi_B^-(\omega^{\prime})}{\omega^{\prime} - \bar n \cdot p- i \, 0}
+ {\cal O}(\alpha_s) \,, \nonumber \\
\Pi(n \cdot p,\bar n \cdot p) &=& {\cal O}(\alpha_s) \,.
\label{factorization of correlator:tree}
\end{eqnarray}
The $B$-meson distribution amplitude (DA) $\phi_B^-(\omega^{\prime})$ is defined as \cite{Beneke:2000wa}
\begin{eqnarray}
&& \langle  0 |\bar d_{\beta} (\tau \, \bar{n}) \, [\tau \bar{n}, 0] \, b_{\alpha}(0)| \bar B(p+q)\rangle \nonumber \\
&& = - \frac{i \tilde f_B(\mu) \, m_B}{4}  \bigg \{ \frac{1+ \! \not v}{2} \,
\left [ 2 \, \tilde{\phi}_{B}^{+}(\tau) + \left ( \tilde{\phi}_{B}^{-}(\tau)
-\tilde{\phi}_{B}^{+}(\tau)  \right )  \! \not n \right ] \, \gamma_5 \bigg \}_{\alpha \beta}\,,
\end{eqnarray}
where the light-cone Wilson line is given by
\begin{eqnarray}
[\tau \bar{n}, 0] = {\rm P}\, \left \{ {\rm Exp} \, \left[ i \, g_s \,
\int_0^{\tau} \, d \lambda \, \bar n \cdot A (\lambda \bar n)  \right ]  \right \} \,,
\end{eqnarray}
with the convention of the covariant derivative in QCD  as $D_{\mu}=\partial_{\mu}- i \, g_s \, T^a \, A_{\mu}^a$,
and the Fourier transformations of $\tilde{\phi}_{B}^{\pm}(\tau)$ lead to
\begin{eqnarray}
\phi_B^{\pm}(\omega^{\prime})= \int_{-\infty}^{+\infty} \, {d \, \tau \over 2 \, \pi}
\, e^{i \, \omega^{\prime} \, \tau }   \, \tilde{\phi}_{B}^{\pm}(\tau-i 0) \,.
\end{eqnarray}
One then can  construct the light-cone projector in momentum space \cite{DeFazio:2007hw}
\begin{eqnarray}
 M_{\beta \alpha}  &=& - \frac{i \tilde f_B(\mu) \, m_B}{4}  \nonumber \\
&&  \times \bigg \{ \frac{1+ \! \not v}{2} \,
\left [ \phi_{B}^{+}(\omega^{\prime}) \, \! \not n + \phi_{B}^{-}(\omega^{\prime}) \, \! \not \bar n
- \frac{2\, \omega^{\prime} }{D-2} \, \phi_{B}^{-}(\omega^{\prime}) \, \gamma_{\perp}^{\rho}
\, \frac{\partial}{\partial k_{\perp \rho}^{\prime}} \right ] \, \gamma_5 \bigg \}_{\alpha \beta}\,
\label{light-cone projector of the B meson}
\end{eqnarray}
in $D$ dimensions. Here, $\tilde f_B(\mu)$ is the $B$-meson decay constant in the static limit
and it can be expressed in terms of  the QCD decay constant
\begin{eqnarray}
f_B  = \tilde f_B(\mu) \, \left [  1 + \frac{\alpha_s \, C_F}{4 \, \pi}
\left (-3 \, \ln {\mu \over m_b}   -2 \right ) \right ]\,.
\label{static B-meson decay constant}
\end{eqnarray}
Note that a single  $B$-meson DA $\phi_B^{-}(\omega^{\prime})$ appears in the tree-level LCSR
(\ref{factorization of correlator:tree}) in contrast to factorization of $B \to \gamma \, \ell \, \nu$
where only $\phi_B^{+}(\omega^{\prime})$ enters the  factorization formulae
of the  form factors $F_{V,A} \, (E_{\gamma})$ at leading power in $\Lambda/m_b$ \cite{Beneke:2011nf}.
The discrepancy can be traced back to the longitudinally polarized interpolating current
for the pion  in the former and to the transversely polarized photon in the latter.

Factorization of $\Pi_{\mu}(n \cdot p,\bar n \cdot p)$ at tree level is straightforward
due to the absence of infrared (soft) divergences. The hard-collinear fluctuation
of the internal $u$-quark guarantees  light-cone expansion  of the non-local matrix element
defining the $B$-meson DAs. For the sake of a clear demonstration  of
factorization of $\Pi_{\mu}(n \cdot p,\bar n \cdot p)$ at one-loop order,
we write down the tree-level approximation
of the  partonic correlation function\footnote{Perturbative matching coefficients entering
the factorization formulae of $\Pi_{\mu}(n \cdot p,\bar n \cdot p)$ are independent of the external partonic state,
and it is a matter of convenience to choose a certain configuration for the practical calculation.
More detailed discussions of this point in the context of factorization
of $B \to \gamma \ell \nu$ can be found in Ref. \cite{DescotesGenon:2002mw}.}
(defined as replacing  $| \bar B(p+q)\rangle$ by $| b (p_B-k) \bar d(k)\rangle$
in Eq. (\ref{correlator: definition}) with $p_B\equiv p+q$)
\begin{eqnarray}
\Pi_{\mu, b \bar d}^{(0)}(n \cdot p,\bar n \cdot p) = \int d \omega^{\prime} \,
T_{\alpha \beta}^{(0)}(n \cdot p,\bar n \cdot p,\omega^{\prime}) \Phi_{b \bar d}^{(0) \, \alpha \beta}(\omega^{\prime}) \,,
\end{eqnarray}
where the superscript $(0)$ indicates the tree level, the Lorenz index ``$\mu$" is suppressed
on the right-hand side, the leading-order hard-scattering kernel is given by
\begin{eqnarray}
T_{\alpha \beta}^{(0)}(n \cdot p,\bar n \cdot p,\omega^{\prime})
={i \over 2} \, \frac{1}{\bar n \cdot p -\omega^{\prime} + i 0} \,
\left [ \! \not n \, \gamma_5 \, \! \not \bar n \, \gamma_{\mu} \right ]_{\alpha \beta}\,,
\label{tree hard kernel}
\end{eqnarray}
and the partonic DA of the $B$-meson reads
\begin{eqnarray}
\Phi_{b \bar d}^{\alpha \beta}(\omega^{\prime}) = \int \frac{d \tau}{2 \pi} \,\, e^{i \, \omega^{\prime} \, \tau}
\langle  0 |\bar d_{\beta} (\tau \, \bar{n}) \,
[\tau \bar{n}, 0] \, b_{\alpha}(0)| b (p_B-k) \bar d(k)\rangle \,
\label{2-particle B-meson DA: partonic}
\end{eqnarray}
with the tree-level contribution
\begin{eqnarray}
\Phi_{b \bar d}^{(0) \, \alpha \beta}(\omega^{\prime}) = \delta(\bar n \cdot k-\omega^{\prime})\,
\bar d_{\beta}(k) \,\, b_{\alpha}(p_B-k) \,.
\end{eqnarray}
It is worthwhile to point out that the variable $\omega^{\prime}$ is not necessarily to be
the same as $\omega \equiv  \bar n \cdot k$ despite the equivalence at tree level.
The partonic light-cone projector can be obtained from Eq. (\ref{light-cone projector of the B meson})
via the replacement $\phi_{B}^{\pm}(\omega^{\prime}) \to \phi_{b \bar d}^{\pm}(\omega^{\prime})$,
and we can write down
\begin{eqnarray}
\Pi_{\mu\,, b \bar d}^{(0)}(n \cdot p,\bar n \cdot p) &=& \Pi_{b \bar d}^{(0)}(n \cdot p,\bar n \cdot p) \, n_{\mu}
+  \widetilde{\Pi}_{b \bar d}^{(0)}(n \cdot p,\bar n \cdot p) \, \bar n_{\mu} \,, \nonumber \\
\widetilde{\Pi}_{b \bar d}^{(0)}(n \cdot p,\bar n \cdot p) &=& \tilde f_B(\mu) \, m_B \,
\frac{\phi_{b \bar d}^-(\omega)}{\omega - \bar n \cdot p - i \, 0} \,,  \qquad
\Pi_{b \bar d}^{(0)}(n \cdot p,\bar n \cdot p) = 0 \,.
\label{tree-level partonic correlator: result}
\end{eqnarray}

With  definitions of the $B \to \pi$ form factors and the pion decay constant
\begin{eqnarray}
\langle \pi(p)|  \bar u \gamma_{\mu} b| \bar B (p_B)\rangle
&=& f_{B \pi}^{+}(q^2) \, \left [ p_B + p -\frac{m_B^2-m_{\pi}^2}{q^2} q  \right ]_{\mu}
+  f_{B \pi}^{0}(q^2) \, \frac{m_B^2-m_{\pi}^2}{q^2} q_{\mu} \,, \nonumber \\
\langle \pi(p)  |\bar d \! \not n \, \gamma_5 \, u |  0 \rangle &=& - i \, n \cdot p \, f_{\pi} \,,
\end{eqnarray}
we obtain the hadronic dispersion relation for the correlation function
\begin{eqnarray}
&& \Pi_{\mu}(n \cdot p,\bar n \cdot p) \nonumber \\
&& = \frac{f_{\pi} \, n \cdot p \, m_B}{2 \, (m_{\pi}^2-p^2)}
\bigg \{  \bar n_{\mu} \, \left [ \frac{n \cdot p}{m_B} \, f_{B \pi}^{+} (q^2) + f_{B \pi}^{0} (q^2)  \right ]
\nonumber \\
&& \hspace{0.4 cm} +   n_{\mu} \, \frac{m_B}{n \cdot p-m_B}  \, \,
\left [ \frac{n \cdot p}{m_B} \, f_{B \pi}^{+} (q^2) -  f_{B \pi}^{0} (q^2)  \right ] \bigg \} \, \nonumber \\
&& \hspace{0.4 cm} + \int_{\omega_s}^{+\infty}  \, d \omega^{\prime} \, \frac{1}{\omega^{\prime} - \bar n \cdot p - i 0} \,
\left [ \rho^{h}(\omega^{\prime}, n \cdot p)  \, n_{\mu} \,
+ \tilde{\rho}^{h}(\omega^{\prime}, n \cdot p)  \, \bar{n}_{\mu}  \right ] \,,
\end{eqnarray}
where $\omega_s$ is the hadronic threshold in the pion channel.
Applying the quark-hadron duality ansatz, the integrals over the hadronic spectral densities can be approximated
by the integrals over the  QCD spectral functions with the threshold parameter reinterpreted as an effective ``internal"
parameter of the sum rule approach. Then,  one can derive the final expressions of
the LCSR after implementing the Borel transformation in the variable $\bar n \cdot p \to \omega_M$
\begin{eqnarray}
f_{B \pi}^{+}(q^2) &=& \frac{\tilde f_B(\mu) \, m_B}{f_{\pi} \,n \cdot p} \, {\rm exp} \left[{m_{\pi}^2 \over n \cdot p \,\, \omega_M} \right]
\int_0^{\omega_s} \, d \omega^{\prime} \, e^{-\omega^{\prime} / \omega_M} \,  \phi_B^{-}(\omega^{\prime}) + {\cal O}(\alpha_s) \,,
\nonumber \\
f_{B \pi}^{0} (q^2) &=&  \frac{n \cdot p}{m_B} \, f_{B \pi}^{+} (q^2) +  {\cal O}(\alpha_s) \,.
\label{the form-factor relation}
\end{eqnarray}
which are in agreement with Ref. \cite{DeFazio:2005dx,Khodjamirian:2006st}.

Albeit with the rather simple structures of the three-level LCSR, some interesting observations can be already made.

\begin{itemize}
\item  {Since the $B$-meson DA $\phi_B^+(\omega^{\prime})$  does not enter the factorization formulae of
$\Pi_{\mu}(n \cdot p,\bar n \cdot p)$ at tree level and $\phi_B^{\pm}(\omega^{\prime})$
do not mix under renormalization at one loop in the massless light-quark limit, the convolution integrals of
$\phi_B^{+}(\omega^{\prime})$ entering the contributions of the one-loop diagrams
of $\Pi_{\mu}(n \cdot p,\bar n \cdot p)$ in QCD must be infrared finite at ${\cal O}(\alpha_s)$ to guarantee the validity of QCD factorization of $\Pi_{\mu}(n \cdot p,\bar n \cdot p)$.   }
\item  {Since only a single invariant function $\widetilde{\Pi}(n \cdot p,\bar n \cdot p)$   survives at tree level,
one concludes that  the one-loop contributions to $\Pi(n \cdot p,\bar n \cdot p)$ in QCD must be infrared finite
due to the vanishing infrared (soft) subtraction at ${\cal O}(\alpha_s)$, provided that factorization of
$\Pi_{\mu}(n \cdot p,\bar n \cdot p)$ holds.}
\item {The Borel mass $\omega_M$ and the threshold parameter $\omega_s$  enter into the LCSR
from the dispersive analysis with respect to the variable $\bar n \cdot p$, indicating that one needs to identify
$\omega_M=M^2/n \cdot p$ and $\omega_s=s_0/n \cdot p$ with $(M^2,s_0)$ from the  dispersive construction
of the LCSR in the variable $p^2$. From the scaling $M^2 \sim s_0 \sim \Lambda^2$, one then finds  the power counting of
$f_{B \pi}^{+}$ and $f_{B \pi}^{0}$ as $\sim ({\rm \Lambda}/m_b)^{3/2}$ at tree level, consistent with the observations of \cite{Beneke:2000wa,Beneke:2003pa}. }
\end{itemize}

\section{Factorization of the correlation function at ${\cal O}(\alpha_s)$}
\label{section: NLO factorization}

The objective of this section is to establish the factorization formulae
 for $\Pi_{\mu}(n \cdot p,\bar n \cdot p)$  in QCD at one-loop level.
We adopt the diagrammatic factorization method expanding the correlator  $\Pi_{\mu,  \, b \bar d}$,
the short-distance function $T$ and the partonic DA of the $B$ meson
 $\Phi_{b \bar d} $ in perturbation theory. Schematically,
\begin{eqnarray}
\Pi_{\mu,  \, b \bar d}&=&\Pi_{\mu,  \, b \bar d}^{(0)}+ \Pi_{\mu,  \, b \bar d}^{(1)} + ...
= \Phi_{b \bar d}  \otimes T \, \nonumber \\
&=&   \Phi_{b \bar d}^{(0)} \otimes  T^{(0)}
+  \left [ \Phi_{b \bar d}^{(0)} \otimes  T^{(1)}
+ \Phi_{b \bar d}^{(1)} \otimes  T^{(0)} \right ]  + ...  \,,
\end{eqnarray}
where $\otimes$ denotes the convolution in the variable $\omega^{\prime}$
defined in Eq. (\ref{2-particle B-meson DA: partonic}),
and the superscripts indicate the order of $\alpha_s$.
The hard-scattering kernel at ${\cal O}(\alpha_s)$ is then determined
by the matching condition
\begin{eqnarray}
\Phi_{b \bar d}^{(0)} \otimes  T^{(1)}
= \Pi_{\mu,  \, b \bar d}^{(1)} -  \Phi_{b \bar d}^{(1)} \otimes  T^{(0)} \,,
\label{matching condition of T1}
\end{eqnarray}
where the second term serves as the infrared (soft) subtraction. One crucial point in the proof of factorization
of $\Pi_{\mu,  \, b \bar d}$ is to demonstrate that the hard-scattering kernel $T$ can be  contributed only
from hard and/or hard-collinear regions at leading power in $\Lambda/m_b$, due to a complete cancelation of
the soft contribution to  $\Pi_{\mu,  \, b \bar d}^{(1)}$ and $\Phi_{b \bar d}^{(1)} \otimes  T^{(0)}$.
In addition, since $B$-meson DAs can only collect the soft QCD dynamics of $\Pi_{\mu,  \, b \bar d}$,
we must show that there is no leading contribution
to the correlation function from the collinear region (with the momentum scaling $l_{\mu} \sim
(1, \lambda^2, \lambda)$) at leading power.

Following Ref. \cite{DescotesGenon:2002mw}, we will evaluate the master formula of $T^{(1)}$ in
Eq. (\ref{matching condition of T1}) diagram by diagram.  However,  we will apply the method of regions
\cite{Beneke:1997zp} to compute the loop integrals in order to obtain the hard coefficient
function ($C$) and the jet function ($J$) simultaneously. To establish the factorization formula
\begin{eqnarray}
\Pi_{\mu,  \, b \bar d} = \Phi_{b \bar d}  \otimes T  = C  \cdot \, J   \otimes  \Phi_{b \bar d}  \,,
\end{eqnarray}
$C$ and $J$ must be well defined in dimensional regularization.
This guarantees that we can adopt dimensional regularization to evaluate the leading-power
contributions of $\Pi_{\mu,  \, b \bar d}$ without introducing an additional ``analytical" regulator.
The strategies of our calculations  are as follows: (i) Identify leading regions of the scalar integral
for each diagram; (ii) Simplify the Dirac algebra in the numerator for a given leading region and evaluate
the relevant integrals using the method of regions;  (iii) Evaluate the hard and hard-collinear contributions with the light-cone
projector of the $B$ meson in momentum space; (iv) Show the equivalence of   the soft subtraction term and
the correlation function in the soft region; (v) Add up  the contributions from the hard and hard-collinear regions separately.

\subsection{Weak vertex diagram}

\begin{figure}
\begin{center}
\includegraphics[width=1.0 \columnwidth]{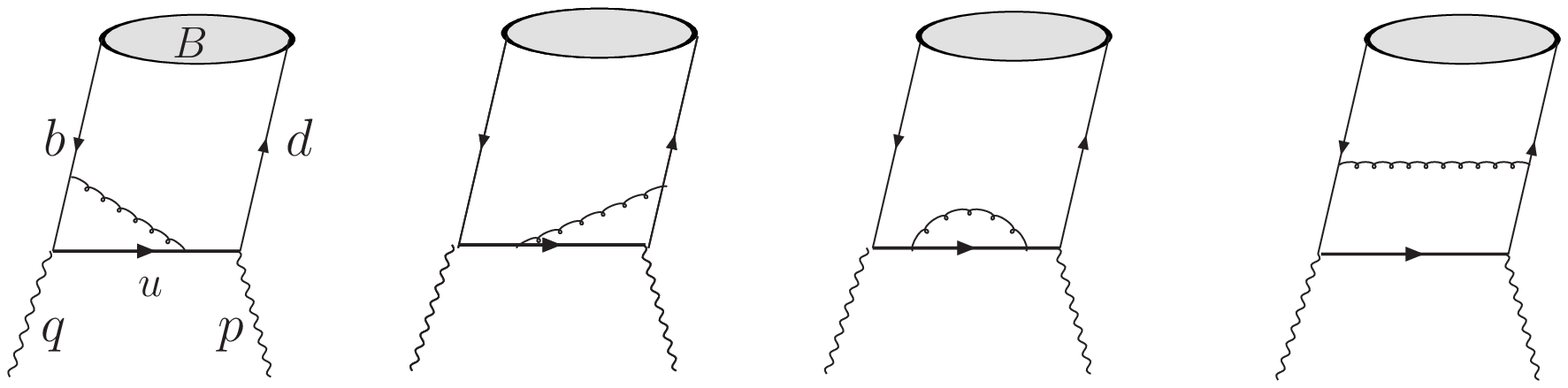}\\
\hspace{-0.5 cm}(a) \hspace{3.0 cm} (b)\hspace{3.5 cm} (c) \hspace{3.5 cm} (d) \\
\vspace*{0.1cm}
\caption{Diagrammatical representation of the correlation function
$\Pi_{\mu}(n \cdot p,\bar n \cdot p)$ at ${\cal O}(\alpha_s)$. }
\label{fig: NLO diagrams of the correlator}
\end{center}
\end{figure}

The contribution to $\Pi_{\mu}^{(1)}$ from the QCD correction to the weak vertex
(the diagram in Fig. \ref{fig: NLO diagrams of the correlator}(a)) is
\begin{eqnarray}
\Pi_{\mu,  \, weak}^{(1)}
&=& \frac{g_s^2 \, C_F}{2 \, (\bar n \cdot p -\omega)} \,
\int \frac{d^D \, l}{(2 \pi)^D} \,   \frac{1}{[(p-k+l)^2 + i 0][(m_b v+l)^2 -m_b^2+ i 0] [l^2+i0]}  \nonumber  \\
&& \bar d(k)  \! \not n \,\gamma_5  \,\! \not {\bar n} \,\, \gamma_{\rho}  \, (\! \not p - \! \not k  + \! \not l)
\, \gamma_{\mu} \, (m_b  \! \not v +  \! \not l+ m_b )\, \gamma^{\rho} \, b(v)  \,,
\label{diagram a: expression}
\end{eqnarray}
where  the label `` $b \bar d$ "  of the partonic correlation function $\Pi_{\mu,  \, b \bar d}$ will be  suppressed from now on
and $D= 4 -2 \, \epsilon$.
Since the perturbative matching coefficients are insensitive to infrared physics, we thus assign the external momenta
$m_b \, v$ to the bottom quark and $k$ (with $k^2=0$) to the light quark.
In accordance with the scaling behaviors
\begin{eqnarray}
n \cdot p \sim m_b\,, \qquad \bar n \cdot p \sim \Lambda\,, \qquad k_{\mu} \sim \Lambda \,,
\end{eqnarray}
we identify the leading-power contributions of the scalar integral
\begin{eqnarray}
I_1= \int [d \, l] \, \frac{1}{[(p-k+l)^2 + i 0][(m_b v+l)^2 -m_b^2+ i 0] [l^2+i0]} \,
\end{eqnarray}
from the hard, hard-collinear and soft regions and the power counting  $I_1 \sim \lambda^0$ implies that only
the leading-power contributions of the numerator in Eq. (\ref{diagram a: expression}) need to be kept
for a given region taking into account the power counting of the tree-level contribution in Eq. (\ref{tree hard kernel}).
We define the integration measure as
\begin{eqnarray}
[d \, l] \equiv \frac{(4 \, \pi)^2}{i} \, \left ( \frac{\mu^2 \, e^{\gamma_E}}{4 \, \pi}\right )^{\epsilon} \,
\frac{d^D \, l}{(2 \pi)^D}\,.
\end{eqnarray}

Inserting the partonic light-cone projector  yields the hard contribution
of $\Pi_{\mu, weak}^{(1)}$ at leading power
\begin{eqnarray}
\Pi_{\mu,  \, weak}^{(1), \, h}&=& i \, g_s^2 \, C_F \, \tilde f_B(\mu) \, m_B \, 
\frac{\phi_{b \bar d}^{-}(\omega)}{\bar n \cdot p -\omega} \int \frac{d^D \, l}{(2 \pi)^D} \, \nonumber \\
&&   \frac{1}{[l^2 + n \cdot p \,\, \bar n \cdot l+ i 0][l^2 + 2 \, m_b \, v \cdot l+ i 0] [l^2+i0]}  \nonumber  \\
&&  \times  \bigg \{ \bar n_{\mu}  \left [ 2 \, m_b \, n \cdot (p+l)  + (D-2)\, l_{\perp}^2 \right ]
- n_{\mu}  \, (D-2) \, (\bar n \cdot l)^2\bigg \} \,,
\label{diagram a: hard region expression}
\end{eqnarray}
where the superscript ``$h$" denotes the hard contribution and we adopt the conventions
\begin{eqnarray}
l_{\perp}^2 \equiv g_{\perp}^{\mu \nu} \, l_{\mu}  \, l_{\nu}\,,  \qquad
g_{\perp}^{\mu \nu} \equiv g^{\mu \nu}-\frac{n^{\mu} \bar n^{\nu}}{2} -\frac{n^{\nu} \bar n^{\mu}}{2} \,.
\end{eqnarray}
Using the results of loop integrals provided in the Appendix \ref{loop integrals}, we obtain
\begin{eqnarray}
\Pi_{\mu,  \, weak}^{(1), \, h}&=& \frac{\alpha_s \, C_F}{4 \, \pi} \, \tilde f_B(\mu) \, m_B \, 
\frac{\phi_{b \, \bar d}^{-}(\omega)}{\bar n \cdot p -\omega} \, 
\bigg \{ \bar n_{\mu}  \bigg [ {1 \over \epsilon^2} +
{1 \over \epsilon} \, \left ( 2 \, \ln {\mu \over  n \cdot p} + 1  \right ) + 2 \, \ln^2 {\mu \over  n \cdot p} \nonumber \\
&& + 2 \,\ln {\mu \over  m_b} -\ln^2 r - 2 \, {\rm Li_2} \left (- {\bar r \over r} \right )  
+{2-r \over r-1} \, \ln r +{\pi^2 \over 12} + 3 \bigg ] \nonumber \\
&& + n_{\mu}  \, \left [ {1 \over r-1} \, \left ( 1 +  {r \over \bar r}  \, \ln r  \right ) \right ]  \,  \bigg \} \,,
\label{diagram a: result of the hard region expression}
\end{eqnarray}
with $r=n \cdot p/m_b$ and $\bar r = 1-r$.

Along the same vein,  one can identify the hard-collinear contribution of $\Pi_{\mu,  \, weak}^{(1)}$ at leading power
\begin{eqnarray}
\Pi_{\mu,  \, weak}^{(1), \, hc}&=& i \, g_s^2 \, C_F \, \tilde f_B(\mu) \, m_B \, 
\frac{\phi_{b \bar d}^{-}(\omega)}{\bar n \cdot p -\omega} \int \frac{d^D \, l}{(2 \pi)^D} \, \nonumber \\
&&   \frac{2 \, m_b \, n \cdot (p+l)}{[ n \cdot (p+l) \,
\bar n \cdot (p-k+l) + l_{\perp}^2  + i 0][ m_b \, n \cdot l+ i 0] [l^2+i0]} \,,
\label{diagram a: hard-collinear region expression}
\end{eqnarray}
where the superscript ``$hc$" indicates the hard-collinear  contribution and the propagators have been expanded
systematically in the hard-collinear region.  Evaluating the integrals with the relations collected in the
Appendix \ref{loop integrals} yields
\begin{eqnarray}
\Pi_{\mu,  \, weak}^{(1), \, hc}&=& \frac{\alpha_s \, C_F}{4 \, \pi} \, \tilde f_B(\mu) \, m_B \, 
\frac{\phi_{b \bar d}^{-}(\omega)}{\omega - \bar n \cdot p} \,\, \bar n_{\mu}  \, \bigg [ {2 \over \epsilon^2} +
{2 \over \epsilon} \, \left (\ln {\mu^2 \over  n \cdot p \, (\omega - \bar n \cdot p)} + 1  \right )  \, \nonumber \\
&&  + \ln^2 {\mu^2 \over  n \cdot p \, (\omega - \bar n \cdot p)}
+ 2 \, \ln {\mu^2 \over  n \cdot p \, (\omega - \bar n \cdot p)}  
-{\pi^2 \over 6} + 4 \bigg ] \,.
\label{diagram a: result of the hard-collinear region expression}
\end{eqnarray}

Applying the method of regions we extract the soft contribution of $\Pi_{\mu, weak}^{(1)}$
\begin{eqnarray}
\Pi_{\mu,  \, weak}^{(1), \, s}&=& \frac{g_s^2 \, C_F}{2 \,( \bar n \cdot p -\omega) } \,
\int \frac{d^D \, l}{(2 \pi)^D} \, \frac{1}{[\bar n \cdot (p-k+l) + i 0][v \cdot l + i 0] [l^2+i0]}  \nonumber  \\
&& \bar d(k) \,\, \! \not n \,\,   \gamma_5  \,\,  \! \not {\bar n} \,\, \gamma_{\mu}  \,\,  b(p_b)  \,  \nonumber \\
&=& \frac{\alpha_s \, C_F}{4 \, \pi} \, \tilde f_B(\mu) \, m_B \, 
\frac{\phi_{b \bar d}^{-}(\omega)}{\bar n \cdot p -  \omega} \,\, \bar n_{\mu}  \,  \nonumber \\
&& \times \bigg [ {1 \over \epsilon^2} + {2 \over \epsilon} \, \ln {\mu \over  \omega - \bar n \cdot p}
+ 2 \, \ln^2 {\mu \over  \omega - \bar n \cdot p}    + {3 \, \pi^2 \over 4} \bigg ] \,,
\label{diagram a: result of the soft region expression}
\end{eqnarray}
where the superscript ``$s$" represents the soft contribution.

\begin{figure}[t]
\begin{center}
\includegraphics[width=0.8  \columnwidth]{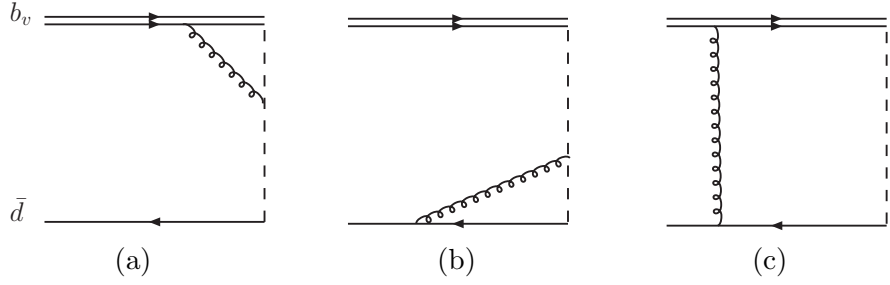}\\
\hspace{0.0 cm}(a) \hspace{3.5 cm} (b)\hspace{3.5 cm} (c)  \\
\vspace*{0.1cm}
\caption{One-loop diagrams for the $B$-meson DA $\Phi_{b \bar u}^{\alpha \beta}(\omega^{\prime})$
defined in (\ref{2-particle B-meson DA: partonic}). }
\label{fig: soft subtraction}
\end{center}
\end{figure}

Now,  we compute the corresponding infrared subtraction term $\Phi_{b \bar d,  \, a}^{(1)} \otimes  T^{(0)}$
as displayed in Fig. \ref{fig: soft subtraction}(a).  With the Wilson-line Feynman rules, we obtain
\begin{eqnarray}
\Phi_{b \bar d, \, a}^{\alpha \beta\,, (1)} (\omega, \omega^{\prime})
&=& i \, g_s^2 \, C_F\, \int \frac{d^D \, l}{(2 \pi)^D} \,
\frac{1}{[\bar n \cdot l + i 0][v \cdot l + i 0] [l^2+i0]}  \nonumber  \\
&& \times [\delta(\omega^{\prime}-\omega-\bar n \cdot l)-\delta(\omega^{\prime}-\omega)] \,
[\bar d(k)]_{\alpha}  \, [b(v)]_{\beta} \,\,,
\label{effective diagram a: wave function}
\end{eqnarray}
from which we can derive the soft subtraction term
\begin{eqnarray}
\Phi_{b \bar d ,  \, a}^{(1)} \otimes T^{(0)}&=& \frac{g_s^2 \, C_F}{2 \,( \bar n \cdot p -\omega) } \,
\int \frac{d^D \, l}{(2 \pi)^D} \, \frac{1}{[\bar n \cdot (p-k+l) + i 0][v \cdot l + i 0] [l^2+i0]}  \nonumber  \\
&& \bar d(k) \,\, \! \not n \,\,   \gamma_5  \,\,  \! \not {\bar n} \,\, \gamma_{\mu}  \,\,  b(v)  \, \,,
\label{diagram a: soft subtraction}
\end{eqnarray}
where the tree-level hard kernel in Eq. (\ref{tree hard kernel}) is used.
We then conclude that
\begin{eqnarray}
\Pi_{\mu,  \, weak}^{(1), \, s}= \Phi_{b \bar d ,  \, a}^{(1)}  \otimes T^{(0)}\,
\end{eqnarray}
at leading power in $\Lambda / m_b$, which is an essential point to prove factorization
of the correlation function $\Pi_{\mu}$.

\subsection{Pion vertex diagram}

Now we turn to compute the QCD correction to the pion vertex
(the diagram in Fig. \ref{fig: NLO diagrams of the correlator}b)
\begin{eqnarray}
\Pi_{\mu,  \, pion}^{(1)}
&=& - \frac{g_s^2 \, C_F}{n \cdot p \, (\bar n \cdot p -\omega)} \,
\int \frac{d^D \, l}{(2 \pi)^D} \,   \frac{1}{[(p-l)^2 + i 0][(l-k)^2 + i 0] [l^2+i0]}  \nonumber  \\
&& \bar d(k)  \, \gamma_{\rho}  \, \! \not l \! \not n \,\gamma_5  \,\,  (\! \not p  - \! \not l)
\, \gamma^{\rho} \, (\! \not p -  \! \not k )\, \gamma_{\mu} \, b(v)  \,.
\label{diagram b: expression}
\end{eqnarray}
One can identify the leading-power contributions of the scalar integral
\begin{eqnarray}
I_2= \int [d \, l] \, \frac{1}{[(p-l)^2 + i 0][(l-k)^2 + i 0] [l^2+i0]} \,
\end{eqnarray}
from the hard-collinear and soft regions, which have the scaling  behaviors
\begin{eqnarray}
I_2^{hc} \sim  I_2^{s} \sim \lambda^{-1} \,,
\end{eqnarray}
by virtue of the power counting analysis. It is evident that the pion vertex
correction would give rise to the power enhanced effect in relative to the tree-level
contribution of Eq. (\ref{tree hard kernel}), provided that no additional suppression factors
come from the spinor structure. Closer inspection shows that expanding the integrand of $I_2$
in the soft region will generate a scaleless integral which vanishes in dimensional regularization.
For the hard-collinear loop momentum, the spinor structure is reduced to
\begin{eqnarray}
\bar d(k)  \, [...] \, b(v) =
\bar d(k)  \, \gamma_5 \, [2 \, (\! \not p  - \! \not l) \, \! \not n \, \! \not l
+(D-4) \,  \! \not l \, \! \not n \,  (\! \not p  - \! \not l)  ] \, (\! \not p -  \! \not k ) \,   b(v)
\end{eqnarray}
which indeed induces a power-suppression factor $\lambda$. It turns out to be less transparent to extract
the leading-power contribution in the hard-collinear region with the insertion of the $B$-meson light-cone projector.
Instead, we first compute the loop integral of Eq. (\ref{diagram b: expression}) exactly without resorting to the method of regions;
then we express $\Pi_{\mu,  \, pion}^{(1)}$ in terms of the partonic DAs by inserting the momentum-space projector.

Employing the expressions of loop integrals in the Appendix A we find
\begin{eqnarray}
\Pi_{\mu,  \, pion}^{(1)}&=& \Pi_{\mu,  \, pion}^{(1),  \,hc}
= \frac{\alpha_s \, C_F}{4 \, \pi} \, \tilde f_B(\mu) \, m_B \, 
\frac{1}{\bar n \cdot p - \omega}  \,\,
\bigg \{ n_{\mu} \, \phi_{b \bar d}^{+}(\omega) \,
\left [ {\bar n \cdot p -\omega \over \omega} \, \ln {\bar n \cdot p -\omega \over \bar n \cdot p}  \right ] \nonumber \\
&& + \bar n_{\mu}  \,\,  \phi_{b \bar d}^{-}(\omega) \, \bigg [
\left ( {1 \over \epsilon} + \ln \left (- {\mu^2 \over p^2} \right )\right )\,
\left ( {2 \, \bar n \cdot p \over \omega} \, \ln {\bar n \cdot p -\omega \over \bar n \cdot p} + 1  \right )  \, \nonumber \\
&&  - {\bar n \cdot p \over \omega} \,  \ln {\bar n \cdot p - \omega \over  \bar n \cdot p}
\, \left ( \ln {\bar n \cdot p - \omega \over  \bar n \cdot p} + {2 \omega \over  \bar n \cdot p}  -4 \right ) + 4 \bigg ]  \bigg \}\,.
\label{diagram b: result of the hard-collinear region expression}
\end{eqnarray}
While the soft contribution of $\Pi_{\mu,  \, pion}^{(1)}$ vanishes in dimensional regularization, it remains   to demonstrate that
the precise cancelation  of $\Pi_{\mu,  \, pion}^{(1),  \, s}$  and $\Phi_{b \bar d ,  \, b} \otimes T^{(0)}$ is independent of
regularization schemes. Applying the method of regions yields
\begin{eqnarray}
\Pi_{\mu,  \, pion}^{(1), \, s}&=& - \frac{g_s^2 \, C_F}{2 \,( \bar n \cdot p -\omega) } \,
\int \frac{d^D \, l}{(2 \pi)^D} \, \frac{1}{[\bar n \cdot (p-l) + i 0][(l-k)^2 + i 0] [l^2+i0]}  \nonumber  \\
&& \bar d(k)  \,\, \! \not \bar n \,\, \! \not l \,\, \! \not n   \,\,   \gamma_5  \,\,  \! \not {\bar n} \,\, \gamma_{\mu}  \,\,  b(v)  \,.
\label{diagram b: result of the soft region expression}
\end{eqnarray}
The corresponding contribution to the partonic DA (the diagram in Fig. \ref{fig: soft subtraction}b)
is  given by
\begin{eqnarray}
\Phi_{b \bar d,  \, b}^{\alpha \beta\,, (1)} (\omega, \omega^{\prime})
&=& i \, g_s^2 \, C_F\, \int \frac{d^D \, l}{(2 \pi)^D} \,
\frac{1}{[\bar n \cdot l + i 0][(k+l) + i 0] [l^2+i0]}  \nonumber  \\
&& \times [\delta(\omega^{\prime}-\omega-\bar n \cdot l)-\delta(\omega^{\prime}-\omega)] \,
[\bar d(k) \, \! \not \bar n \,\, (\! \not k + \! \not l)  ]_{\alpha}  \, [b(v)]_{\beta} \,\,.
\label{effective diagram b: wave function}
\end{eqnarray}
One then deduces the soft subtraction term
\begin{eqnarray}
\Phi_{b \bar d,  \, b}^{(1)}  \otimes T^{(0)}&=& - \frac{g_s^2 \, C_F}{2 \,( \bar n \cdot p -\omega) } \,
\int \frac{d^D \, l}{(2 \pi)^D} \, \frac{1}{[\bar n \cdot (p-k-l) + i 0][(k+l)^2 + i 0] [l^2+i0]}  \nonumber  \\
&& \bar d(k)  \,\, \! \not \bar n \,\, (\! \not k + \! \not l)\,\, \! \not n   \,\,   \gamma_5
\,\,  \! \not {\bar n} \,\, \gamma_{\mu}  \,\,  b(v)  \,,
\label{diagram b: soft subtraction}
\end{eqnarray}
which coincides with $\Pi_{\mu,  \, pion}^{(1), \, s}$ exactly after the shift of the loop momentum  $l \rightarrow l-k$.

\subsection{Wave function renormalization}

The self-energy correction to the intermediate quark propagator
(the diagram in Fig. \ref{fig: NLO diagrams of the correlator}c) can be written as
\begin{eqnarray}
\Pi_{\mu,  \, wfc}^{(1)}
&=& \frac{g_s^2 \, C_F}{(n \cdot p)^2 \, (\bar n \cdot p -\omega)^2} \,
\int \frac{d^D \, l}{(2 \pi)^D} \,   \frac{1}{[(p-k+l)^2 + i 0] [l^2+i0]}  \nonumber  \\
&& \bar d(k)  \! \not n \,\gamma_5  \,(\! \not  p-\! \not k) \,\, \gamma_{\rho}
\,(\! \not p -\! \not k +\! \not l)  \, \gamma^{\rho} \, (\! \not p - \! \not k)
\, \gamma_{\mu} \, b(v)  \,.
\label{diagram c: expression}
\end{eqnarray}
Apparently, $\Pi_{\mu,  \, wfc}^{(1)}$ is free of soft and collinear divergences and a straightforward
calculation gives
\begin{eqnarray}
\Pi_{\mu,  \, wfc}^{(1)}&=&  \frac{\alpha_s \, C_F}{4 \, \pi} \, \tilde f_B(\mu) \, m_B \, 
\frac{\phi_{b \bar d}^{-}(\omega)}{\bar n \cdot p - \omega} \,\, \bar n_{\mu}  \,\,
\bigg [{1 \over \epsilon} + \ln {\mu^2 \over n \cdot p \, (\omega-\bar n \cdot p)}+ 1 \bigg ] \,.
\label{diagram c: result}
\end{eqnarray}

Now we evaluate the perturbative matching coefficient from the wave function renormalization
of the external quark fields. It is evident that the wave function renormalization of a massless quark
does not contribute to the matching coefficient when  dimensional regularization is applied to regularize
both ultraviolet and infrared divergences, i.e.,
\begin{eqnarray}
\Pi_{\mu,  \, dwf}^{(1)} -\Phi_{b \bar d,  \, dwf}^{(1)}  \otimes T^{(0)} =0 \,.
\label{u-quark wf in QCD}
\end{eqnarray}
The wave function renormalization of the $b$-quark in QCD gives
\begin{eqnarray}
\Pi_{\mu,  \, bwf}^{(1)}&=&  - \frac{\alpha_s \, C_F}{8 \, \pi} \,
\bigg [{3 \over \epsilon} + 3 \, \ln {\mu^2 \over m_b^2} + 4 \bigg ] \, \Pi_{\mu}^{(0)} \,,
\label{b-quark wf in QCD}
\end{eqnarray}
with $\Pi_{\mu}^{(0)}$ displayed in Eq. (\ref{tree-level partonic correlator: result}).
The wave function renormalization of the $b$-quark in HQET is
\begin{eqnarray}
\Phi_{b \bar d,  \, bwf}^{(1)}  \otimes T^{(0)}&=& 0 \,,
\label{b-quark wf in HQET}
\end{eqnarray}
due to the scaleless integral,  we then find
\begin{eqnarray}
\Pi_{\mu,  \, bwf}^{(1)} -\Phi_{b \bar d,  \, bwf}^{(1)}  \otimes T^{(0)} &=& - \frac{\alpha_s \, C_F}{8 \, \pi} \,
\bigg [{3 \over \epsilon} + 3 \, \ln {\mu^2 \over m_b^2} + 4 \bigg ] \, \Pi_{\mu}^{(0)} \,.
\label{matching coefficient from b-quark wf}
\end{eqnarray}

\subsection{Box diagram}

The one-loop contribution to $\Pi_{\mu}$ from the box diagram is given by
\begin{eqnarray}
\Pi_{\mu,  \, box}^{(1)}
&=& g_s^2 \, C_F \,
\int \frac{d^D \, l}{(2 \pi)^D} \,   \frac{-1}{[(m_b v+l)^2 -m_b^2+ i 0][(p-k+l)^2 + i 0] [(k-l)^2+i0][l^2+i0]}  \nonumber  \\
&& \bar d(k)  \, \gamma_{\rho}  \,  \,\,  (\! \not k - \! \not l)
\! \not n \,\gamma_5  \,\,  (\! \not p - \! \not k  + \! \not l) \, \gamma_{\mu} \,\!
\, (m_b  \! \not v +  \! \not l+ m_b )\, \gamma^{\rho} \, b(v)  \,.
\label{diagram d: expression}
\end{eqnarray}
This is the only diagram at one-loop level without a hard-collinear propagator outside of the loop,
hence we must identify the enhancement factor $m_b/\Lambda$ from the corresponding scalar integral
so that it can give rise to the leading-power contribution compared to the tree-level amplitude in
Eq. (\ref{tree-level partonic correlator: result}). With the scaling behaviors of the external momenta,
one can establish the scaling of
\begin{eqnarray}
I_4= \int [d \, l] \, \frac{1}{[(m_b v+l)^2 -m_b^2+ i 0][(p-k+l)^2 + i 0] [(k-l)^2+i0][l^2+i0]} \,
\end{eqnarray}
as $\lambda^{-1}$ ($\lambda^{-2}$) in the hard-collinear and semi-hard (soft) regions \footnote{No power enhanced
factor can be induced for $I_4$ in other regions by the power counting analysis, which are therefore irrelevant here.}.
It is straightforward to verify that the semi-hard contribution will be reduced to a scaleless integral,
since there is no external semi-hard mode in the box diagram. We are only left with the hard-collinear and
soft regions at leading power in $\Lambda/m_b$.  The term $(\not \!  k - \not \! l)$ in the spinor structure
will give a suppression factor $\lambda$ in the soft region so that both the hard-collinear and the soft contributions
are of the same power. One might be curious about the observation that the box diagram contributes to the jet function
entering the factorization formulae of the $B$-meson-to-vacuum correlation function $\Pi_{\mu}$ at one-loop level while
the hard-collinear contribution  of the box diagram vanishes in the radiative leptonic decay $B \to \gamma \ell \nu$ \cite{DescotesGenon:2002mw,Bosch:2003fc}.  The crucial discrepancy  attributes to the longitudinally polarized pion
interpolating current in the former and  the transversely polarized photon in the latter.
As a consequence,  one is not able to pick up the large components of two intermediate up-quark propagators
\begin{eqnarray}
(\! \not k - \! \not l) \! \not \epsilon_{\gamma}^{\ast}  \,\,  (\! \not p - \! \not k  + \! \not l) \,
\end{eqnarray}
simultaneously in the case of  $B \to \gamma \ell \nu$, while this is possible in  the contribution of the box diagram
for $\Pi_{\mu}$ as indicated in Eq. (\ref{diagram d: expression}).

Evaluating the hard-collinear contribution of $\Pi_{\mu,  \, box}^{(1)}$ with the partonic momentum-space projector  yields
\begin{eqnarray}
\Pi_{\mu,  \, box}^{(1), \, hc} &=& i \, g_s^2 \, C_F \, \tilde f_B(\mu) \,
{m_B \over m_b} \, \bar n_{\mu} \, \int \frac{d^D \, l}{(2 \pi)^D} \,
 \left [ (2-D)\, n \cdot l \,\phi_{b \bar d}^{+}(\omega) + 2\, m_b \, \phi_{b \bar d}^{-}(\omega)    \right ] \nonumber \\
&&  \,  \times \frac{n \cdot (p+l) }{[ n \cdot (p+l) \, \bar n \cdot (p-k+l) + l_{\perp}^2  + i 0]
[n \cdot l \, \bar n (l-k)+ l_{\perp}^2  + i 0][l^2+i0]} \,. \nonumber \\
\label{diagram d: hard-collinear region expression}
\end{eqnarray}
Using the expressions of  loop integrals collected in the Appendix \ref{loop integrals} we obtain
\begin{eqnarray}
\Pi_{\mu,  \, box}^{(1), \, hc}&=& \frac{\alpha_s \, C_F}{4 \, \pi} \, \tilde f_B(\mu) \,  
\frac{m_B}{\omega} \, \bar n_{\mu}  \, 
\bigg \{\phi_{b \, \bar d}^{+}(\omega) \,  \bigg [ r\, \ln (1+\eta) \bigg ]
- 2 \, \phi_{b \, \bar d}^{-}(\omega) \,  \ln(1+\eta) \, \nonumber \\
&& \times \left [ {1 \over \epsilon}  + \ln {\mu^2 \over n \cdot p \, (\omega-\bar n \cdot p)}
+ {1 \over 2} \, \ln (1+\eta) + 1 \right ]  \,  \bigg \} \,,
\label{diagram d: result of the hard-collinear region}
\end{eqnarray}
with $\eta=-\omega/ \bar n \cdot p$.

Extracting the soft contribution of $\Pi_{\mu,  \, box}^{(1)}$ with the method of regions gives
\begin{eqnarray}
\Pi_{\mu,  \, box}^{(1), \,s}
&=& - \frac{g_s^2 \, C_F}{2}\,
\int \frac{d^D \, l}{(2 \pi)^D} \,   \frac{1}{[ v \cdot l+ i 0]
[\bar n \cdot (p-k+l) + i 0] [(k-l)^2+i0][l^2+i0]}  \nonumber  \\
&& \bar d(k)  \, \! \not v  \,  (\! \not k - \! \not l)  \,\,
\! \not n \,\gamma_5  \,\,  \! \not \bar n \,\, \gamma_{\mu} \, b(v)  \,.
\label{diagram d: soft region expression}
\end{eqnarray}
Now we compute the corresponding NLO contribution to the partonic DA
(the diagram in Fig. \ref{fig: soft subtraction}c)
\begin{eqnarray}
\Phi_{b \bar d,  \, c}^{\alpha \beta\,, (1)} (\omega, \omega^{\prime})
&=& - i \, g_s^2 \, C_F\, \int \frac{d^D \, l}{(2 \pi)^D} \,
\frac{1}{[(l-k)^2 + i 0][v \cdot l + i 0] [l^2+i0]}  \nonumber  \\
&& \times \delta(\omega^{\prime}-\omega+\bar n \cdot l) \,
[\bar d(k) \, \! \not v \,  (\! \not l - \! \not k) ]_{\alpha}  \, [b(v)]_{\beta} \,\,,
\label{effective diagram c: wave function}
\end{eqnarray}
from which one can deduce the soft subtraction term
\begin{eqnarray}
\Phi_{b \bar d,  \, c}^{(1)} \otimes T^{(0)} &=&  \frac{g_s^2 \, C_F}{2}\,
\int \frac{d^D \, l}{(2 \pi)^D} \,   \frac{1}{[ v \cdot l+ i 0]
[\bar n \cdot (p-k+l) + i 0] [(l-k)^2+i0][l^2+i0]}  \nonumber  \\
&& \bar d(k)  \, \! \not v  \,  (\! \not l - \! \not k)  \,\,
\! \not n \,\gamma_5  \,\,  \! \not \bar n \,\, \gamma_{\mu} \, b(v)  \,,
\label{diagram d: soft subtraction term}
\end{eqnarray}
which cancels out  the soft contribution of the correlation function $\Pi_{\mu,  \, box}^{(1) ,\, s}$ completely.
The absence of such soft contribution to the perturbative matching coefficient is particularly important for the box diagram,
since the relevant  loop integrals in the soft region depend  on {\it two} components of the soft spectator momentum
$\bar n \cdot k$ and $ v \cdot k$, and the light-cone OPE fails in the soft region \footnote{The bottom and down quarks
entering the $B$-meson state is {\rm not}  light-cone separated for the soft exchanged gluon in
Fig. \ref{fig: NLO diagrams of the correlator}d, therefore one is not allowed to use  $B$-meson DAs to
absorb the  long-distance physics (i.e., non-perturbative QCD dynamics). The construction of QCD factorization  itself requires
decoupling of  soft contributions from  perturbative fluctuations in general.}.

\subsection{The hard-scattering kernel at ${\cal O}(\alpha_s)$ }

The one-loop hard-scattering kernel of the correlation function $\Pi_{\mu}$ can be readily computed
from the matching condition in Eq. (\ref{matching condition of T1}) by collecting different pieces together
\begin{eqnarray}
\Phi_{b \bar d}^{(0)} \otimes  T^{(1)}  &=& \left [ \Pi_{\mu , \, weak}^{(1)}  + \Pi_{\mu , \, pion}^{(1)}
+ \Pi_{\mu , \, wfc}^{(1)}   + \Pi_{\mu , \, box}^{(1)}
+ \Pi_{\mu , \, bwf}^{(1)} + \Pi_{\mu , \, dwf}^{(1)} \right ] \nonumber \\
&& - \left [  \Phi_{b \bar d  , \, a}^{(1)} + \Phi_{b \bar d , \, b}^{(1)} + \Phi_{b \bar d , \, c}^{(1)}
+ \Phi_{b \bar d  , \, bwf}^{(1)}  + \Phi_{b \bar d  , \, dwf}^{(1)}  \right ]  \otimes  T^{(0)}  \, \nonumber \\
&=& \left [  \Pi_{\mu , \, weak}^{(1) , \, h}
+ \left ( \Pi_{\mu, \, bwf}^{(1)} -  \Phi_{b \bar d  , \, bwf}^{(1)} \right )\right ]
\nonumber \\
&& + \left [\Pi_{\mu , \, weak}^{(1) , \, hc} + \Pi_{\mu , \, pion}^{(1) , \, hc} + \Pi_{\mu , \, wfc}^{(1) , \, hc}
+  \Pi_{\mu , \, box}^{(1) , \, hc}  \right ] \,, 
\label{schematic form of NLO hard kernel}
\end{eqnarray}
where the terms in   the first and second square brackets of the second equality
correspond to the hard matching coefficients and  the jet functions at  ${\cal O}(\alpha_s)$.
Finally,  one can derive the  factorization formulae of $\Pi$ and $\widetilde{\Pi}$ defined
in Eq. (\ref{correlator: definition})
\begin{eqnarray}
\Pi &=& \tilde{f}_B(\mu) \, m_B \sum \limits_{k=\pm} \,
C^{(k)}(n \cdot p, \mu) \, \int_0^{\infty} {d \omega \over \omega- \bar n \cdot p}~
J^{(k)}\left({\mu^2 \over n \cdot p \, \omega},{\omega \over \bar n \cdot p}\right) \,
\phi_B^{(k)}(\omega,\mu)  \,, \nonumber \\
\widetilde{\Pi} &=& \tilde{f}_B(\mu) \, m_B \sum \limits_{k=\pm} \,
\widetilde{C}^{(k)}(n \cdot p, \mu) \, \int_0^{\infty} {d \omega \over \omega- \bar n \cdot p}~
\widetilde{J}^{(k)}\left({\mu^2 \over n \cdot p \, \omega},{\omega \over \bar n \cdot p}\right) \,
\phi_B^{(k)}(\omega,\mu)  \,, \nonumber \\
\label{NLO factorization formula of correlator}
\end{eqnarray}
at leading power in $\Lambda/m_b$, where we keep  the factorization-scale dependence explicitly,
the hard coefficient functions are given by
\begin{eqnarray}
C^{(+)}  &=& \tilde{C}^{(+)}=1, \nonumber \\
C^{(-)} &=& \frac{\alpha_s \, C_F}{4 \, \pi}\, {1 \over \bar r} \,
\left [ {r \over \bar r} \, \ln r + 1 \right ]\,, \nonumber \\
\tilde{C}^{(-)} &=& 1 - \frac{\alpha_s \, C_F}{4 \, \pi}\,  \bigg [ 2 \, \ln^2 {\mu \over n \cdot p}
+ 5 \, \ln {\mu \over m_b} - \ln^2 r  - 2 \, {\rm Li_2} \left ( - {\bar r \over r} \right ) \nonumber \\
&& + {2-r \over r-1} \, \ln r  +  {\pi^2 \over 12} + 5 \bigg ] \,,
\label{results of hard coefficients}
\end{eqnarray}
and the jet functions are
\begin{eqnarray}
J^{(+)}&=& {1 \over r} \, \tilde{J}^{(+)}
= \frac{\alpha_s \, C_F}{4 \, \pi} \, \left (1- { \bar n \cdot p \over \omega } \right ) \,
\ln \left (1- { \omega  \over \bar n \cdot p } \right ) \,, \nonumber \\
\qquad  J^{(-)} &=& 1 \,, \nonumber \\
\tilde{J}^{(-)}&=& 1 + \frac{\alpha_s \, C_F}{4 \, \pi} \,
\bigg [ \ln^2 { \mu^2 \over  n \cdot p (\omega- \bar n \cdot p) }
- 2 \ln {\bar n \cdot p -\omega \over \bar n \cdot p } \, \ln { \mu^2 \over  n \cdot p (\omega- \bar n \cdot p) }
\,  \nonumber \\
&& - \ln^2 {\bar n \cdot p -\omega \over \bar n \cdot p }
- \left ( 1 +  {2 \bar n \cdot p \over \omega} \right )  \ln {\bar n \cdot p -\omega \over \bar n \cdot p }
-{\pi^2 \over 6} -1 \bigg ] \,.
\label{results of jet functions}
\end{eqnarray}

Now,  we verify the factorization-scale independence of $\Pi$ and $\widetilde{\Pi}$ as a consequence of
QCD factorization by construction. Note that the correlation function $\Pi_{\mu}$ is defined by the conserved currents
in QCD, hence the ultraviolet renormalization-scale dependence of $\Pi_{\mu}$ is determined by the renormalization constant
of the strong coupling constant $\alpha_s$ and no additional QCD  operator renormalization (ultraviolet subtraction) is needed
in obtaining the renormalized hard coefficients and jet functions. It is straightforward to write down the following evolution
equations
\begin{eqnarray}
&& {d \over d \ln \mu} \tilde{C}^{(-)}(n \cdot p, \mu) = - \frac{\alpha_s \, C_F}{4 \, \pi}
\left [ \Gamma_{\rm cusp}^{(0)} \ln { \mu \over n \cdot p} + 5\right ]  \tilde{C}^{(-)}(n \cdot p, \mu)\,,
\label{RGE of tildeC1}\\
&& {d \over d \ln \mu} \tilde{J}^{(-)}\left({\mu^2 \over n \cdot p \, \omega},{\omega \over \bar n \cdot p}\right)
= \frac{\alpha_s \, C_F}{4 \, \pi} \left [ \Gamma_{\rm cusp}^{(0)} \ln { \mu^2 \over n \cdot p\, \omega} \right ]
\tilde{J}^{(-)}\left({\mu^2 \over n \cdot p \, \omega},{\omega \over \bar n \cdot p}\right) \nonumber \\
&& \hspace{3 cm} +  \frac{\alpha_s \, C_F}{4 \, \pi} \, \int_0^{\infty} \, d \omega^{\prime}  \, \omega \,\,
\Gamma(\omega,\omega^{\prime},\mu) \,\,
\tilde{J}^{(-)} \left({\mu^2 \over n \cdot p \, \omega^{\prime}},{\omega^{\prime} \over \bar n \cdot p}\right)  \,,  \\
&& {d \over d \ln \mu} \left [ \tilde{f}_B(\mu) \, \phi_B^-(\omega, \mu) \right ] =  - \frac{\alpha_s \, C_F}{4 \, \pi}
\left [ \Gamma_{\rm cusp}^{(0)} \ln { \mu \over \omega} - 5\right ] \left [ \tilde{f}_B(\mu) \, \phi_B^-(\omega, \mu) \right ]  \nonumber \\
&& \hspace{3.5 cm} -  \frac{\alpha_s \, C_F}{4 \, \pi} \, \int_0^{\infty} \, d \omega^{\prime}  \, \omega \,\,
\Gamma(\omega,\omega^{\prime},\mu) \,\, \left [ \tilde{f}_B(\mu) \, \phi_B^-(\omega^{\prime}, \mu) \right ] \,,
\end{eqnarray}
where the function $\Gamma$ is given by \cite{Bell:2008er}
\begin{eqnarray}
\Gamma(\omega,\omega^{\prime},\mu)= - \Gamma_{\rm cusp}^{(0)} \,
\frac{\theta(\omega^{\prime}-\omega)}{\omega \, \omega^{\prime}}
-  \Gamma_{\rm cusp}^{(0)} \,
\left [ \frac{\theta(\omega^{\prime}-\omega)}{\omega^{\prime} \, (\omega^{\prime}-\omega)}
+ \frac{\theta(\omega-\omega^{\prime})}{\omega \, (\omega-\omega^{\prime})}  \right ]_{\oplus} \,
\end{eqnarray}
at one-loop order,  with the $\oplus$ function defined as
\begin{eqnarray}
\int_0^{\infty} \, d \omega^{\prime} \, \left [ f(\omega,\omega^{\prime}) \right ]_{\oplus} \, g(\omega^{\prime})
= \int_0^{\infty} \, d \omega^{\prime} \, f(\omega,\omega^{\prime})  \,
\left [ g(\omega^{\prime}) - g(\omega) \right ] \,,
\end{eqnarray}
and $\Gamma_{\rm cusp}^{(0)}=4$ determined by the geometry of Wilson lines.
The renormalization kernel of $\phi_B^-(\omega, \mu)$ at one-loop level was first computed in
\cite{Bell:2008er} and then confirmed in \cite{DescotesGenon:2009hk}.
We also mention in passing that the RG equations of both the $B$-meson DAs
and  the jet functions  take a particularly  simple  form in the ``dual" momentum space where the Lange-Neubert kernel
\cite{Lange:2003ff} at one loop is diagonalized. More details can be found in Ref. \cite{Bell:2013tfa}
(see also \cite{Braun:2014owa}) and we will not pursue the discussions along this line further.
With the evolution equations displayed above, it is evident that
\begin{eqnarray}
{d \over d \ln \mu} \left [\Pi(n \cdot p, \bar n \cdot p) \,,
\widetilde{\Pi}(n \cdot p, \bar n \cdot p)  \right ]
= {\cal O}(\alpha_s^2).
\end{eqnarray}

Inspection of  Eqs. (\ref{results of hard coefficients}), (\ref{results of jet functions})
and (\ref{static B-meson decay constant}) indicates that one cannot avoid the parametrically large
logarithms of order $\ln (m_b/\Lambda_{\rm QCD})$ in the hard functions, the jet functions,
$\tilde{f}_B(\mu)$ and the $B$-meson DAs concurrently,  by choosing a common value of $\mu$.
Resummation of these logarithms to all orders of $\alpha_s$ can be achieved by solving the three RG equations shown above.
Since the hadronic scale entering the initial conditions of the $B$-meson DAs
$\phi_B^{\pm}(\omega, \mu_0)$, $\mu_0  \simeq 1 \,\, {\rm GeV}$, is quite close
to the hard-collinear scale $\mu_{hc} \simeq  \sqrt{m_b \, \Lambda_{\rm QCD}} \approx 1.5 \,\, {\rm GeV}$, we will not sum
logarithms of $\mu_{hc}/\mu_0$ due to the minor evolution effect \cite{Beneke:2011nf}.
Because the hard scale $\mu_{h1} \sim  n \cdot p$ in the hard function $\tilde{C}^{(-)}(n \cdot p, \mu)$ differs from
the one $\mu_{h2} \sim  m_b$ in $\tilde f_B(\mu)$,  the resulting evolution functions due to running of the renormalization scale
from $\mu_{h1}$($\mu_{h2}$) to $\mu_{hc}$ in $\tilde{C}^{(-)}(n \cdot p, \mu)$ ($\tilde f_B(\mu)$) are
\begin{eqnarray}
\tilde{C}^{(-)}(n \cdot p, \mu) &=& U_1(n \cdot p,\mu_{h1},\mu ) \, \tilde{C}^{(-)}(n \cdot p, \mu_{h1}) \,, \nonumber \\
\tilde f_B(\mu) &=& U_2(\mu_{h2},\mu ) \, \tilde f_B(\mu_{h2}) \,.
\end{eqnarray}
To achieve NLL resummation of large logarithms in the hard coefficient $\tilde{C}^{(-)}$
we need to generalize  the RG equation (\ref{RGE of tildeC1}) to
\begin{eqnarray}
{d \over d \ln \mu} \tilde{C}^{(-)}(n \cdot p, \mu) =
\left [ - \Gamma_{\rm cusp}(\alpha_s) \ln { \mu \over n \cdot p} + \gamma(\alpha_s) \right ]  \tilde{C}^{(-)}(n \cdot p, \mu)\,,
\label{general RGE of tildeC1}
\end{eqnarray}
where the cusp anomalous dimension, $\gamma(\alpha_s)$ and the QCD $\beta$-function are expanded as
\begin{eqnarray}
\Gamma_{\rm cusp}(\alpha_s) &=& {\alpha_s \, C_F \over 4 \pi} \, \left [ \Gamma_{\rm cusp}^{(0)}
+ \left ({\alpha_s \over 4 \pi} \right )\, \Gamma_{\rm cusp}^{(1)}
+ \left ({\alpha_s \over 4 \pi} \right )^2 \, \Gamma_{\rm cusp}^{(2)} + ... \right ] \,, \nonumber \\
\gamma(\alpha_s)&=& {\alpha_s \, C_F \over 4 \pi} \, \left [ \gamma^{(0)}
+ \left ({\alpha_s \over 4 \pi} \right )\, \gamma^{(1)} + ... \right ] \,, \nonumber \\
\beta(\alpha_s)&=& -8 \, \pi\, \left [ \left ({\alpha_s \over 4 \pi} \right )^2 \,  \beta_0
+ \left ({\alpha_s \over 4 \pi} \right )^3 \,  \beta_1
+ \left ({\alpha_s \over 4 \pi} \right )^4 \,  \beta_2   + ... \right ] \,.
\end{eqnarray}
The  cusp anomalous dimension at the three-loop order and the remanning anomalous dimension
$\gamma(\alpha_s)$  determining  renormalization of the SCET heavy-to-light current at two loops
will enter $U_1(n \cdot p,\mu_{h1},\mu )$ at NLL accuracy. The manifest expressions of $\Gamma_{\rm cusp}^{(i)}$,
$\gamma^{(i)}$ and $\beta_i$ can be found in \cite{Beneke:2011nf} and references therein
\footnote{Note that there is a factor $C_F$ difference of  our conventions of $\Gamma_{\rm cusp}^{(i)}$ and
$\gamma^{(i)}$ compared with   \cite{Beneke:2011nf}.},
the evolution function $U_1(n \cdot p,\mu_{h1},\mu )$ can be read from Eq. (A.3) in \cite{Beneke:2011nf}
with the replacement rules $E_{\gamma} \to {n \cdot p / 2}$ and $\mu_h \to \mu_{h1}$.
The three-loop evolution of the strong coupling $\alpha_s$ in the ${\rm \overline{MS}}$ scheme
\begin{eqnarray}
\alpha_s(\mu) &=& {2 \pi \over \beta_0} \, \bigg \{ 1- {\beta_1 \over 2 \, \beta_0^2 } \, {\ln (2 L) \over L }
+{\beta_1^2 \over 4 \beta_0^4 \, L^2} \, \left [ \left (\ln (2 L) -{1 \over 2} \right )^2
+{\beta_2 \beta_0 \over \beta_1^2} - {5 \over 4} \right ] \bigg \}  \,, \nonumber \\
L &=& \ln \left ( { \mu  \over  \Lambda^{(n_f)}_{\rm QCD}} \right ) \,
\end{eqnarray}
is used with $\Lambda^{(4)}_{\rm QCD} =229 \, {\rm MeV}$.

The RG equation of $\tilde{f}_B(\mu)$ at the two-loop order is given by
\begin{eqnarray}
{d \over d \ln \mu} \, \tilde{f}_B(\mu) =\tilde{\gamma}(\alpha_s)\, \tilde{f}_B(\mu) \,,
\end{eqnarray}
with
\begin{eqnarray}
\tilde{\gamma}(\alpha_s) &=& {\alpha_s \, C_F \over 4 \pi} \, \left [ \tilde{\gamma}^{(0)} +
\left ({\alpha_s \over 4 \pi} \right )\, \tilde{\gamma}^{(1)} + ...  \right ] \,, \nonumber \\
\tilde{\gamma}^{(0)} &=& 3 \,, \qquad
\tilde{\gamma}^{(1)}= {127 \over 6} + {14\, \pi^2 \over 9} - {5 \over 3}\, n_f \,,
\end{eqnarray}
where $n_f=4$ is the number of light quark flavors.
Solving this RG equation yields
\begin{eqnarray}
U_2(\mu_{h2},\mu ) &=& {\rm Exp}  \bigg [ \int_{\alpha_s(\mu_{h2})}^{\alpha_s(\mu)} \,
d \alpha_s \, \frac{\tilde{\gamma}(\alpha_s)}{\beta(\alpha_s)} \bigg ] \, \nonumber \\
&=& z^{- \frac{\tilde{\gamma}_0 }{2 \,\beta_0} \, C_F} \bigg [1+ \frac{\alpha_s(\mu_{h2}) \, C_F}{4 \pi}  \,
\left (  {\tilde{\gamma}^{(1)} \over 2 \, \beta_0} - {\tilde{\gamma}^{(0)} \, \beta_1 \over 2 \, \beta_0^2 } \right ) (1-z)
+{\cal O}(\alpha_s^2) \bigg ]\,,
\end{eqnarray}
with $z=\alpha_s(\mu)/\alpha_s(\mu_{h2})$.

The final factorization formulae of $\Pi$ and $\widetilde{\Pi}$ with RG improvement at NLL accuracy can be written as
\begin{eqnarray}
\Pi &=& m_B    \, \left [U_2(\mu_{h2},\mu ) \, \tilde{f}_B(\mu_{h2}) \right ]
\int_0^{\infty} {d \omega \over \omega- \bar n \cdot p}~
J^{(+)}\left({\mu^2 \over n \cdot p \, \omega},{\omega \over \bar n \cdot p}\right) \,
\phi_B^{(+)}(\omega,\mu)  \, \nonumber \\
&& + m_B  \,\left [U_2(\mu_{h2},\mu ) \, \tilde{f}_B(\mu_{h2}) \right ] \, C^{(-)}(n \cdot p, \mu) \,
\int_0^{\infty} {d \omega \over \omega- \bar n \cdot p}~ \,
\phi_B^{(-)}(\omega,\mu)  \,, \nonumber \\
\widetilde{\Pi} &=& m_B  \, \left [U_2(\mu_{h2},\mu ) \, \tilde{f}_B(\mu_{h2}) \right ]
\int_0^{\infty} {d \omega \over \omega- \bar n \cdot p}~
\widetilde{J}^{(+)}\left({\mu^2 \over n \cdot p \, \omega},{\omega \over \bar n \cdot p}\right) \,
\phi_B^{(+)}(\omega,\mu)  \, \nonumber \\
&& + m_B   \, \left [U_1(n \cdot p,\mu_{h1},\mu ) \, U_2(\mu_{h2},\mu ) \right ] \,
\left [ \tilde{f}_B(\mu_{h2}) \, \widetilde{C}^{(-)}(n \cdot p, \mu_{h1})  \right ] \,  \nonumber \\
&& \hspace{0.3 cm} \times  \int_0^{\infty} {d \omega \over \omega- \bar n \cdot p}~
\widetilde{J}^{(-)}\left({\mu^2 \over n \cdot p \, \omega},{\omega \over \bar n \cdot p}\right) \,
\phi_B^{(-)}(\omega,\mu)  \,,
\label{resummation improved factorization formula}
\end{eqnarray}
where $\mu$ should be taken as a hard-collinear scale of order $\sqrt{m_b \, \Lambda}$.


\subsection{Comparison with previous approaches}

The aim of this subsection is to develop a better understanding of the factorization structures
of $\Pi$ and $\widetilde{\Pi}$  obtained above. Inspecting Eq. (\ref{schematic form of NLO hard kernel})
shows that the hard-scale fluctuation of the correlation function $\Pi_{\mu}(n \cdot p,\bar n \cdot p)$
comes solely from the contributions of the weak vertex diagram and the $b$-quark wave function renormalization.
This demonstrates that the hard matching coefficients $C^{(-)}$ and $\tilde{C}^{(-)}$ can be also extracted
from the one-loop hard matching coefficients of the QCD current $\bar q \, \gamma_{\mu}\, b$ in SCET \cite{Bauer:2000yr}
\begin{eqnarray}
\bar q \,\, \gamma_{\mu} \,\, b \rightarrow \left [ C_4 \, \bar n_{\mu} + C_5 \, v_{\mu} \right ] \,
\bar \xi_{\bar n} \,W_{hc } \, Y_{s}^{\dag} \, b_v + ... \,,
\label{SCET-I matching}
\end{eqnarray}
where $W_{hc } $ and $ Y_{s}^{\dag}$ denote the  hard-collinear and soft Wilson lines,
the ellipses represent terms with different Dirac structures and sub-leading power contributions.
Inserting (\ref{NLO factorization formula of correlator}) into (\ref{correlator: definition}) and comparing
with (\ref{SCET-I matching}) gives \footnote{$C^{(-)}$ and $\tilde{C}^{(-)}$ correspond to the hard matching coefficients
of $A$-type SCET currents. This can be understood from the fact that factorization of the associated SCET matrix elements
involve the same DA $\phi_{B}^{(-)}(\omega)$ as in the tree-level approximation.
$C^{(+)}$ and $\tilde{C}^{(+)}$ are the hard matching coefficients of $B$-type SCET currents whose matrix elements
start at the first order of $\alpha_s$, therefore only the tree-level contributions of $C^{(+)}$ and $\tilde{C}^{(+)}$
enter the factorization formulae of the correlation function $\Pi_{\mu}(n \cdot p,\bar n \cdot p)$ at one loop. }
\begin{eqnarray}
C^{(-)} = {1 \over 2} \, C_5, \qquad \tilde{C}^{(-)} = C_4 + {1 \over 2} \, C_5 \,.
\label{hard coefficient relations}
\end{eqnarray}
The explicit expressions of $C_4$ and $C_5$ can be found in \cite{Bauer:2000yr,Beneke:2004rc}
\begin{eqnarray}
C_4 &=& 1 - \frac{\alpha_s \, C_F}{4 \, \pi}\,  \bigg [ 2 \, \ln^2 {\mu \over m_b}
- ( 4 \, \ln r - 5 ) \, \ln {\mu \over m_b} + 2\,  \ln^2 r  + 2 \, {\rm Li_2} \left (1-r \right ) \nonumber \\
&& +  {\pi^2 \over 12} + \left ( {r^2 \over \bar r^2} - 2 \right ) \, \ln r  + {r \over 1-r } + 6 \bigg ] \,, \\
C_5 &=& {2 \over r} + {2 \, r \over \bar r^2} \, \ln r \,,
\end{eqnarray}
from which one can readily verify the relations in Eq. (\ref{hard coefficient relations}).
Conceptually, this is just an example to show that  perturbative coefficient functions entering
QCD factorization formulae are independent of the external partonic configurations
used in the matching procedure.

The jet functions $\tilde{J}^{(\pm)}$ also confront with  the earlier calculations in \cite{DeFazio:2007hw}
with SCET Feynman rules. It is a straightforward task to show that $\tilde{J}^{(-)}$ coincides with
(2.23) in \cite{DeFazio:2007hw} while $\tilde{J}^{(+)}$ ($J^{(+)}$) is in agreement with (3.9) of \cite{DeFazio:2007hw}.
A final remark is devoted to  $J^{(-)}$. Because the corresponding hard coefficient $C^{(-)}$ starts at ${\cal O}(\alpha_s)$,
only the tree-level jet function $J^{(-)}$ enters the one-loop factorization of $\Pi_{\mu}$.

\section{The LCSR for $B \to \pi$ form factors at ${\cal O}(\alpha_s)$}
\label{section: NLL resummation improved sum rules}

Now, we are ready to construct the sum rules of $f_{B \pi}^{+}(q^2)$ and $f_{B \pi}^{0}(q^2)$ including
the  radiative corrections at ${\cal O}(\alpha_s)$. Following the prescriptions to construct the tree-level sum rules
in Section \ref{Recapitulation of the LCSR method} and expressing the correlation function $\Pi_{\mu}$ in a dispersion
form with the relations in the Appendix \ref{appendix B}, we obtain
\begin{eqnarray}
&& f_{\pi} \,\, e^{-m_{\pi}^2/(n \cdot p \, \omega_M)} \,\,\,
\left \{ \frac{n \cdot p} {m_B} \, f_{B \pi}^{+}(q^2)
\,, \,\,\,   f_{B \pi}^{0}(q^2)  \right \}  \,  \nonumber \\
&& = \left [U_2(\mu_{h2},\mu ) \, \tilde{f}_B(\mu_{h2}) \right ]
\,\,\, \int_0^{\omega_s} \,\, d \omega^{\prime} \,\, 
\, e^{-\omega^{\prime} / \omega_M}  \, \bigg [  r\, \phi_{B, \rm eff}^{+}(\omega^{\prime}, \mu) \nonumber \\
&& \hspace{0.4 cm} + \left [ U_1(n \cdot p,\mu_{h1},\mu ) \, \widetilde{C}^{(-)}(n \cdot p, \mu_{h1})  \right ]
\, \phi_{B, \rm eff}^{-}(\omega^{\prime}, \mu)  \nonumber \\
&& \hspace{0.4 cm}  \pm \,\,\, \frac{n \cdot p -m_B} {m_B}   \,\,\,
\left ( \phi_{B, \rm eff}^{+}(\omega^{\prime}, \mu)
+ C^{(-)}(n \cdot p, \mu) \, \phi_{B}^{-}(\omega^{\prime}, \mu) \right )  \bigg ] \,,
\label{NLO sum rules of form factors}
\end{eqnarray}
where the functions $\phi_{B, \rm eff}^{\pm}(\omega^{\prime}, \mu)$ are defined as
\begin{eqnarray}
\phi_{B, \rm eff}^{+}(\omega^{\prime}, \mu) &=&  \frac{\alpha_s \, C_F}{4 \, \pi} \,\,
\int_{\omega^{\prime}}^{\infty} \,\, {d \omega \over \omega} \,\, \phi_{B}^{+}(\omega, \mu) \,\,\,,
\label{effective B-meson plus}  \\
\phi_{B, \rm eff}^{-}(\omega^{\prime}, \mu) &=& \phi_{B}^{-}(\omega^{\prime}, \mu)
+  \frac{\alpha_s \, C_F}{4 \, \pi} \,\, \bigg \{ \int_0^{\omega^{\prime}} \,\, d \omega \,\,\,
\left [ {2 \over \omega - \omega^{\prime}}  \,\,\, \left (\ln {\mu^2 \over n \cdot p \, \omega^{\prime}}
- 2 \, \ln {\omega^{\prime} - \omega \over \omega^{\prime}} \right )\right ]_{\oplus} \,\, \nonumber \\
&& \times \phi_{B}^{-}(\omega, \mu)  - \int_{\omega^{\prime}}^{\infty} \,\, d \omega \,\,\,
\bigg [ \ln^2 {\mu^2 \over n \cdot p \, \omega^{\prime}} - \left ( 2 \, \ln {\mu^2 \over n \cdot p \, \omega^{\prime}}  + 3 \right ) \,\,
\ln {\omega - \omega^{\prime} \over \omega^{\prime}} \,\, \nonumber \\
&& + \, 2 \,\, \ln {\omega \over \omega^{\prime}}   + {\pi^2 \over 6} - 1 \bigg ]
\,\,\, {d \phi_{B}^{-}(\omega, \mu) \over d \omega}  \bigg \} \,\,.
\label{effective B-meson minus}
\end{eqnarray}

Several comments on the structures of the sum rules are  in order.

\begin{itemize}
\item {The symmetry-breaking effects of the form-factor relation (\ref{the form-factor relation}) can be immediately read from
the last line of (\ref{NLO sum rules of form factors}). The first term comes from the hard-collinear fluctuation and the
corresponding integral is infrared finite  in the heavy quark limit. One can readily confirm  that this term gives an identical
result of the spectator-interaction induced  symmetry-breaking  correction shown in Eq. (56) of \cite{Beneke:2000wa} in the
leading approximation, provided that the tree-level sum rules of $f_{\pi}$ in the Appendix \ref{Appendix C}
and the asymptotic expression of the twist-2 pion DA are implemented \cite{DeFazio:2005dx}.
The second term corresponds  to the symmetry-breaking
effect induced by the hard fluctuation and it also coincides with the second term in the bracket of  Eq. (30) in \cite{Beneke:2000wa}. }
\item {The scaling behavior of $\omega^{\prime}$ in  (\ref{NLO sum rules of form factors}) is
$\omega^{\prime} \sim \Lambda^2/m_b$ due to the bounds of
the integration, while the power counting of $\omega$ in (\ref{effective B-meson minus}) is ${\cal O} (\Lambda)$ determined by
the canonical behaviors of the $B$-meson DAs $\phi_{B}^{\pm}(\omega, \mu)$.
It is then evident that $\ln \left [{\left (\omega - \omega^{\prime} \right ) / \omega^{\prime}} \right ]$
and $\ln \left ({\omega / \omega^{\prime}} \right )$ appeared in $\phi_{B, \rm eff}^{-}(\omega^{\prime}, \mu)$
are counted  as $\ln (m_b / \Lambda)$ in the heavy quark limit.
Such large logarithms are identified as the end-point divergences in QCD factorization approach  (see also the discussions in \cite{DeFazio:2007hw}). However, we should also  keep in mind that the NLL resummation improved hard coefficient $\left [ U_1(n \cdot p,\mu_{h1},\mu ) \,\,  \widetilde{C}^{(-)}(n \cdot p, \mu_{h1})  \right ]$  vanishes in the heavy quark limit.}
\end{itemize}

\section{Numerical analysis}
\label{section: numerical analysis}

In this section we aim at exploring phenomenological implications of the sum rules for $ f_{B \pi}^{+,0}(q^2)$
in Eq. (\ref{NLO sum rules of form factors}) including the shapes of the two form factors,  the normalized $q^2$ spectra
of $B \to \pi \ell \nu $ for $\ell = \mu , \tau$ as well as  the determinations of the CKM matrix element $|V_{ub}|$.
We will first discuss the theory inputs (the $B$-meson DAs, the ``internal" sum rule parameters,
the decay constants of the $B$-meson and  pion, etc) entering the sum rule analysis,  compute the form factors at zero momentum
transfer, and then predict the shapes of $ f_{B \pi}^{+,0}(q^2)$ in the small $q^2$ region and extrapolate the sum rule computations to
the full kinematic  region with the $z$-series parametrization.

\subsection{Theory input parameters}

The $B$-meson DAs serve as fundamental ingredients for  the LCSR of the $B \to \pi$ form factors $f_{B \pi}^{+,0}(q^2)$.
Albeit with the encouraging progresses in understanding their  properties at large $\omega$ in perturbative  QCD \cite{Lee:2005gza,Feldmann:2014ika},  our knowledge of the behaviors of $\phi_{B}^{\pm}(\omega, \mu)$
at small $\omega$ is still rather limited due to the poor
understanding of non-perturbative QCD dynamics (see \cite{Braun:2003wx} for discussions in the context of
the QCD sum rule method). To achieve a better understanding of the model dependence of $\phi_{B}^{\pm}(\omega, \mu)$
in the sum rule analysis, we consider the following four different parameterizations for the shapes of
the $B$-meson DA $\phi_{B}^+(\omega,\mu_0) $:
\begin{eqnarray}
&&  \phi_{B,\rm I}^+(\omega,\mu_0) = \frac{\omega}{\omega_0^2} \, e^{-\omega/\omega_0} \,,
\nonumber \\
&&  \phi_{B,\rm II}^+(\omega,\mu_0)= \frac{1}{4 \pi \,\omega_0} \, {k \over k^2+1} \,
\left[ {1 \over k^2+1} - \frac{2 (\sigma_B^{(1)} -1)}{\pi^2}  \, \ln k \right ] \,, \hspace{0.5 cm}
k= \frac{\omega}{1 \,\, \rm GeV} \,, \,
\nonumber \\
&&  \phi_{B,\rm III}^+(\omega,\mu_0)= \frac{2 \omega^2}{\omega_0 \omega_1^2} \, e^{-(\omega/\omega_1)^2} \,, \hspace{0.5 cm}
\omega_1= {2 \, \omega_0 \over \sqrt{\pi}} \,,
\nonumber \\
&&   \phi_{B,\rm IV}^+(\omega,\mu_0)= \frac{\omega }{\omega_0 \omega_2} \,
{\omega_2 - \omega \over \sqrt{\omega(2 \omega_2-\omega)} } \,\, \theta(\omega_2-\omega)\,,
\hspace{0.5 cm} \omega_2= {4 \, \omega_0 \over 4- \pi} \,.
\label{four models of B-meson DAs}
\end{eqnarray}
$\phi_{B,\rm I}^+(\omega,\mu_0)$ was originally proposed in \cite{Grozin:1996pq}
inspired by a tree-level QCD sum rule analysis. $\phi_{B,\rm II}^+(\omega,\mu_0)$ suggested in \cite{Braun:2003wx}
was motivated from  the QCD sum rule calculations at ${\cal O}(\alpha_s)$ with the parameter $\sigma_B^{(1)}$ defined as
\begin{eqnarray}
\sigma_B^{(n)}(\mu) &=& \lambda_B(\mu) \, \int_0^{\infty} \, {d \omega \over \omega} \,
\ln^n {\mu \over \omega} \, \phi_{B}^+(\omega,\mu) \,, \nonumber \\
\lambda_B^{-1}(\mu) &=& \int_0^{\infty} \, {d \omega \over \omega} \,
\,\phi_{B}^+(\omega,\mu) \,.
\end{eqnarray}
$\phi_{B,\rm III}^+(\omega,\mu_0)$ and $\phi_{B,\rm IV}^+(\omega,\mu_0)$ are deduced from the two models
of $\phi_{B}^-(\omega,\mu_0)$ \cite{DeFazio:2007hw} with the Wandzura-Wilczek approximation (i.e., neglecting  contributions of
$B$-meson three-particle DAs) to maximize the model dependence of $\phi_{B}^{\pm}(\omega,\mu_0)$
in theory predictions, because these two models result in the same value of $\lambda_B$ as $\phi_{B,\rm I}^{+}(\omega,\mu_0)$
while the derivative $d \phi_{B}^{+}(\omega,\mu_0)/d \omega$ at $\omega=0$ takes extreme values 0 and $\infty$.
The corresponding expression of $\phi_{B}^-(\omega,\mu_0)$ for each model is determined by the equation-of-motion constraint
in the absence of contributions from  three-particle DAs \cite{Beneke:2000wa}
\begin{eqnarray}
\phi_{B}^-(\omega,\mu_0) = \int_0^1 \, { d \xi \over \xi} \,
\phi_{B}^+ \left ({\omega \over \xi}\,, \mu_0 \right ) \,.
\label{WW relation}
\end{eqnarray}
We emphasize that the above models  can only provide a reasonable description  of $\phi_{B}^{\pm}(\omega,\mu_0)$
at small $\omega$ due to the radiative tail developed from QCD corrections (except the second model)
and the mismatch of large $\omega$ behaviors predicted from  the perturbative QCD analysis \cite{Feldmann:2014ika}.
Nevertheless, the dominant contributions of  $f_{B \pi}^{+,0}(q^2)$ in the LCSR (\ref{NLO sum rules of form factors})
come from the small $\omega$ region due to the strong suppression of $\phi_{B}^{\pm}(\omega,\mu_0)$ at large $\omega$.
This is also an essential prerequisite to validate QCD factorization of the correlation function $\Pi_{\mu}$ whose
qualifications rely on the power counting scheme $\omega \sim \Lambda$  by construction.

\begin{figure}[t!bph]
\begin{center}
\includegraphics[width=0.45  \columnwidth]{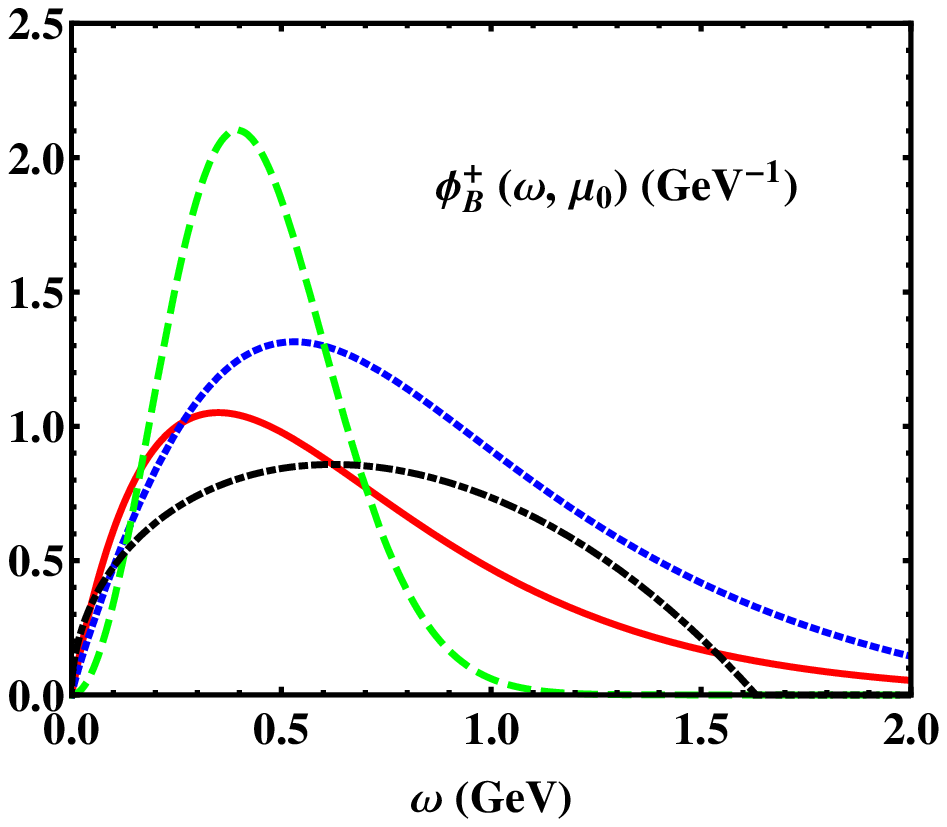} \qquad
\includegraphics[width=0.45  \columnwidth]{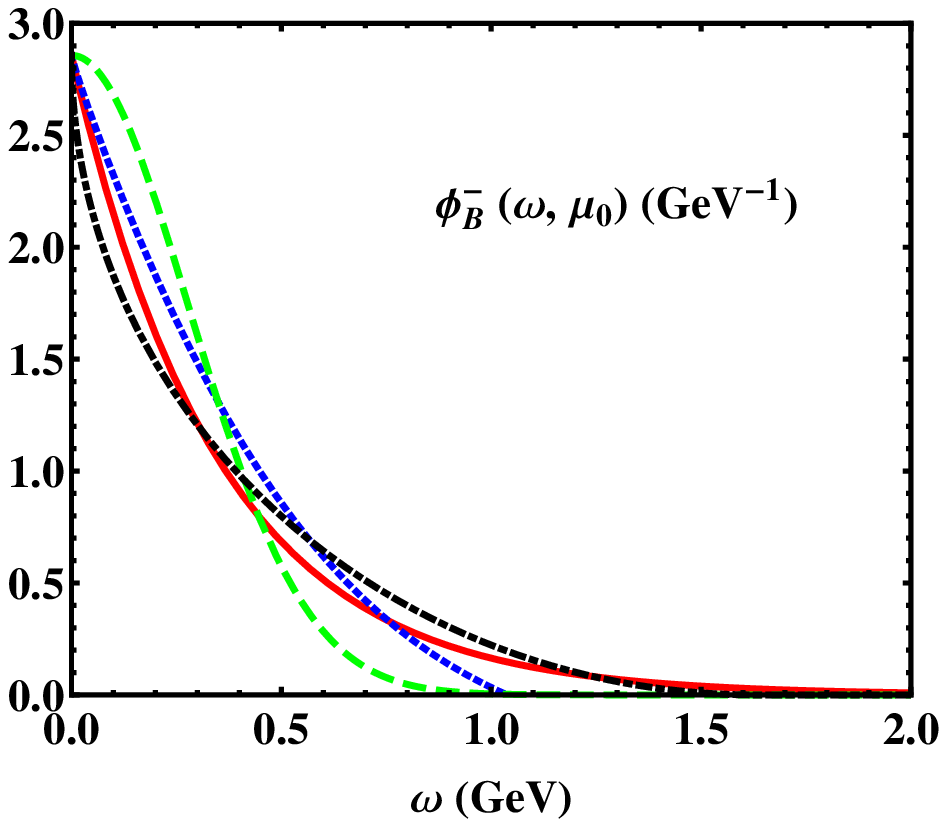}
\vspace*{0.1cm}
\caption{Four different models of $\phi_B^{+}(\omega,\mu_0)$ (left plot) and $\phi_B^{-}(\omega,\mu_0)$ (right plot).
A reference value of $\omega_0(\mu_0)=350 \, {\rm MeV}$ is taken for all the models. Solid (red), dotted (blue),
dashed (green) and dot-dashed (black) curves correspond to $\phi_{B,\rm I}^{\pm}$, $\phi_{B,\rm II}^{\pm}$,
$\phi_{B,\rm III}^{\pm}$ and $\phi_{B,\rm IV}^{\pm}$, respectively.}
\label{the shape of B-meson DAs}
\end{center}
\end{figure}

As a default value, we take the factorization scale $\mu=1.5 \, {\rm GeV}$ with a variation between $1.0 \, {\rm GeV}$
and $2.0 \, {\rm GeV}$ for the estimate of theory uncertainty. The scale dependence of  $\lambda_B^{-1}(\mu) $
and of $\sigma_B^{(1)}(\mu)$ are governed by the following evolution equations \cite{Bell:2008er,Bell:2013tfa}
\begin{eqnarray}
\frac{d}{d \ln \mu} \, \lambda_B^{-1}(\mu) &=& - \, \lambda_B^{-1}(\mu) \,
\left [\Gamma_{\rm cusp}(\alpha_s) \, \sigma_B^{(1)}(\mu)
+ \gamma_{+}(\alpha_s) \, \,  \right ] \,, \nonumber \\
\frac{d}{d \ln \mu} \, \left [\sigma_B^{(1)}(\mu) \right ]
&=& 1 + \Gamma_{\rm cusp}(\alpha_s) \, \left [(\sigma_B^{(1)}(\mu))^2
- \sigma_B^{(2)}(\mu)  \right ] \,,
\end{eqnarray}
at ${\cal O}(\alpha_s)$, where the anomalous dimension $\gamma_{+}(\alpha_s)$ is
\begin{eqnarray}
\gamma_{+}(\alpha_s) = {\alpha_s \, C_F \over 4 \, \pi} \, \left [ \gamma_{+}^{(0)}
+ \left ({\alpha_s \over 4 \, \pi} \, \right ) \gamma_{+}^{(1)}  + ... \right ]\,,
\qquad  \gamma_{+}^{(0)} =-2 \,.
\end{eqnarray}
Solving these equations yields
\begin{eqnarray}
\frac{\lambda_B(\mu_0)}{\lambda_B(\mu)} &=&
1 + {\alpha_s(\mu_0) \, C_F \over 4 \, \pi} \, \ln {\mu \over \mu_0} \,
\left [2 - 2\, \ln {\mu \over \mu_0} - 4 \, \sigma_B^{(1)}(\mu_0) \right ] + {\cal O}(\alpha_s^2)\,,
\label{lambdab evolution} \\
\sigma_B^{(1)}(\mu) &=& \sigma_B^{(1)}(\mu_0) + \ln {\mu \over \mu_0}   \,
\left ( 1 + {\alpha_s(\mu_0) \, C_F \over \pi} \,
\left [(\sigma_B^{(1)}(\mu_0))^2 - \sigma_B^{(2)}(\mu_0)  \right ]  \right ) + {\cal O}(\alpha_s^2)\,,
\hspace{0.5 cm}
\label{sigmab evolution: first log moment}
\end{eqnarray}
where we need the evolution equation of $ \sigma_B^{(2)}(\mu)$ \cite{Bell:2008er}
\begin{eqnarray}
\frac{d}{d \ln \mu} \, \left [ \sigma_B^{(2)}(\mu) \right ] &=&
2 \,  \sigma_B^{(1)}(\mu)  +  \Gamma_{\rm cusp}(\alpha_s) \,
\left [ \sigma_B^{(1)}(\mu) \, \sigma_B^{(2)}(\mu) - \sigma_B^{(3)}(\mu)
+ 4 \, \zeta_3 \,  \sigma_B^{(0)}(\mu)  \right ]  \nonumber \\
&& + {\cal O}(\alpha_s^2) \,
\label{sigmab evolution: second log moment}
\end{eqnarray}
to derive the second relation (\ref{sigmab evolution: first log moment})
with $\zeta_3 $ being the Riemann zeta function.
As mentioned before we are not aiming at the resummation of $\ln \left ({\mu / \mu_0} \right )$ here.
The two logarithmic moments  will be taken as
$\sigma_B^{(1)}( 1 \, \rm GeV) = 1.4 \pm 0.4$ \cite{Braun:2003wx} and $\sigma_B^{(2)}(1 \, \rm GeV)= 3 \pm 2$
\cite{Beneke:2011nf}. The determination of $\lambda_B(\mu_0)$, which constitutes the most important
theory uncertainty in the $B$-meson LCSR approach,  will be discussed later.
Note also that we will presume the  validity of the parameterizations of $\phi_B^{\pm}(\omega, \mu_0)$
in (\ref{four models of B-meson DAs}) at a ``hard-collinear" scale of order $1.5 \, {\rm GeV}$
to avoid a complicated RG evolution of $\phi_B^{\pm}(\omega,\mu)$ in the momentum space.
We will first determine $\lambda_B(\mu_0)$ at a ``hard-collinear" scale and then convert it
to $\lambda_B(1 \, {\rm GeV})$, using the relation in (\ref{lambdab evolution}), for a comparison
of values determined in other approaches.
To illustrate the features of four models displayed in  (\ref{four models of B-meson DAs}),
numerical examples for the small $\omega$ behaviors
of $\phi_B^{\pm}(\omega, \mu_0)$ at $\mu_0=1.5 \, {\rm GeV}$  are plotted in Fig.  \ref{the shape of B-meson DAs}
with a reference value of $\omega_0(\mu_0)=350 \, {\rm MeV}$, where $\sigma_B^{(1)}(\mu_0)$ is evaluated from
$\sigma_B^{(1)}( 1 \, \rm GeV)$ with the relation in  (\ref{sigmab evolution: first log moment}).

Now we turn to discuss the determinations of the  Borel parameter $\omega_M$  and the effective threshold $\omega_s$.
We first recall the power counting
\begin{eqnarray}
\omega_s \sim \omega_M \sim {\Lambda^2 /  m_b} \,,
\label{slaing of internal SR parameters}
\end{eqnarray}
in addition to which the following requirements
\begin{itemize}
\item {The continuum contributions in the dispersion integrals of $\Pi$ and $\tilde{\Pi}$ need to be less than 50 \%.}
\item {The sum rules for $ f_{B \pi}^{+,0}(q^2)$ are insensitive to the variation of the Borel mass $\omega_M$.
For definiteness, we impose the constraint proposed in \cite{DeFazio:2005dx}
\begin{eqnarray}
\frac{\partial \ln f_{B \pi}^{+,0}}{ \partial \ln \omega_M} \leq  35  \, \%.
\end{eqnarray}}
\item {The effective threshold needs to be close to that determined from the two-point correlation function
with pion interpolating currents:
\begin{eqnarray}
s_0 \simeq  4 \, \pi^2 \, f_{\pi}^2  \,,
\end{eqnarray}
indicated by the parton-hadron duality.
}
\end{itemize}
are implemented to determine these ``internal" sum rule parameters.
Proceeding with the above-mentioned  procedure yields
\begin{eqnarray}
M^2 \equiv n \cdot p \,\, \omega_M  = (1.25 \pm 0.25) \, {\rm GeV^2} \,, \qquad
s_0 \equiv n \cdot p \,\, \omega_s = (0.70 \pm 0.05) \, {\rm GeV^2} \,,
\end{eqnarray}
in agreement with the intervals in \cite{Khodjamirian:2006st}.

The static decay constant $\tilde{f}_B(\mu)$ entering  the sum rules (\ref{NLO sum rules of form factors})
will be traded into the QCD decay constant $f_B$ with the relation (\ref{static B-meson decay constant}),
which is evaluated from the two-point QCD sum rules at ${\cal O}(\alpha_s)$ as presented in the Appendix \ref{Appendix C}.
The Borel parameter and the effective duality threshold are taken as $\overline{M}^2=5.0 \pm 1.0  \, {\rm GeV^2}$ and
$\bar s_0=35.6^{+2.1}_{-0.9} \, {\rm GeV^2}$ \cite{Duplancic:2008ix}. The pion decay constant $f_{\pi}$ determined  from
the sum of branching ratios of $\pi^{-} \to \mu \bar \nu$ and $\pi^{-} \to \mu \bar \nu \, \gamma$
is $f_{\pi} = (130.41 \pm 0.03 \pm 0.02) \, {\rm MeV}$ \cite{Agashe:2014kda}.
To reduce the theory uncertainties induced by the ``internal" sum rule parameters
we will instead use the two-point sum rules
of $f_{\pi}$ presented in the Appendix \ref{Appendix C} for the numerical analysis.
We will return to this point later on.

A ``reasonable" choice of the factorization scale is $\mu=1.5 \,  {\rm GeV}$ with the variation
in the interval $1 \, {\rm GeV} \leq \mu \leq  2 \, {\rm GeV} $ and the hard scales $\mu_{h1}$
and $\mu_{h2}$ will be set to be equal and varied in $[m_b/2, 2\, m_b]$ around the default value $m_b$.
We take the bottom-quark mass in the ${\rm \overline{MS}}$ scheme $\bar{m}_b(\bar{m}_b)= (4.16 \pm 0.03) \, {\rm GeV}$
as adopted in \cite{Khodjamirian:2010vf}, which is still in agreement with the most recent determinations from
the non-relativistic sum rules at next-to-next-to-next-to-leading order (NNNLO) \cite{Beneke:2014pta} and
from the relativistic sum rules at ${\cal O}(\alpha_s^3)$ \cite{Dehnadi:2015fra}.

\subsection{Numerical results of the form factors $f_{B \pi}^{+,0}(q^2)$}

Now we are in a position to discuss the inverse moment $\lambda_B(1 \, {\rm GeV})$
whose determination is also of central importance in the theoretical description of  the radiative leptonic $B$-meson decays
as well as the semi-leptonic and  charmless hadronic $B$ decays. Unfortunately, the favored values of $\lambda_B(1 \, {\rm GeV})$
implied by the hadronic $B$-decay data in QCD factorization \cite{Beneke:2003zv} are not supported by the NLO
QCD sum rule calculation \cite{Braun:2003wx} (see also \cite{Beneke:2006mk} for a discussion).
Recent searches of the radiative leptonic $B \to \ell \nu \, \gamma$ ($\ell=e, \, \mu$) decays from the Belle collaboration
\cite{Heller:2015vvm} only set a boundary $\lambda_B(1 \, {\rm GeV}) > 238 \, {\rm GeV}$
\footnote{We were informed by M. Beneke that a slightly different constraint
$\lambda_B(1 \, {\rm GeV}) > 217 \, {\rm GeV}$ is obtained with the formulae presented  in  \cite{Beneke:2011nf}.}.
Given the poor knowledge of $\lambda_B(1 \, {\rm GeV})$ we will attempt to determine this parameter by matching the $B$-meson
LCSR of $f_{B \pi}^{+}(q^2)$ at zero momentum transfer to a given input value computed from a different method.
Taking $f_{B \pi}^{+} (0)=0.28 \pm 0.03$ \cite{Khodjamirian:2011ub} evaluated from the LCSR with pion DAs
(see  \cite{Imsong:2014oqa} for a recent update with somewhat larger values)  and proceeding with the matching procedure yields
\begin{eqnarray}
\omega_0(1 \, {\rm GeV}) &=& 354^{+38}_{-30} \, {\rm MeV} \,, \qquad {\rm (Model-I)} \, \nonumber  \\
\omega_0(1 \, {\rm GeV}) &=& 368^{+42}_{-32} \, {\rm MeV} \,, \qquad {\rm (Model-II)} \, \nonumber  \\
\omega_0(1 \, {\rm GeV}) &=& 389^{+35}_{-28} \, {\rm MeV} \,, \qquad {\rm (Model-III)} \, \nonumber \\
\omega_0(1 \, {\rm GeV}) &=& 303^{+35}_{-26} \, {\rm MeV} \,, \qquad {\rm (Model-IV)} \,
\label{fitted values of omega0 from sum rules}
\end{eqnarray}
where the four models correspond to that shown in (\ref{four models of B-meson DAs}).
It is evident that the extracted values of $\omega_0(1 \, {\rm GeV})$ are sensitive to the specific models
of $\phi_{B}^{\pm}(\omega,\mu_0)$ entering the LCSR of   $f_{B \pi}^{+,0}(q^2)$ in (\ref{NLO sum rules of form factors}),
because these sum rules cannot be controlled by the inverse moment $\lambda_B(1 \, {\rm GeV})$  of the DA
$\phi_{B}^{+}(\omega,\mu_0)$ to a good approximation and the precise shapes of $B$-meson DAs at small $\omega$ are
in demand for the sum rule analysis \cite{DeFazio:2007hw}. In other words,
\begin{eqnarray}
\int_0^{\omega_s} \, d \omega^{\prime} \, e^{-\omega^{\prime}/\omega_M} \, \phi_{B}^{-}(\omega,\mu_0)
\simeq \phi_{B}^{-}(\omega=0,\mu_0) \, \int_0^{\omega_s} \, d \omega^{\prime} \, e^{-\omega^{\prime}/\omega_M} \,
\label{rough approximation of LO B-meson SR}
\end{eqnarray}
should {\it not} be taken seriously as one would expect at first sight.
Mathematically, the precision of such approximation depends on the fluctuant rapidity of
$\phi_{B}^{-}(\omega,\mu_0)$ at small $\omega$.  A similar observation was already made by inspecting the
LCSR with pion DAs in the heavy quark limit \cite{DeFazio:2005dx}, where the knowledge of the two lowest-order
Gegenbauer moments is not sufficient to determine the key non-perturbative object $\phi^{\prime}_{\pi}(1)$
which is highly dependent on  the exact form of $\phi_{\pi}(u)$.
We stress that the quantity $\lambda_B(\mu_0)$ itself is well defined
at the operator level and is independent of the specific models of $\phi_{B}^{+}(\omega,\mu_0)$.
A precision determination of $\lambda_B(\mu_0)$ by other means (e.g., Lattice QCD simulation) would be of great value
to discriminate certain models of  the $B$-meson DAs.

\begin{figure}[t]
\begin{center}
\includegraphics[width=0.60  \columnwidth]{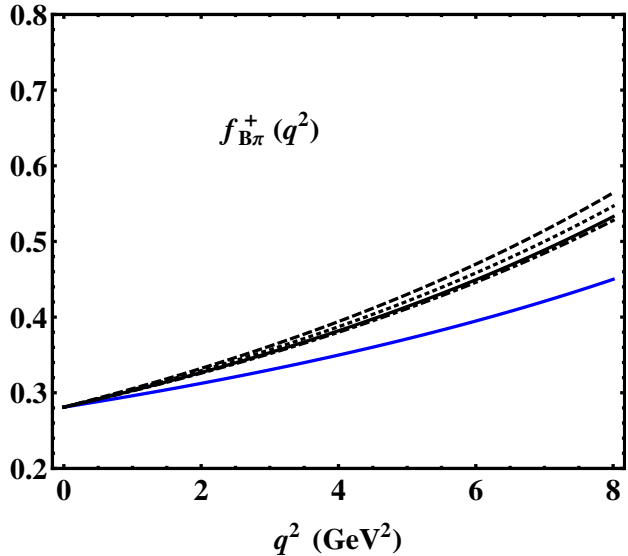}
\vspace*{0.1cm}
\caption{The shape of $f_{B \pi}^{+}(q^2)$ with the value at zero momentum transfer fixed to
the prediction from the LCSR with pion DAs, from which the parameter $\omega_0(1 \, {\rm GeV})$
is determined for a given model of $\phi_B^{\pm}(\omega,\mu_0)$.
Solid (blue), solid (black), dotted (black),  dashed (black) and dot-dashed (black)
curves are obtained from the pion LCSR  and from the ones with the $B$-meson DAs  $\phi_{B,\rm I}^{\pm}$,
$\phi_{B,\rm II}^{\pm}$, $\phi_{B,\rm III}^{\pm}$ and $\phi_{B,\rm IV}^{\pm}$, respectively.}
\label{insensitivity of the shape of form factors}
\end{center}
\end{figure}

To reduce the sizeable uncertainty from modeling the $B$-meson DAs, we will merely aim at predicting the shape
of $f_{B \pi}^{+}(q^2)$ which is insensitive to the precise behaviors of $\phi_{B}^{\pm}(\omega,\mu_0)$ at small $\omega$,
as displayed in Fig. \ref{insensitivity of the shape of form factors}, due to a large cancelation of the model dependence
in the form-factor ratio $f_{B \pi}^{+}(q^2)/f_{B \pi}^{+}(0)$.
We also find that the results of $f_{B \pi}^{+}(q^2)$ evaluated from different models of $\phi_{B}^{\pm}(\omega,\mu_0)$
are systematically lower than that obtained from the LCSR with pion DAs confirming an earlier observation from the
tree-level calculations \cite{Khodjamirian:2006st}. The underlying mechanism responsible for such discrepancy might be
due to the yet unaccounted sub-leading power corrections and/or the different ansatz  of the parton-hadron duality in the
constructions of sum rules, and we will return to this point later on.
Hereafter, we will take  $\phi_{B,\rm I}^{\pm}(\omega,\mu_0)$ as the default model to study the implications of the sum rules
in (\ref{NLO sum rules of form factors}) and the systemic uncertainty from the model dependence of the $B$-meson DAs
will be included in the final predictions of the two form factors $f_{B \pi}^{+,0}(q^2)$.

\begin{figure}[t]
\begin{center}
\includegraphics[width=0.46  \columnwidth]{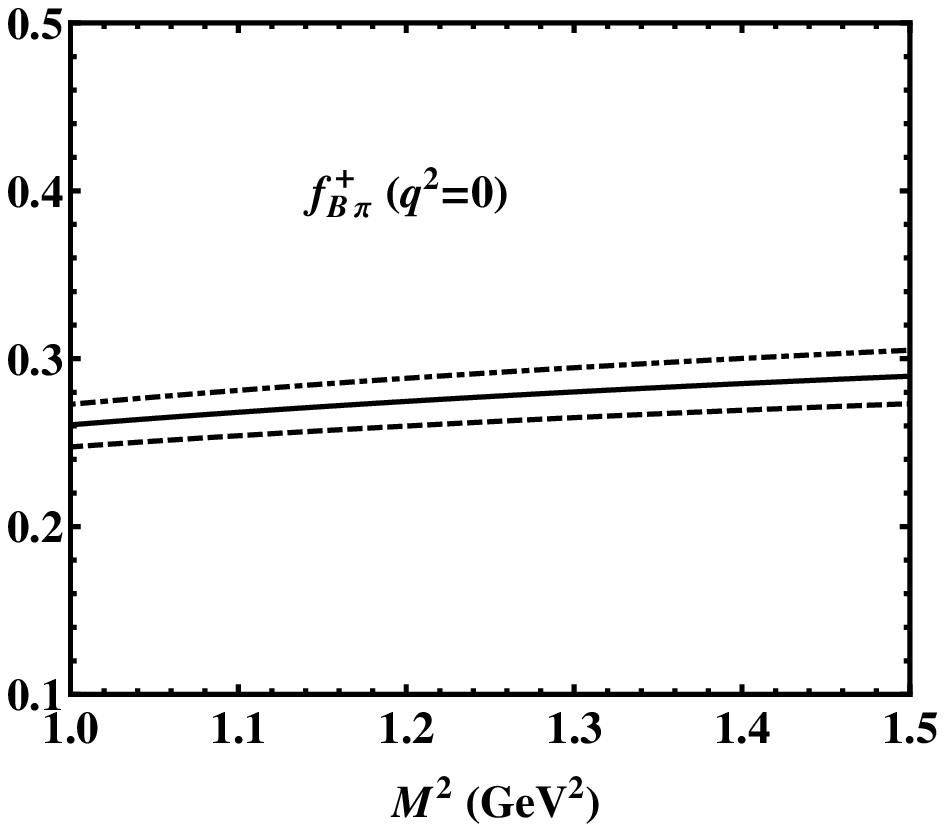} \qquad
\includegraphics[width=0.46  \columnwidth]{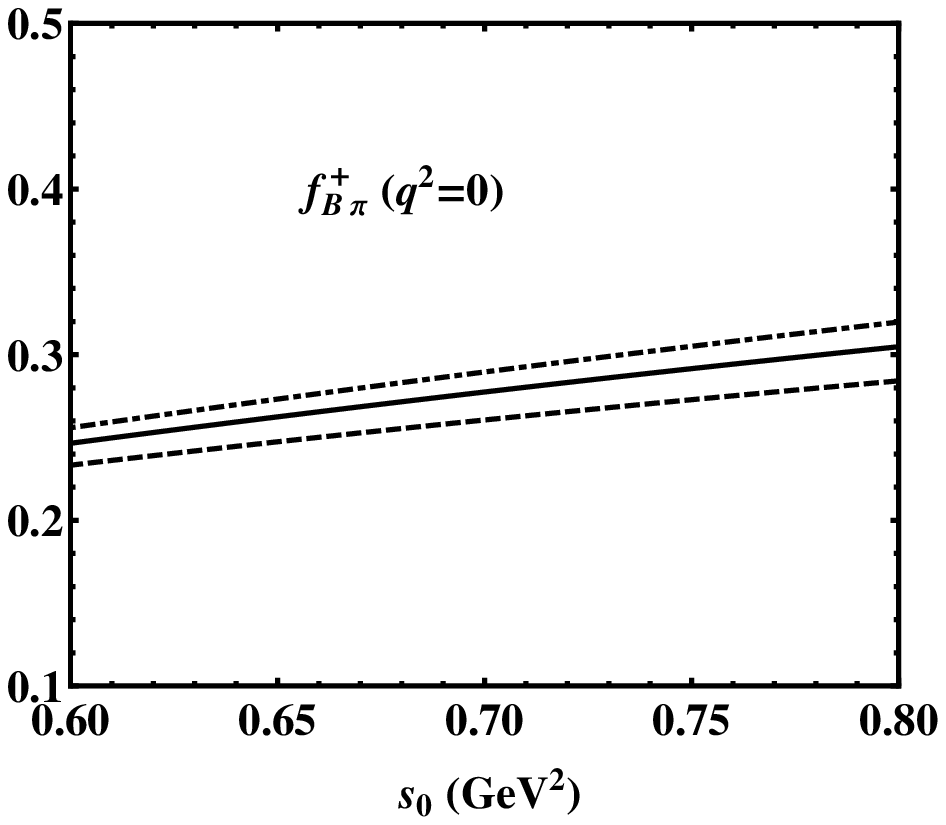}\\
\includegraphics[width=0.46  \columnwidth]{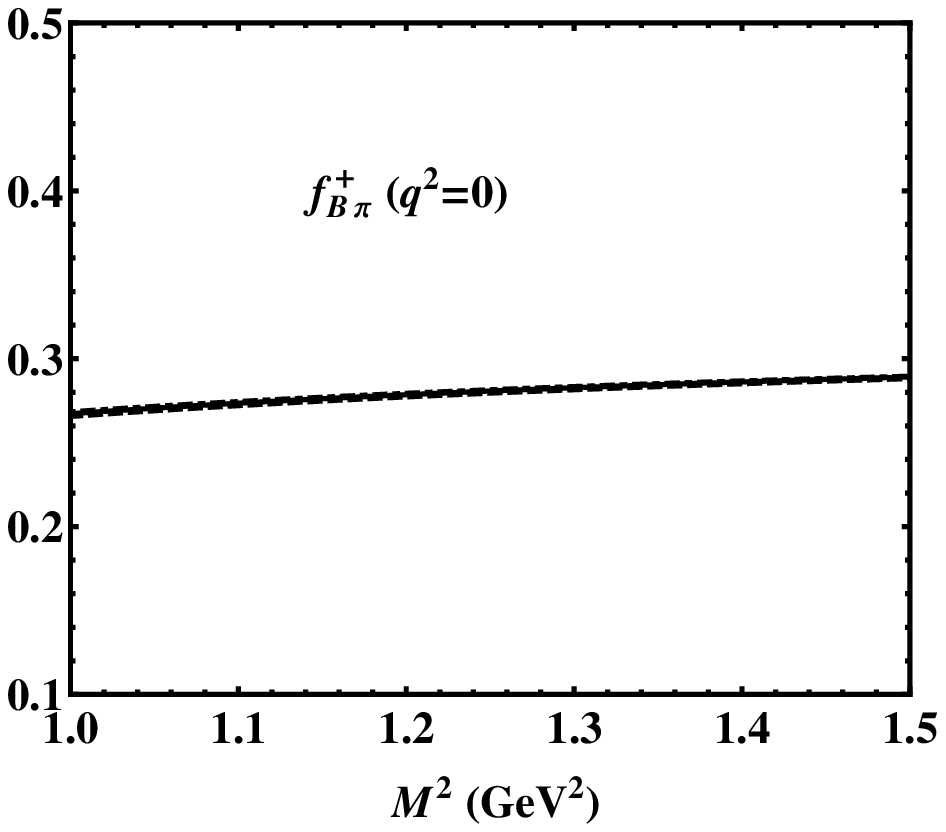} \qquad
\includegraphics[width=0.46  \columnwidth]{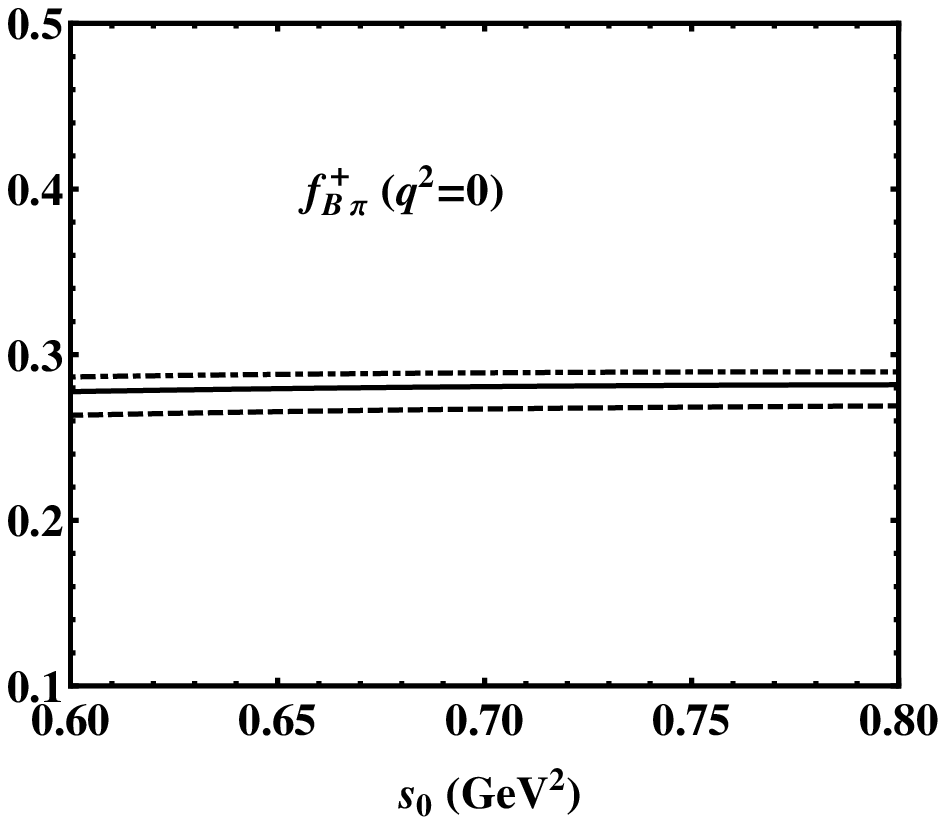}
\vspace*{0.1cm}
\caption{Dependence  of the form factor $f_{B \pi}^{+}(0)$ computed from the NLL resummation improved
sum rules (\ref{NLO sum rules of form factors}) on the Borel parameter (left panel)
and on the effective threshold (right panel). The solid, dashed and dot-dashed  curves correspond to
$s_0=0.70 \, {\rm GeV^2} \,,  \, 0.65 \, {\rm GeV^2} \,, 0.75 \, {\rm GeV^2}$
(left panel) and $M^2=1.25 \, {\rm GeV^2} \,,  \, 1.0 \, {\rm GeV^2} \,, 1.25 \, {\rm GeV^2}$ (right panel), respectively.}
\label{internal SR parameter dependences}
\end{center}
\end{figure}

To demonstrate the stability of the LCSR predictions we show the dependencies of $f_{B \pi}^{+}(q^2)$
on the ``internal" sum rule parameters $M^2$ and $s_0$ in Fig. \ref{internal SR parameter dependences}
where the two plots on the top are obtained from  NLL resummation improved sum rules (\ref{NLO sum rules of form factors})
with $f_{\pi}$ extracted from the experimental data as explained before; while the two-point QCD sum rules
of $f_{\pi}$ are substituted in the LCSR to produce the two plots on the bottom.
One can readily find that the systematic uncertainties induced by the Borel parameter and the effective threshold
are significantly reduced in the latter case, albeit with the absence of a model-independent justification of correlating
the ``internal" parameters in the two types of sum rules.

\begin{figure}[t!bph]
\begin{center}
\includegraphics[width=0.46  \columnwidth]{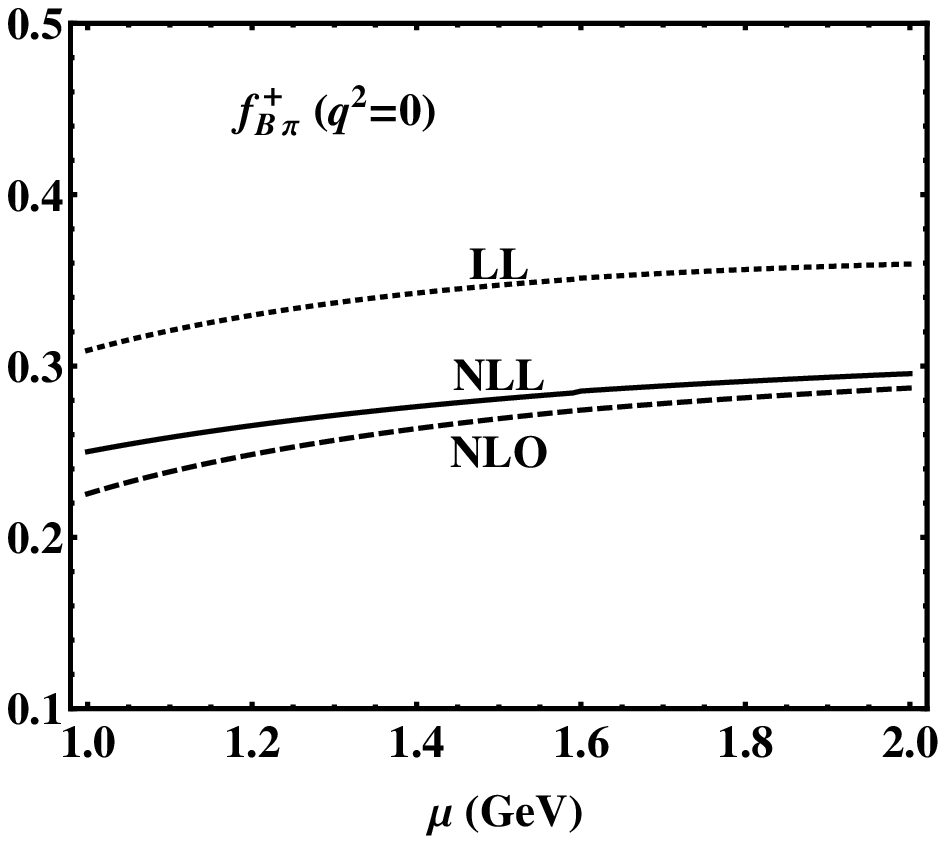} \qquad
\includegraphics[width=0.46  \columnwidth]{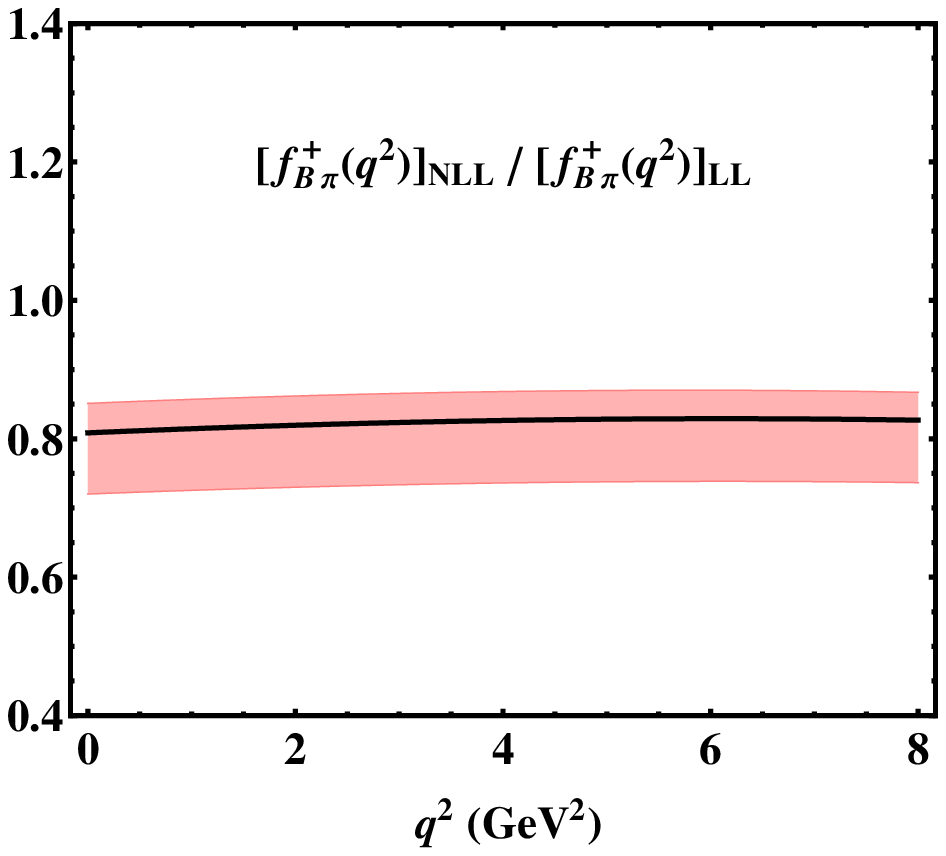}
\vspace*{0.1cm}
\caption{Hard-collinear scale dependence of the form factor $f_{B \pi}^{+}(0)$
(left panel) and  $q^2$ dependence of the NLO radiative correction to $f_{B \pi}^{+}(q^2)$
with both hard and hard-collinear scales varied in the allowed regions
as explained in the text (right panel).}
\label{radiative correction and scale dependence}
\end{center}
\end{figure}

Now we come to investigate the factorization-scale dependence of the NLL and the leading-logarithmic (LL) resummation improved LCSR
for  $f_{B \pi}^{+,0}(q^2)$, where the LL predictions can be achieved by employing the cusp anomalous dimension
at ${\cal O}(\alpha_s^2)$ as well as  $\gamma(\alpha_s)$ and $\tilde{\gamma}(\alpha_s)$ at the one-loop order
in the evolution functions $U_1(n \cdot p,\mu_{h1},\mu )$ and  $U_2(\mu_{h2},\mu )$ of  (\ref{NLO sum rules of form factors}).
Figure \ref{radiative correction and scale dependence} shows that the scale dependence of the NLL predictions
is {\it not} significantly reduced compared to the LL approximation
for the hard-collinear scale varied in the interval $[1.0 \,, 2.0] \, {\rm GeV}$
and the discrepancy of  the scale dependency for  the NLL and LL predictions
will be more visible for a somewhat ``unrealistic" hard-collinear scale $\mu < 1.0 \, {\rm GeV}$ which is therefore excluded in the plot.
The dominant radiative effect arises from the NLO QCD corrections to  perturbative matching coefficients instead of
resummation of the parametrically  large logarithms in the heavy quark limit. However, the resummation improvement  stabilizes the
factorization-scale dependence in the allowed region and strengthens the predictive power of the LCSR method.
One can also find that the NLO QCD correction is stable against the momentum-transfer dependence of  $f_{B \pi}^{+}(q^2)$
in contrast to the case of $B \to \gamma \ell \nu$ \cite{Beneke:2011nf}.

\begin{figure}[t!bph]
\begin{center}
\includegraphics[width=0.46  \columnwidth]{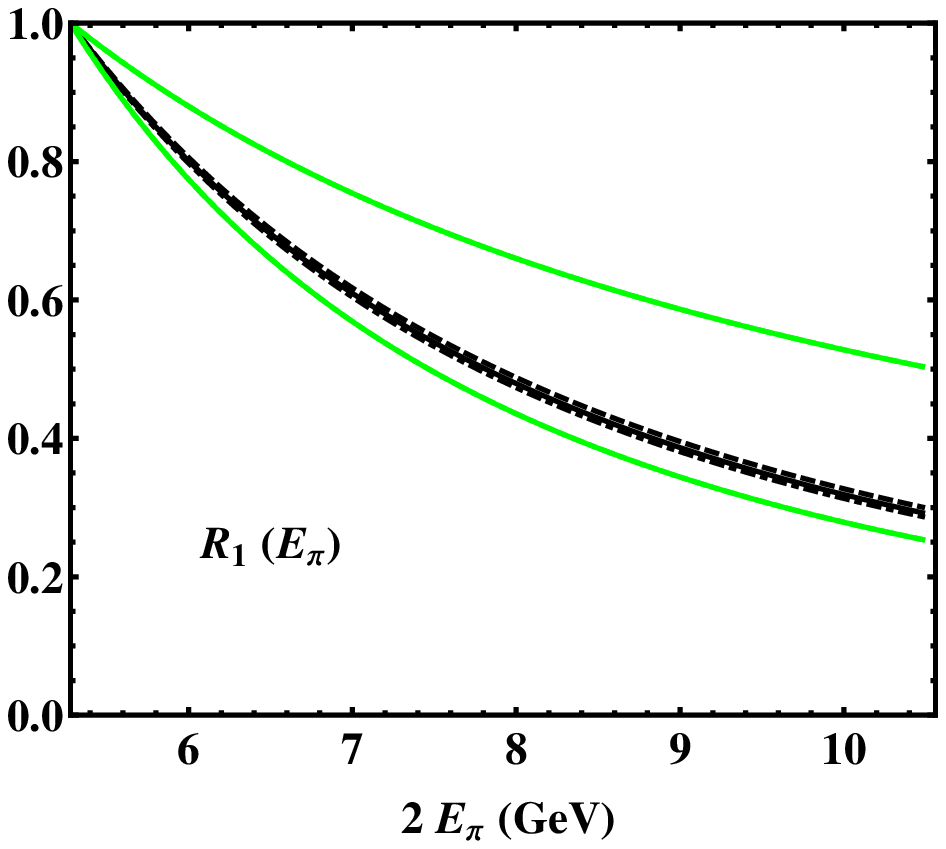} \qquad
\includegraphics[width=0.46  \columnwidth]{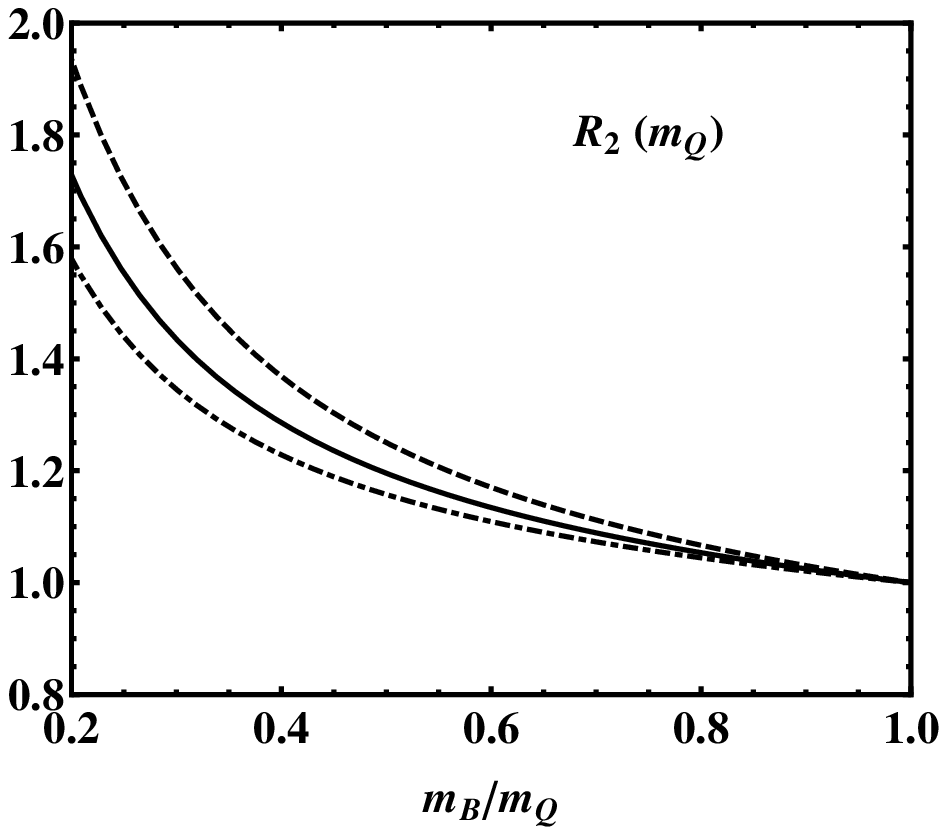}
\vspace*{0.1cm}
\caption{{\it Left:} The  pion energy dependence of the ratio $R_1(E_{\pi})$.
The black curves correspond to the sum rule predictions with the Borel mass  taken as $1.25 \, {\rm GeV}$ (solid),
$1.0 \, {\rm GeV}$ (dashed) and $1.5 \, {\rm GeV}$ (dot-dashed).  The two green curves illustrate a  pure
$1/E_{\pi}$ and a pure $1/E_{\pi}^2$ dependence.
{\it Right:} The heavy-quark mass dependence of the quantity $R_2(m_Q)$.
The three curves are predicted from the sum rules with the Borel mass varied between $1.0 \, {\rm GeV}$
and  $1.5 \, {\rm GeV}$ around the default value $1.25 \, {\rm GeV}$. Note that in both plots we take the pion decay constant
$f_{\pi}= 130.41  \, {\rm MeV}$ \cite{Agashe:2014kda} instead of using the two-point QCD sum rules in the Appendix \ref{Appendix C}
as done in the remainder of this paper. }
\label{R1 and R2}
\end{center}
\end{figure}

Understanding  the pion energy  and  the heavy quark mass dependencies of the form factors $f_{B \pi}^{+,0}(q^2) $
are of both theoretical and phenomenological interest in that different competing mechanisms appear in the
theory description of heavy-to-light form factors in the large recoil region and a better control of the form factor shapes
can be achieved by incorporating the energy-scaling laws and the Lattice (sum-rule) calculations of  form factors
at high (low) $q^2$. In accordance with the factorization formulae \cite{Beneke:2005gs}
\begin{eqnarray}
f_{B \pi}^{i} (E_{\pi})&=& C_i(E_{\pi}) \, \xi_{\pi}(E_{\pi}) + \int \, d \tau \, C_i^{(B1)}(E_{\pi}\,,\tau) \,
\Xi_a(\tau \,, E_{\pi}) \,, \nonumber \\
\Xi_a(\tau \,, E_{\pi}) &=& \int_0^{\infty} \, d \omega \, \int_0^ 1 d u \, J_{\|}(\tau, \, u \,, \omega) \,
\tilde{f}_B(\mu) \, \phi_B^{+}(\omega,\mu) \, f_{\pi} \, \phi_{\pi}(u,\mu) \,,
\label{QCDF for heavy-to-light FFs}
\end{eqnarray}
one can readily deduce that both terms in the first line of (\ref{QCDF for heavy-to-light FFs})
scale as $1/E_{\pi}^2$ in the large energy limit and as $(\Lambda/m_b)^{3/2}$
in the heavy quark limit \cite{Beneke:2000wa,Hill:2005ju}. It is our objective to verify such scaling behaviors from
the  NLL resummation improved sum rules (\ref{NLO sum rules of form factors}).
In doing so we define the following two ratios \cite{DeFazio:2007hw}
\begin{eqnarray}
R_1(E_{\pi}) \equiv  \frac{f_{B \pi}^{+}(E_{\pi})}{f_{B \pi}^{+}(m_B/2)} \,, \qquad
R_2(m_Q) \equiv \frac{m_Q \, \tilde{f}_B(\mu)}{m_B \, \tilde{f}_Q(\mu)} \, \frac{f_{Q \pi}^{+}(m_Q/2)}{f_{B \pi}^{+}(m_B/2)}\,,
\end{eqnarray}
where the argument of the form factor  refers to $n \cdot p / 2$ different from that ($q^2$)
used in the remaining of this paper, the pre-factors in the definition of $R_2(m_Q)$ is introduced
to achieve a simple scaling $R_2(m_Q) \to 1$ in the heavy quark limit.
The expression of $f_{Q \pi}^{+}(n \cdot p / 2)$ can be obtained from
Eq. (\ref{NLO sum rules of form factors}) via the replacement $(m_b \,, m_B) \to (m_Q \,, m_Q)$.
One should also  keep in mind  that the scalings of the ``internal" sum rule parameters shown in
(\ref{slaing of internal SR parameters}) need to be respected when deriving the power-counting laws of  the large energy
and  the heavy quark mass dependencies. We present the sum rule predictions for the two ratios $R_1(E_{\pi})$ and $R_2(m_Q)$
in Fig. \ref{R1 and R2}, where we observe that the yielding energy dependence is indeed close to the $1/E_{\pi}^2$ behavior
and the heavy-quark mass scaling is also justified from the LCSR with $B$-meson DAs.
However, the sum rule results become more and more instable at $m_Q > 2 \, m_B$ where the Borel parameter dependence is not
under control any more as displayed in Fig. \ref{R1 and R2}, and one can also find that the continuum effect dominates over the ground state
contribution in the dispersion integral of the correlation function $\Pi_{\mu}$ (see also the discussions in \cite{DeFazio:2007hw}).

One more comment concerns the ratio $R_2(m_Q)$ which allows to estimate  $D \to \pi$ form factors from the corresponding $B$-meson
cases in the leading-power approximation. However, this statement needs to be taken with a grain of salt in reality
in view of the sizeable power correction in the decay-constant ratio $f_B/f_D$ which is determined as
\begin{eqnarray}
\frac{f_B}{f_D} &=& \left [{m_D \over m_B} \right ]^{1/2} \,
\left [ {\alpha_s(m_c) \over \alpha_s(m_b)} \right ]^{\frac{\tilde{\gamma}_0 }{2 \,\beta_0} \, C_F} \,
\bigg \{1 +{[\alpha_s(m_b)-\alpha_s(m_c)] \, C_F \over 4 \, \pi}\,  \nonumber \\
&& \hspace{4.5 cm} \times \left [- 2 + \left (  {\tilde{\gamma}^{(1)} \over 2 \, \beta_0}
- {\tilde{\gamma}^{(0)} \, \beta_1 \over 2 \, \beta_0^2 } \right )  \right ] \bigg \}\, \nonumber \\
& \simeq & 0.69 \,,
\end{eqnarray}
significantly lower than the QCD sum rule prediction $0.93 \leq f_B/f_D \leq 1.19$ \cite{Gelhausen:2013wia}
and the Lattice QCD result computed from $f_B=(190.5 \pm 4.2) \, {\rm MeV}$ and $f_D=(209.3 \pm 3.3) \, {\rm MeV}$ with
$N_f=2+1$ \cite{Aoki:2013ldr}.

To validate the light-cone expansion of the correlation function $\Pi_{\mu}$ in the region
$|\bar n \cdot p| \sim {\cal O} (\Lambda)$ we need to keep the photon energy as a hard scale,
above the practical value of a hard-collinear scale $\sim 1.5 \, {\rm GeV}$, then the LCSR with $B$-meson
DAs can be trusted at $q^2 \leq  q_{max}^2 = 8 \, {\rm GeV^2}$ (see \cite{Khodjamirian:2006st} for more detailed discussions)
on the conservative side. To extrapolate the computed form factors from the LCSR method at large recoil toward
large momentum transfer $q^2$ we apply the  $z$-series parametrization based upon the analytical and asymptotic properties
of the form factors, where the entire cut $q^2$-plane is mapped onto the unit disk $|z(q^2, t_0)|<1$ via the conformal transformation
\begin{eqnarray}
z(q^2, t_0) = \frac{\sqrt{t_{+}-q^2}-\sqrt{t_{+}-t_0}}{\sqrt{t_{+}-q^2}+\sqrt{t_{+}-t_0}}\,,
\end{eqnarray}
where $t_{+}=(m_B + m_{\pi})^2$ denotes the threshold of continuum states in the $B^{\ast}(1^{-})$ meson channel.
The free parameter $t_0 \in ( - \infty \,,  t_{+})$ determines the value of $q^2$ mapped onto the origin in the $z$ plane
and can be adjusted to minimize the $z$ interval from mapping the LCSR region
$q_{min}^2 \leq q^2 \leq q_{max}^2$. For definiteness, we follow \cite{Khodjamirian:2011ub}
\begin{eqnarray}
t_0 = t_{+}^2 - \sqrt{t_{+}-t_{-}} \, \sqrt{t_{+} -q_{min}^2}\,,
\end{eqnarray}
with $q_{min}^2= -6.0 \, {\rm GeV^2}$ and $t_{-} \equiv (m_B - m_{\pi})^2$,
and we also refer to \cite{Bourrely:2008za,Khodjamirian:2011ub}
and references therein for more discussions on different versions of the $z$-parametrization
and to \cite{Imsong:2014oqa} for a new implementation of the unitary bounds for the vector $B \to \pi$
form factor.

Employing the $z$-series expansion and taking into account the threshold $t_{+}$ behavior
implies the following parametrization of the vector form factor \cite{Khodjamirian:2011ub}
\begin{eqnarray}
f^+_{B\pi}(q^2) = \frac{f^+_{B\pi}(0)}{1-q^2/m_{B^*}^2}
\Bigg\{1+\sum\limits_{k=1}^{N-1}b_k\,\Bigg(z(q^2,t_0)^k- z(0,t_0)^k \nonumber\\
-(-1)^{N-k}\frac{k}{N}\bigg[z(q^2,t_0)^N- z(0,t_0)^N\bigg]\Bigg)\Bigg\}
\label{z parametrization of fplus}\,,
\end{eqnarray}
where the expansion coefficients $b_k$ can be determined by matching the computed $f^+_{B\pi}(q^2)$
at low $q^2$ onto Eq. (\ref{z parametrization of fplus}) and we truncate the $z$-series  at $N=2$
in the practical calculation. One can keep more terms of the $z$ expansion in the fitting program
to quantify the systematic uncertainty induced by the truncation, however, one could also run the risk of
increasingly unconstrained fit when introducing too many parameters \cite{Hill:2010yb}, and we will leave a refined
statistic analysis for the future. 
Along this line, one can further parameterize the scalar form factor as
\begin{eqnarray}
f^0_{B\pi}(q^2) = f^0_{B\pi}(0) \Bigg\{ 1+\sum\limits_{k=1}^{N} \tilde{b}_k\,
\Bigg(z(q^2,t_0)^k- z(0,t_0)^k\Bigg)\Bigg\}
\label{z parametrization of f0}\,,
\end{eqnarray}
where the pole factor is removed because the lowest scalar $B(0^+)$ meson is located above the continuum cut $t_{+}$,
and the series is truncated at $N=1$ with $f^0_{B\pi}(0)=f^+_{B\pi}(0)$  by definition.
We also implement the unitary bound constraints on the coefficients of $b_k$ and $\tilde{b}_k$
in the fitting program, which are however too weak to take effect for the truncation at $N=2$
for $f^{+}_{B\pi}(q^2)$ and at  $N=1$ for $f^{0}_{B\pi}(q^2)$.

\begin{figure}[t!bph]
\begin{center}
\vspace{1.0 cm}
\includegraphics[width=1.0  \columnwidth]{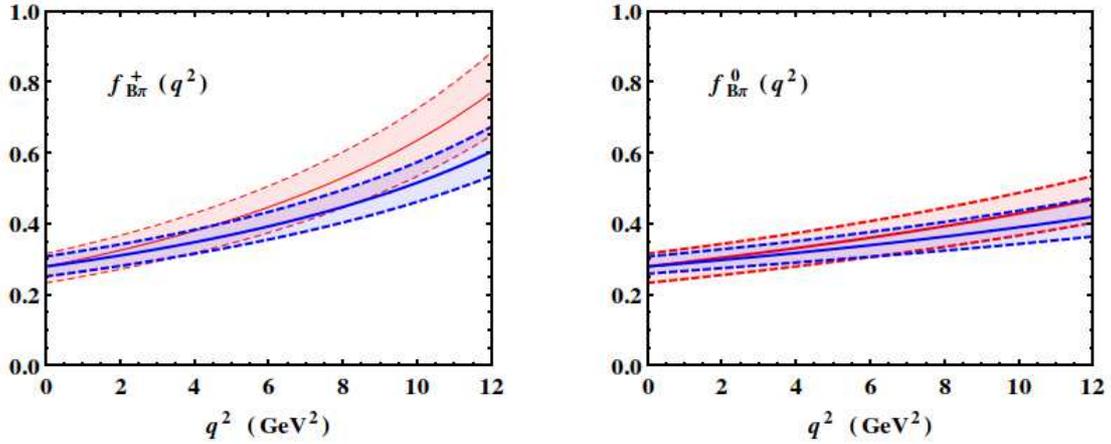}
\vspace*{0.1cm}
\caption{$q^2$ dependence of the vector form factor $f^{+}_{B\pi}(q^2)$ ({\it left}) and of the scalar
form factor $f^{0}_{B\pi}(q^2)$ ({\it right}). The pink (solid) and the blue (solid) curves  are computed
from the LCSR with $B$-meson DAs and with pion DAs, respectively, and the shaded regions indicate the estimated
uncertainties. }
\label{final form factor shape}
\end{center}
\end{figure}

Figure \ref{final form factor shape} shows  the $q^2$ dependence of the two form factors $f^{+,0}_{B\pi}(q^2)$ computed from
the LCSR with $B$-meson DAs at $q^2 < 8 \, {\rm GeV^2}$ with an extrapolation to $q^2=12 \, {\rm GeV^2}$
(pink band) using the $z$ expansion, and theoretical predictions from the LCSR with pion DAs \cite{Khodjamirian:2011ub}
{\it without} any extrapolation at $q^2 < 12 \, {\rm GeV^2}$ (blue band) are also presented for a comparison.
It is evident that the predict shape of $f^{0}_{B\pi}(q^2)$ is in good agreement with that computed from the sum rules
with pion DAs while a similar comparison for the vector form factor  $f^{+}_{B\pi}(q^2)$ reveals perceptible discrepancies
in particular at high $q^2$  as already observed before.

\begin{figure}[t!bph]
\begin{center}
\includegraphics[width=0.60  \columnwidth]{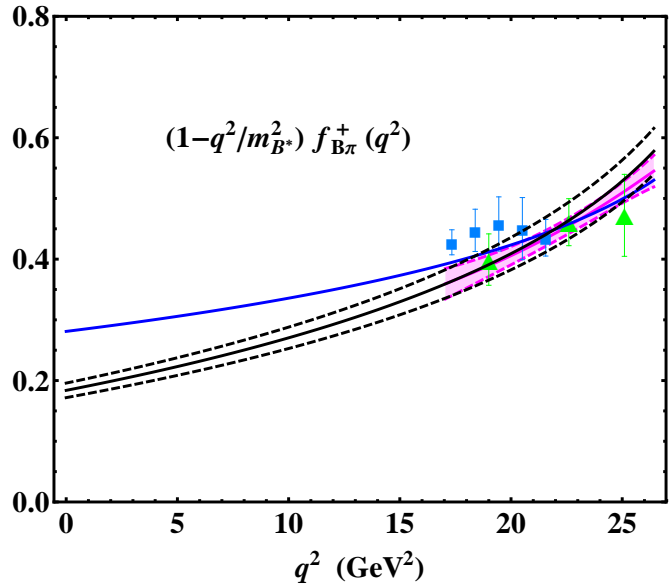}
\vspace*{0.1cm}
\caption{$q^2$ dependence of the re-scaled form factor $(1-q^2/m^2_{B^{\ast}})\, f^{+}_{B\pi}(q^2)$
predicted from the sum rules with $B$-meson DAs and the parameter $\omega_0(1 \, {\rm GeV})$ determined by
matching the Lattice point at $q^2= 17.34 \, {\rm GeV^2}$ \cite{Lattice:2015tia}.
The Lattice data are taken from Fermilab/MILC \cite{Lattice:2015tia} (pink band),
HPQCD \cite{Dalgic:2006dt} (blue squares), RBC/UKQCD \cite{Flynn:2015mha}
(green triangles).  The blue curve is again obtained from the sum rules with pion DAs \cite{Khodjamirian:2011ub}
with central inputs.}
\label{fplus shape from the new fit}
\end{center}
\end{figure}

As the first attempt to understand this issue  it would be interesting
to inspect influence of the matching condition of $\omega_0(1 \, {\rm GeV})$, described before
Eq. (\ref{fitted values of omega0 from sum rules}), on the final predictions of the form-factor shapes.
Taking $f_{B \pi}^{+} (17.34 \, {\rm GeV^2})=0.94^{+0.06}_{-0.07}$ from Fermilab/MILC collaborations
\cite{Lattice:2015tia} as an input and proceeding with the matching procedure we obtain
$\omega_0(1 \, {\rm GeV}) = 525 \pm 29 \, {\rm MeV}$ for the default model of $\phi_B^{\pm}(\omega,\mu_0)$, which is significantly
larger than the determinations displayed in  (\ref{fitted values of omega0 from sum rules}).
The resulting shape of  the re-scaled form factor $(1-q^2/m^2_{B^{\ast}})\, f^{+}_{B\pi}(q^2)$ is presented in
Fig. \ref{fplus shape from the new fit} where the Lattice data from HPQCD Collaboration \cite{Dalgic:2006dt},
RBC/UKQCD Collaborations \cite{Flynn:2015mha} and Fermilab/MILC Collaborations
\cite{Lattice:2015tia} are also displayed for a comparison. One can readily observe that the higher
$q^2$ shape of $f^{+}_{B\pi}(q^2)$  predicted by  Fermilab/MILC  \cite{Lattice:2015tia}
lies in between that obtained from the LCSR with $B$-meson DAs and the one with pion DAs.
In fact, the recent Lattice calculations \cite{Lattice:2015tia} (see figure 24 there) already revealed a faster growing
form factor $f^{+}_{B\pi}(q^2)$ in the momentum  transfer squared compared to that computed from
the LCSR with pion DAs. We should stress that the new matching procedure discussed here needs to be interpreted more carefully,
because extrapolating the sum rule computations toward large momentum transfer with the $z$ expansion
is also implemented for the sake of determining  $\omega_0(1 \, {\rm GeV})$ from the Lattice input
at a high $q^2=17.34 \, {\rm GeV^2}$.

\begin{table}[t!bph]
\begin{center}
\begin{tabular}{|c|c|c|c|c|c|c|c|c|}
  \hline
  \hline
&&&&&&&& \\[-3.5mm]
 Parameter & default & $\omega_0$ & $\sigma_B^{(1)}$ & $\mu$ & $\mu_{h1(2)}$ &
$\{M^2, s_0\}$ &
$\{\overline{M}^2, \overline{s}_0\}$ &  $\phi_B^{\pm}(\omega)$ \\
  \hline
    &&&&&&&& \\[-1mm]
  $f^+_{B\pi}(0)$ &$0.281$ & $^{-0.029}_{+0.027}$ & $^{-0.008}_{+0.008}$ &$^{+0.015}_{-0.031}$&
$^{+0.005}_{-0.004}$&
$^{+0.008}_{-0.014}$ & $^{+0.012}_{-0.007}$ & - \\
  &&&&&&&& \\[-2mm]
\hline
&&&&&&&& \\[-3mm]
  $b_1$ & $-3.92$ & $^{-0.10}_{+0.09}$ & $^{-0.03}_{+0.03}$ & $^{+0.06}_{-0.00}$ &
$^{-0.01}_{+0.06}$ & $^{+0.08}_{-0.09}$ & -& $^{+0.14}_{-0.95}$ \\[2mm]
\hline
&&&&&&&& \\[-2mm]
   $\tilde{b}_1$ &$-5.37$ &$^{-0.13}_{+0.12}$  & $^{-0.03}_{+0.03}$ & $^{+0.21}_{-0.41}$ &
$^{+0.05}_{-0.00}$  & $^{+0.11}_{-0.12}$ & - & $^{+0.17}_{-1.15}$ \\[2mm]
    \hline
    \hline
\end{tabular}
\end{center}
\caption{Fitted values of the form factor $f^{+}_{B \pi}(q^2)$
at zero momentum transfer and of the slop parameters $b_1$, $\tilde{b}_1$
entering the $z$ expansions (\ref{z parametrization of fplus}) and
(\ref{z parametrization of f0}). The notation ``default"   means that
all the parameters are taken as the central values in the numerical evaluation.
Note that the central value of $f^{+}_{B \pi}(0)$
is taken from \cite{Khodjamirian:2011ub} to determine $\omega_0(1 \, {\rm GeV})$
from the matching condition, whose variations induce the combined uncertainty estimated
in \cite{Khodjamirian:2011ub} by construction. Negligible uncertainties induced by
variations of the remaining parameters are not shown  but are taken into account
in the combined uncertainty.}
\label{tab of fitted parameters}
\end{table}

We present the fitted values of $f^{+}_{B\pi}(0)$ and of the slop parameters $b_1$ and $\tilde{b}_1$
in Table \ref{tab of fitted parameters} where breakdown of the numerically important uncertainties
is also shown. One observes that the very limited information of $\phi_B^{\pm}(\omega)$ (indicated
by the variations of $\omega_0$ and by the model dependence of $\phi_B^{\pm}(\omega)$ in the table)
remains the  most significant source of theory uncertainties.
Comparing the new predictions in Table \ref{tab of fitted parameters} with that  of  \cite{Khodjamirian:2011ub}
we  notice again the greater slop parameters for both the vector and scalar form factors determined by the
LCSR with $B$-meson DAs. We would be  benefited from precision measurements of the
normalized $q^2$ distributions in $B \to \pi \ell \nu_l$ at the Belle-II experiment to gain some insight
into this problem on the phenomenological side.

\subsection{$|V_{ub}|$ and the normalized $q^2$ distributions of $B \to \pi \ell \nu_{\ell}$}

The CKM matrix element $|V_{ub}|$ can be determined from the (partial) branching fraction of
$B \to \pi \ell \nu_{\ell}$
\begin{eqnarray}
\frac{d \Gamma}{d q^2} \, (B \to \pi \ell \nu_{\ell}) &=& \frac{G_F^2 |V_{ub}|^2}{24 \pi^3 \, q^4 \, m_B^2}
(q^2-m_l^2)^2 \, |\vec{p}_{\pi}| \, \, 
 \bigg [  \left (1 + {m_l^2 \over 2 \, q^2} \right) \, m_B^2 \,|\vec{p}_{\pi}|^2  \,
|f_{B \pi}^{+}(q^2)|^2  \nonumber \\
&& + {3 \, m_l^2 \over 8 q^2} \, (m_B^2-m_\pi^2)^2 \, |f_{B \pi}^{0}(q^2)|^2  \bigg ]\,,
\label{differential distribution formula}
\end{eqnarray}
where $|\vec{p}_{\pi}|$ is the magnitude of the pion three-momentum in the $B$-meson rest frame,
and in the massless lepton limit the above equation can be reduced to
\begin{eqnarray}
\frac{d \Gamma}{d q^2} \, (B \to \pi \mu \nu_{\mu}) &=& \frac{G_F^2 |V_{ub}|^2}{24 \pi^3}
\, |\vec{p}_{\pi}|^3 \, \, |f_{B \pi}^{+}(q^2)|^2 \,.
\label{differential distribution formula: muon}
\end{eqnarray}
Following \cite{Khodjamirian:2011ub} we define the following quantity
\begin{eqnarray}
\Delta \zeta (0, q_0^2)=  \frac{G_F^2}{24 \pi^3}
\, \int_{0}^{q_0^2} \, d q^2 \,
|\vec{p}_{\pi}|^3 \, \, |f_{B \pi}^{+}(q^2)|^2  \,,
\end{eqnarray}
which allows a straightforward extraction of $|V_{ub}|$ when compared to experimental measurements
for the partial branching ratio of $B \to \pi \mu \nu_{\mu}$ integrated over the same kinematic region.
Implementing the computed form factor $f_{B \pi}^{+}(q^2)$ from the sum rules with $B$-meson DAs
and performing the extrapolation to $q^2=12 \, {\rm GeV^2}$ with the  $z$-series parametrization  yield
\begin{eqnarray}
\Delta \zeta (0, 12 \, {\rm GeV^2}) &=& 5.89 \,\,{}^{+1.12}_{-1.10}\Big|_{\omega_0}
\,\,{}^{+0.30}_{-0.29}\Big|_{\sigma_B^{(1)}}
\,\,{}^{+0.60}_{-1.22}\Big|_{\mu}
\,\,{}^{+0.21}_{-0.21}\Big|_{\mu_{h1(2)}}
\,\,{}^{+0.34}_{-0.53}\Big|_{M,s_0}
\,\,{}^{+0.52}_{-0.25}\Big|_{\overline{M},\overline{s}_0}
\,\, \mbox{ps}^{-1}\nonumber\\
&=&5.89^{+1.63}_{-1.82}~\mbox{ps}^{-1}\,,
\label{Delta Zeta 0 to 12}
\end{eqnarray}
where the negligibly small uncertainties from variations of the remaining parameters
are not presented but are included in the final combined uncertainty.

\begin{figure}
\begin{center}
\vspace{0.5 cm}
\includegraphics[width=0.70  \columnwidth]{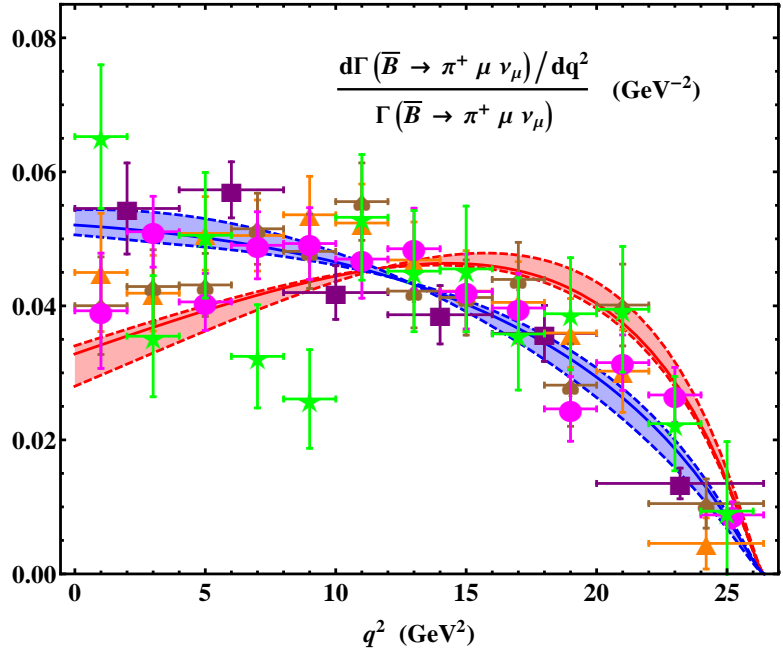} \\
\vspace{1.5 cm}
\includegraphics[width=0.70  \columnwidth]{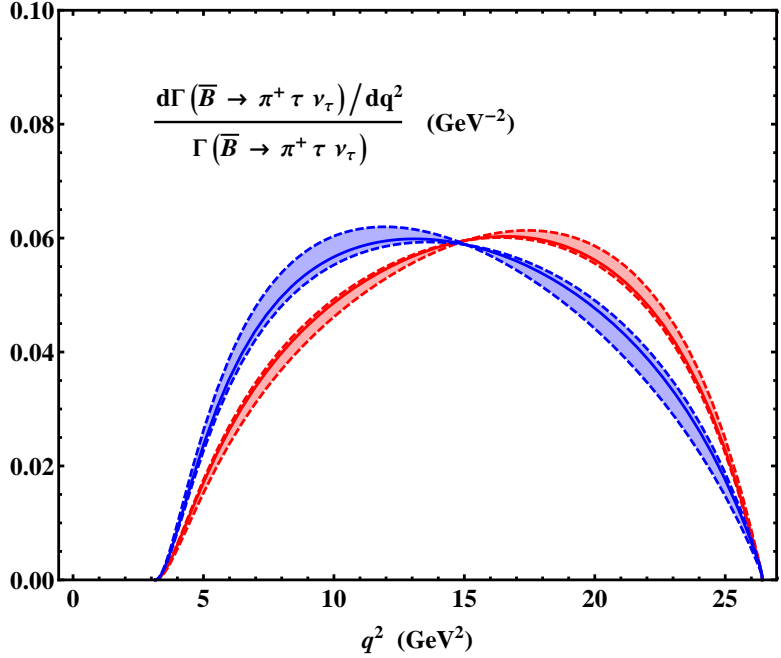}
\vspace*{0.1cm}
\caption{{\it Top:} The normalized  differential $q^2$ distribution of $B \to \pi \mu \nu_{\mu}$
computed from (\ref{differential distribution formula: muon}) with the form factor $f_{B \pi}^{+}(q^2)$
predicted from the sum rules with $B$-meson DAs and fitted to the  $z$-parametrization (red band),
and that predicted from the sum rules with pion DAs and  $z$-parametrization (blue band).
The experimental data bins are taken from \cite{Lees:2012vv} (brown spade suits),
\cite{Sibidanov:2013rkk} (green five-pointed stars),
\cite{delAmoSanchez:2010af} (purple squares),
\cite{delAmoSanchez:2010zd} (orange triangles),  \cite{Ha:2010rf}
(magenta full circles).
{\it Bottom:} The normalized differential distribution of $B \to \pi \tau \nu_{\tau}$.
The red and blue bands are obtained with the form factors computed from the $B$-meson LCSR
and from the pion LCSR, respectively. }
\label{normalized q2 distribution of B to pi l nu}
\end{center}
\end{figure}

Employing experimental measurements of the integrated branching ratio
\begin{eqnarray}
\Delta {\cal BR} (0, q_0^2)= |V_{ub}|^2 \,\Delta \zeta (0, q_0^2)  \,
\end{eqnarray}
of the semi-leptonic  $\bar B^0 \to \pi^{+} \, \mu \, \nu_{\mu}$  decay
\cite{Lees:2012vv,Sibidanov:2013rkk}:
\begin{eqnarray}
\Delta {\cal BR} (0,  12 \, {\rm GeV^2}) &=& (0.83 \pm 0.03 \pm 0.04) \times 10^{-4} \,,
\hspace{0.9 cm} {\rm [BaBar \,\, 2012]}
\nonumber \\
\Delta {\cal BR} (0,  12 \, {\rm GeV^2}) &=& (0.808 \pm 0.062) \times 10^{-4} \,,
\hspace{1.7 cm}  {\rm [Belle \,\,\,\,\,\, 2013]}
\end{eqnarray}
and taking the mean lifetime $\tau_{B^0} =(1.519 \pm 0.005) \, {\rm ps}$ \cite{Agashe:2014kda}
we obtain
\begin{eqnarray}
|V_{ub}|= \left(3.05^{+0.54}_{-0.38} |_{\rm th.} \pm 0.09 |_{\rm exp.}\right)  \times 10^{-3} \,,
\end{eqnarray}
where the reduction of $|V_{ub}|$ compared to \cite{Khodjamirian:2011ub} is attributed to the rapidly increasing
form factor $f_{B \pi}^{+}(q^2)$, with respect to $q^2$, computed from the sum rules with $B$-meson DAs,
and the diminishing $\Delta {\cal BR} (0,  12 \, {\rm GeV^2})$ from the new measurements
\cite{Lees:2012vv,Sibidanov:2013rkk} in relative to the previous BaBar measurements
\cite{delAmoSanchez:2010af,delAmoSanchez:2010zd}; the theoretical uncertainty is from the computation of
$\Delta \zeta (0, 12 \, {\rm GeV^2})$ as displayed in  (\ref{Delta Zeta 0 to 12}).

Now we turn to compute the normalized differential $q^2$ distributions of $B \to \pi \ell \nu_{\ell}$ using the form
factors obtained with the $B$-meson LCSR and extrapolated with the $z$-series parametrization.
Our predictions of the normalized $q^2$ distribution are plotted in Fig. \ref{normalized q2 distribution of B to pi l nu}
where the available data from BaBar and Belle Collaborations are also shown for a comparison.
We observe a reasonable agreement of our predictions for the $q^2$ distribution of $B \to \pi \mu \nu_{\mu}$
and the new Belle and BaBar data points \cite{Sibidanov:2013rkk,Lees:2012vv}, but a poor agreement  when confronted
with the previous measurements \cite{delAmoSanchez:2010af,delAmoSanchez:2010zd,Ha:2010rf} in particular in the
low $q^2$ region. It is  evident that  the theory uncertainty of the normalized differential distribution of  $B \to \pi \mu \nu_{\mu}$
is somewhat smaller than that of the form factors shown in Fig. \ref{final form factor shape} because of the partial cancelation
in the ratio of the differential and the total branching ratio with respect to  the variations of theory inputs.
The $q^2$ shape of the normalized distribution for  $B \to \pi \mu \nu_{\mu}$ is also confronted with the prediction
from the pion LCSR in \cite{Khodjamirian:2011ub}. As a by-product, we further plot the normalized differential distribution of
$B \to \pi \tau \nu_{\tau}$ in Fig. \ref{normalized q2 distribution of B to pi l nu}, which provides an independent way
to extract $|V_{ub}|$ with the aid of future measurements at the Belle-II experiment.

\section{Three-particle DAs of the $B$ meson}
\label{section: three-particle DAs}

We have not touched three-parton Fock-state contributions  to the form factors $f_{B \pi}^{+,0}(q^2)$
in the context of the LCSR with $B$-meson DAs.
This topical problem  has triggered ``sophisticated"  discussions  in the literature from different perspectives,
see \cite{Beneke:2003pa,Lange:2003jz,Ball:2003bf,Hardmeier:2003ig,Chen:2011gv} for an incomplete list.
We will first make some general comments on  non-valence Fock state contributions to $B \to \pi$ form factors,
and then discuss how  $B$-meson three-particle DAs could contribute to the sum rules presented in this work
briefly.

\begin{itemize}

\item  {The representation of the heavy-to-light currents in the context of ${\rm SCET \, (c,s)}$
indicates that three-parton Fock-state contributions already appear  at leading power in ${\Lambda / m_b}$
\cite{Beneke:2003pa} and these contributions preserve the large-recoil symmetry relations at leading power albeit with the emergence
of endpoint divergences \cite{Beneke:2003pa,Lange:2003jz}. This observation was confirmed independently by QCD sum rule calculations
of $B \to \pi$ form factors with pion DAs \cite{Ball:2003bf}.  }

\item {The tree-level contribution of three-particle DAs in $B \to \pi$ form factors
is of minor importance numerically (at percent level) confirmed by two different types of
sum rules with $B$-meson DAs \cite{Khodjamirian:2006st} and with pion DAs \cite{Duplancic:2008ix},
respectively. The insignificant tree-level effect can be understood transparently from the
sum rules with pion DAs, where the collinear gluon emission from the $b$-quark propagator yields
power suppression in ${\Lambda / m_b}$. However,  this power-suppression mechanism
will be removed at ${\cal O}(\alpha_s)$, because the radiative gluon can be emitted from the
(hard)-collinear light-quark propagators in the evaluation of the corresponding correlation function at NLO
(for a concrete example, see \cite{Ball:2003bf}). A complete calculation of three-parton Fock-state contributions
to $B \to \pi$ form factors is unfortunately not available in the framework of both sum rule approaches at present.
In the following we will sketch  this absorbing and  challenging calculation in the context of the LCSR with $B$-meson DAs.}

\begin{figure}[t]
\begin{center}
\includegraphics[width=1.0  \columnwidth]{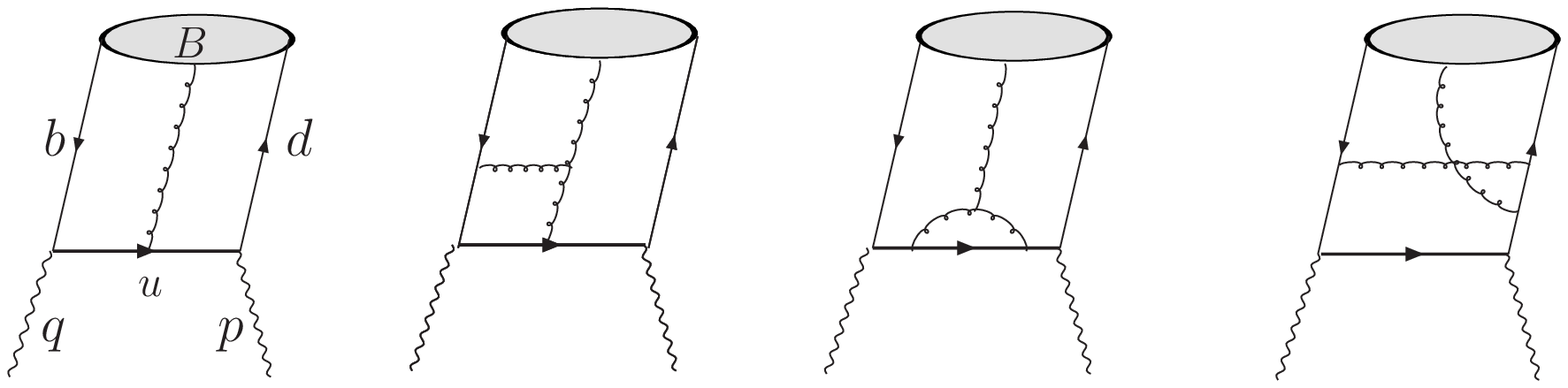}
\hspace{-0.5 cm}(a) \hspace{3.0 cm} (b)\hspace{3.5 cm} (c) \hspace{3.5 cm} (d) \\
\vspace*{1.0 cm}
\includegraphics[width=0.65  \columnwidth]{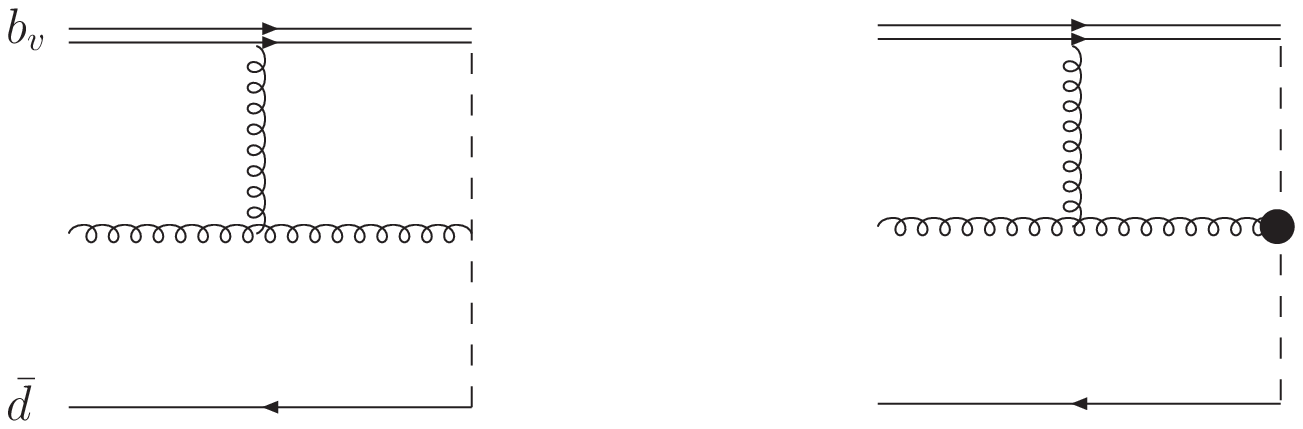} \\
\hspace{4.0 cm}(e) \hspace{5.0 cm} (f)\hspace{3.5 cm}
\vspace*{0.1cm}
\caption{{\it Top:} Contributions of $B$-meson three-particle DAs to the correlation function
$\Pi_{\mu}$ at tree level (the diagram (a)) and at one-loop order (typical diagrams
displayed in (b), (c) and (d)). {\it Bottom:} Typical diagrams for renormalization of two-parton (the diagram (e))
and three-parton (the diagram (f)) DAs at ${\cal O}(g_s^3)$.
The black blob in the diagram (f) indicates the external gluon field in the
string operator defining three particle DAs of the $B$-meson. }
\label{correlator with 3-particle DAs}
\end{center}
\end{figure}

\item {$B$-meson three-particle DAs  could manifest themselves in the NLO sum rules in a variety of ways.
First,  these contributions are essential to compensate the factorization-scale dependence of $\phi_B^{-}(\omega,\mu)$
entering the factorization formulae of $\Pi(n \cdot p \,, \bar n \cdot p)$ and $\tilde{\Pi}(n \cdot p \,, \bar n \cdot p)$
at  ${\cal O}(g_s^3)$ recalling the evolution equation \cite{DescotesGenon:2009hk}
\begin{eqnarray}
{d \over d \ln \mu} \, \phi_B^-(\omega, \mu) =  - \frac{\alpha_s \, C_F}{4 \, \pi}\bigg \{
\left [ \Gamma_{\rm cusp}^{(0)} \ln { \mu \over \omega} - 2\right ] \, \phi_B^-(\omega, \mu)  \nonumber \\
+ \int_0^{\infty} \, d \omega^{\prime}  \, \omega \,\,
\Gamma(\omega,\omega^{\prime},\mu) \,\, \phi_B^-(\omega, \mu) \, \nonumber \\
+ \int_0^{\infty} d \omega^{\prime} \, \int_0^{\infty} d \xi^{\prime} \,
\gamma_{-,3}^{(1)}(\omega,\omega^{\prime},\xi^{\prime},\mu) \,
\left [ \Psi_A -\Psi_V \right ](\omega^{\prime},\xi^{\prime},\mu) \bigg \}\,,
\end{eqnarray}
where the mixing term is solely governed by the light degrees of freedom in the composite operator \cite{Knodlseder:2011gc}.
A sample diagram is shown in Fig. \ref{correlator with 3-particle DAs} (b) whose soft divergences can be reproduced by
adding up amplitudes of the two effective diagrams displayed in Figs. \ref{correlator with 3-particle DAs} (e)
and \ref{correlator with 3-particle DAs}  (f)
convoluted with the corresponding tree-level  hard-scattering kernel.
Second,  $B$-meson three-particle DAs can induce leading-power contributions without recourse to
the above-mentioned mixing pattern and one also expects that the soft subtraction is not needed here due to
power suppression of the  tree-level contribution from three-particle DAs.
One should notice that renormalization of  $B$-meson three-particle DAs will not generate the inverse
mixing into two-particle DAs, at least, at ${\cal O}(\alpha_s)$, while a similar statement holds to all orders of $\alpha_s$
for  pion DAs due to conformal symmetry \cite{Braun:2009vc}.
Third, three-particle DAs can revise the Wandzura-Wilczek relation (\ref{WW relation})
which needs to be generalized into \cite{Bell:2013tfa}
\begin{eqnarray}
\omega \, \phi_B^-(\omega) - \int_0^\omega d\eta \left[ \phi_B^-(\eta)-\phi_B^+(\eta) \right ] 
&=& 2 \, \int_0^\omega d\eta \int_{\omega-\eta}^\infty \frac{d\xi}{\xi} \, \frac{\partial}{\partial \xi}
\left [  \Psi_A(\eta,\xi)-\Psi_V(\eta,\xi)\right ] \,. \nonumber \\
\end{eqnarray}
}

\end{itemize}

\section{Conclusions and discussions}
\label{section: summaries}

We have carried out, for the first time,  perturbative corrections to $B \to \pi$ form factors
from the QCD LCSR with $B$-meson  DAs proposed in \cite{Khodjamirian:2005ea,Khodjamirian:2006st}
where the sum rules for heavy-to-light form factors were established at tree level including
contributions from both two-particle and three-particle DAs. We placed particular emphasis on the demonstration
of factorization of the vacuum-to-$B$-meson correlation function $\Pi_{\mu}(n \cdot p,\bar n \cdot p)$
at ${\cal O}(\alpha_s)$ taking advantage of the method of regions which allows a transparent separation
of different leading regions with the aid of the power counting scheme.
Precise cancelation of  the soft contribution to the correlation function $\Pi_{\mu}$ and the infrared subtraction
was perspicuously shown at the diagrammatic level.  The short-distance function obtained with integrating out the
hard-scale fluctuation receives the contribution from the weak-vertex diagram solely because the loop integrals
from the remaining diagrams do not involve any external invariant of order $1$.
The resulting hard coefficients coincide with the corresponding matching coefficients of the vector QCD weak current
in ${\rm SCET_I}$ indicating that  perturbative coefficients in the OPE
are independent of the external partonic configuration chosen in the  matching procedure as expected.
The computed jet functions from integrating out dynamics of the hard-collinear scale  are also in agreement
with the expressions derived from the SCET Feynman rules \cite{DeFazio:2005dx}.
We further verified  factorization-scale independence of the correlation function at ${\cal O}(\alpha_s)$
employing evolution equations of the hard function $\tilde{C}^{(-)}$, the jet function $\tilde{J}^{(-)}$
and the $B$-meson DA $\phi_B^{-}(\omega, \mu)$,
then summed up the large logarithms due to the appearance of distinct energy scales by the standard RG
approach in the momentum space. We left out  resummation of the parametrically large logarithms of  $\mu_{hc} / \mu_0$
due to the insignificance numerically.  However, there is no difficulty to achieve this resummation whenever such theory
precision is in demand and an elegant way to perform resummation of large logarithms in the presence of the cusp anomalous
dimension in the evolution equations is to work in the ``dual" momentum space where the Lange-Neubert kernel
of the $B$-meson DAs are diagonalized \cite{Bell:2013tfa}.

With the resummation improved sum rules (\ref{NLO sum rules of form factors}) at hand,
we explored their phenomenological implications on $B \to \pi$ form factors at large hadronic recoil in detail.
Due to our poor knowledge of the inverse moment of the $B$-meson DA $\phi_B^{+}(\omega, \mu)$
we first determine this parameter by matching the $B$-meson LCSR prediction of $f_{B \pi}^{+} (q^2)$ at zero momentum
transfer to the result obtained from the sum rules with pion DAs at NLO, utilizing four different models
of $\phi_B^{\pm}(\omega, \mu)$ displayed in (\ref{four models of B-meson DAs}).
While these models do not capture the features of large $\omega$ behaviors from  perturbative QCD analysis,
the power counting rule $\omega \sim \Lambda$, thanks to the canonical picture of the $B$-meson bound state,
implemented in the construction of QCD factorization for the correlation function $\Pi_{\mu}$ requires that
the dominant contribution  in the factorized amplitude must be  from the small $\omega$ region.
We then found that  the obtained values of the shape parameter $\omega_0(1 \, {\rm GeV})$ from the matching procedure
are rather sensitive to the shapes of $\phi_B^{\pm}(\omega, \mu)$ at small $\omega$, pointing to the poor ``local"
approximation (\ref{rough approximation of LO B-meson SR}) and confirming an earlier observation
made in \cite{DeFazio:2007hw}. However,  the $q^2$ shape of $f_{B \pi}^{+} (q^2)$  predicted from LCSR
is insensitive to the specific model of the $B$-meson DAs after determining $\omega_0(1 \, {\rm GeV})$ from the
above-described matching condition. This is not surprising because of a large cancelation of the theory uncertainty
in the form-factor ratio $f_{B \pi}^{+}(q^2)/f_{B \pi}^{+}(0)$.
Moreover, we showed that the dominating radiative effect originates from the NLO QCD correction instead of
the NLL resummation of large logarithms in the heavy quark limit, which does however improve the stability of varying
the factorization scale.  Proceeding with the resummation improved sum rules (\ref{NLO sum rules of form factors})
in the large recoil region,  $q^2 \leq 8 \, {\rm GeV^2}$, and extrapolating the computed from factors toward large momentum
transfer, we found that the obtained scalar form factor  $f_{B \pi}^{0}(q^2)$ is in a reasonable agreement with that
from the LCSR with pion DAs at  $q^2 \leq 12 \, {\rm GeV^2}$; while  a similar comparison of the vector form factor
$f_{B \pi}^{+}(q^2)$ calculated from two different sum rule approaches reveals noticeable discrepancies particularly
at higher $q^2$. We made a first step towards understanding this intriguing problem by determining the parameter
$\omega_0(1 \, {\rm GeV})$ from matching the form factor $f_{B \pi}^{+} (13.74 \, {\rm GeV^2})$ to the Lattice data
from Fermilab/MILC Collaborations. It was then shown that the shape of $f_{B \pi}^{+} (q^2)$ at high $q^2$ predicted
from Lattice QCD simulation lies in between that obtained from the two sum rule approaches.
It remains unclear whether the observed discrepancies arise from the sub-leading power contributions to both LCSR
and/or from the systematic uncertainties induced by different kinds of parton-hadron duality relations  in the constructions
of sum rules and/or from the yet unknown leading-power contribution from three-particle DAs in both  approaches.
As a result of the rapidly growing form factor $f_{B \pi}^{+}(q^2)$ we obtain lower values of
$|V_{ub}|=\left(3.05^{+0.54}_{-0.38} |_{\rm th.} \pm 0.09 |_{\rm exp.}\right)  \times 10^{-3}$,
in contrast to the predictions of the pion LCSR \cite{Khodjamirian:2011ub,Imsong:2014oqa}, by
comparing BaBar and Belle measurements of the integrated branching ratio of
$B \to \pi \, \mu \, \nu_{\mu}$ in the region
$ 0< q^2 < 12 \, {\rm GeV^2}$ with the computed quantity $\Delta \zeta (0, 12 \,{\rm GeV^2} )$
in (\ref{Delta Zeta 0 to 12}).  The theory uncertainty is dominated by the rather limited information of
the $B$-meson DAs at small $\omega$ encoding formidable (non-perturbative) strong-interaction dynamics.
Precision measurements of the differential $q^2$ distributions of $B \to \pi \ell \nu_{\ell}$ ($\ell=\mu, \tau$)
at the Belle-II experiment might shed light on the promising orientation to resolve the tension of the form factor
shapes and subsequently to put meaningful constraints on the small $\omega$ behaviors of  $\phi_B^{\pm}(\omega)$.

We further turned to discuss non-valence Fock-state contributions to the form factors $f_{B \pi}^{+,0} (q^2)$,
which are the missing ingredients of our computations, and to illustrate modifications of the sum rule analysis
in the presence of  three-particle DAs of the $B$ meson. Given the fact that non-valence Fock-state contributions
to the form factors $f_{B \pi}^{+,0} (q^2)$ are either suppressed in powers of $\Lambda/m_b$ at tree level or
suppressed by the QCD coupling constant $\alpha_s$  at NLO,  one may expect that including three-particle DAs of the $B$ meson
may  generate a modest effect on the form-factor predictions albeit with a high demand of computing their
contributions at ${\cal O}(\alpha_s)$ in the conceptual aspect.

Developing the LCSR of $B \to \pi$ form factors with $B$-meson DAs beyond this work can be pursued further by
including the sub-leading power contributions of the considered correlation function, which requires
better knowledge of the  sub-leading DAs of the $B$ meson (e.g., off-light-cone corrections)
and demonstrations of factorization for the correlation function at sub-leading power in $\Lambda/m_b$.
Computing yet higher order QCD corrections of the correlation function  would be also interesting conceptually,
but one would need the two-loop evolution equation of $\phi_B^{-}(\omega, \mu)$, which can be complicated by more
involved mixing of string operators under renormalization
\footnote{The new technique developed in \cite{Braun:2014vba} using exact conformal symmetry
of QCD at the critical point would be powerful in this respect.}.
Moreover, we also expect phenomenological  extensions of this work to compute NLO QCD corrections of
many other hadronic matrix elements from the LCSR with bottom-hadron DAs.
First, it is of interest to perform a comprehensive analysis of $B \to P\,, V$ form factors with $P=\pi, K$
and $V=\rho \,, K^{\ast}$ from the $B$-meson LCSR at ${\cal O}(\alpha_s)$,  which serve as fundamental theory inputs
for QCD factorization of the electro-weak penguin $B \to K^{(\ast)} \, \ell \ell$ decays  and the charmless hadronic
$B \to P \, P$ and $B \to P \, V$ decays. In particular, such analysis could be of value to understand
the tension of  form-factor ratios between the traditional LCSR and QCD factorization
firstly observed in \cite{Beneke:2000wa}, keeping in mind that these ratios are less sensitive to the shapes of
$B$-meson DAs at small $\omega$.
Second,  a straightforward extension of this work to compute form factors describing
the exclusive $B \to D^{(\ast)} \, \tau \, \nu_{\tau}$ decays will deepen our understanding towards
the topical $R(D^{(\ast)})$ puzzles, referring to the $2 \, \sigma$ ($2.7 \, \sigma$) deviations of the measured ratios
of the corresponding  branching fractions in  muon and tauon channels.
Such computations will enable us to pin down perturbative uncertainties in the tree-level predictions
of $B \to D^{(\ast)}$ form factors \cite{Faller:2008tr} and also allow for a better comparison of LCSR and heavy-quark expansion
in a different testing ground.
A complete discussion of radiative corrections to $B \to D^{(\ast)}$ form factors from the $B$-meson LCSR
including a proper treatment of the charm-quark mass will be presented elsewhere.
Third, the techniques discussed in this work can be applied to compute $\Lambda_b \to p \,, \Lambda$ transition form factors
from the sum rules with the $\Lambda_b$-baryon DAs \cite{Wang:2009hra,Feldmann:2011xf},
which are of phenomenological interest for an alternative determination
of $|V_{ub}|$  exclusively and for a complementary  search of physics beyond the Standard Model.
However, one should be aware of the fact that constructing baryonic sum rules are more involved than the mesonic counterpart
due to various ways to choose baryonic interpolating currents and potential  contaminations from negative-parity baryons
in the hadronic dispersion relations \cite{Khodjamirian:2011jp}.
To summarize, we foresee straightforward extensions of this work into different directions.

\section*{Acknowledgements}

We are grateful to M.~Beneke, N. Offen and J. Rohrwild for illuminating discussions,
to M.~Beneke for many valuable comments on the manuscript,
and to D.P. Du for providing us with the Lattice data points in \cite{Lattice:2015tia}.
YMW  is supported in part by the Gottfried Wilhelm Leibniz programme
of the Deutsche Forschungsgemeinschaft (DFG).

\appendix

\section {Loop integrals}
\label{loop integrals}

In this appendix,  we collect some useful one-loop integrals in our calculations.
\begin{eqnarray}
I_0 &=& \int [d \, l]
\frac{1}{[(p-k+l)^2 + i 0][l^2 + 2 \, m_b \, v \cdot l+ i 0]} \, \nonumber \\
&=& {1 \over \epsilon}  + 2 \, \ln {\mu \over m_b}  + {r \over \bar r} \, \ln r + 2\,,  \\
I_1^{h} &=& \int [d \, l] \,
\frac{1}{[l^2 + n \cdot p \,\, \bar n \cdot l+ i 0][l^2 + 2 \, m_b \, v \cdot l+ i 0] [l^2+i0]} \, \nonumber \\
&=& \frac{-1}{2 \, m_b \, n \cdot p} \, \left [{1 \over \epsilon^2} +
{2 \over \epsilon} \, \ln {\mu \over  n \cdot p} + 2\, \ln^2 {\mu \over  n \cdot p}
-\ln^2 r -2 \, {\rm Li_2} \left (- {\bar r \over r} \right )  +{\pi^2 \over 12}\right ] \,, \,\, \\
I_1^{hc} &=&  \int \frac{d^D \, l}{(2 \pi)^D} \,  \frac{1}{[ n \cdot (p+l) \,
\bar n \cdot (p-k+l) + l_{\perp}^2  + i 0][ \, n \cdot l+ i 0] [l^2+i0]}  \nonumber \\
&=& {1 \over n \cdot p} \, \left [ {1 \over \epsilon^2} + {1 \over \epsilon} \,
\ln {\mu^2 \over n \cdot p \, (\omega-\bar n \cdot p)}  \,
+ {1 \over 2}\,  \ln^2 {\mu^2 \over n \cdot p \, (\omega-\bar n \cdot p)} - {\pi^2 \over 12} \right ]\,, \\
I_{1 \, \mu} &=& \int [d \, l]
\frac{l_{\mu}}{[(p-k+l)^2 + i 0][l^2 + 2 \, m_b \, v \cdot l+ i 0] [l^2+i0]} \, \nonumber \\
&\equiv& C_1 \, (p-k)_{\mu}  +  C_2 \, m_b \, v_{\mu} \,,   \\
C_1^{h}&=& - \frac{1}{m_b^2 \, r} \, \left [ {1 \over \epsilon}  + 2 \, \ln {\mu \over m_b}
-{r-2 \over r-1} \, \ln r  + 2 \right ] \,, \\
C_2&=&C_2^{h}= - \frac{1}{m_b^2 \, \bar r} \, \ln r \,, \\
C_1^{hc}&=& C_1-C_1^{h}=\frac{1}{m_b^2 \, r} \, \left [ {1 \over \epsilon}
+ \ln {\mu^2 \over n \cdot p \, (\omega-\bar n \cdot p)}+ 2 \right ] \,,  \\
I_{1, \, a} &=& \int [d \, l]
\frac{n \cdot l \,\, \bar n \cdot l}{[(p-k+l)^2 + i 0][l^2 + 2 \, m_b \, v \cdot l+ i 0] [l^2+i0]} \, \nonumber \\
&=& {1 \over 2} \, \left [ {1 \over \epsilon} + 2 \, \ln {\mu \over  m_b} + {r \over \bar r} \, \ln r + 2 \right ] \,, \\
I_{1, \,b} &=& \int [d \, l]
\frac{ (\bar n \cdot l)^2 }{[(p-k+l)^2 + i 0][l^2 + 2 \, m_b \, v \cdot l+ i 0] [l^2+i0]} \, \nonumber \\
&=& - {1 \over 2 \, \bar r^2} \, \left [ r \, \ln r + \bar r \right ] \,,   \\
I_2 &=& \int [d \, l] \,   \frac{l_{\alpha} \,\, (p-l)_{\beta}}{[(p-l)^2 + i 0][(l-k)^2 + i 0] [l^2+i0]} \nonumber \\
&\equiv& - {g_{\alpha \beta} \over 2} \, I_{2, a}
- {1 \over  p^2} \, \left [ k_{\alpha} \, k_{\beta} \,  I_{2, b} - p_{\alpha} \, p_{\beta} \,  I_{2, c}
- k_{\alpha} \, p_{\beta} \,  I_{2, d} + p_{\alpha} \, k_{\beta} \,  I_{2, e} \right ]  \,,   \\
I_{2, a} &=& {1 \over 2} \left [{1 \over \epsilon} +\ln \left (-{\mu^2 \over p^2} \right )
-{1+\eta \over \eta} \, \ln (1+\eta) + 3 \right ] \,,  \\
I_{2, b} &=&  \frac{2 \, \eta - \eta^2 - 2 \, \ln (1+\eta)}{2\, \eta^3} \,
\left [{1 \over \epsilon} +\ln \left (-{\mu^2 \over p^2} \right ) - \, \ln (1+\eta) + 3 \right ] \nonumber \\
&& + \frac{\eta^2 - \ln^2 (1+\eta)}{2\, \eta^3}   \,, \\
I_{2, c} &=& \frac{\ln (1+\eta)}{2 \, \eta} \,, \\
I_{2, d} &=&  \frac{\ln (1+\eta)-\eta}{\eta^2} \,
\left [{1 \over \epsilon} +\ln \left (-{\mu^2 \over p^2} \right ) - \, \ln (1+\eta) + {5 \over 2} \right ]
+ \frac{\ln^2 (1+\eta)}{2\, \eta^2} \,,  \\
I_{2, e} &=&  \frac{\eta - \ln (1+\eta)}{2 \, \eta^2}  \,,  \\
I_3&=& \int [d \, l] \frac{1}{[(p-k+l)^2 + i 0][l^2 +  i 0]}
= {1 \over \epsilon}  +  \ln {\mu^2 \over n \cdot p \, \bar n \cdot (k-p)}  + 2\,, \\
I_{3 \, \mu}&=& \int [d \, l]
\frac{l_{\mu}}{[(p-k+l)^2 + i 0][l^2 +  i 0]}  = - \frac{I_3}{2}\, (p-k)_{\mu}  \,,   \\
I_{4,a} &=&  \int \frac{d^D \, l}{(2 \pi)^D} \,
\frac{n \cdot (p+l) }{[ n \cdot (p+l) \, \bar n \cdot (p-k+l) + l_{\perp}^2  + i 0]
[n \cdot l \, \bar n (l-k)+ l_{\perp}^2  + i 0][l^2+i0]} \, \nonumber \\
&=& {\ln (1+\eta) \over \omega} \, \left [ {1 \over \epsilon} + \ln {\mu^2 \over n \cdot p \, (\omega-\bar n \cdot p)}
+ {1 \over 2} \, \ln (1+\eta) + 1 \right ] \,,  \\
I_{4,b} &=&  \int \frac{d^D \, l}{(2 \pi)^D} \,
\frac{n \cdot l \,\, n \cdot (p+l) }{[ n \cdot (p+l) \, \bar n \cdot (p-k+l) + l_{\perp}^2  + i 0]
[n \cdot l \, \bar n (l-k)+ l_{\perp}^2  + i 0][l^2+i0]} \, \nonumber \\
&=& {n \cdot p \over 2 \, \omega} \, \, \ln (1+\eta)  \,.
\end{eqnarray}
with $r=n \cdot p/m_b$, $\bar r = 1-r$,  $\omega=\bar n \cdot k$ and $\eta=-\omega/ \bar n \cdot p$.

\section {Spectral representations}
\label{appendix B}

We collect the spectral functions of convolution integrals entering the factorization formulae of
$\Pi$ and $\widetilde{\Pi}$ in (\ref{resummation improved factorization formula}).
These expressions were first derived in \cite{DeFazio:2007hw}, we confirmed  these spectral functions
independently and also verified the corresponding dispersion integrals.
\begin{eqnarray}
&& {1 \over \pi} \,\, {\rm Im_{\omega^{\prime}}}  \, \int_0^{\infty} \, {d \omega \over \omega} \,
\ln {\omega^{\prime} - \omega \over \omega^{\prime}} \, \phi_B^{+}(\omega, \mu)
= \int_{\omega^{\prime}}^{\infty} \,\, {d \omega \over \omega} \,\, \phi_{B}^{+}(\omega, \mu) \,,  \\
&& {1 \over \pi} \,\, {\rm Im_{\omega^{\prime}}}  \, \int_0^{\infty}
\, {d \omega \over \omega  - \omega^{\prime} - i 0} \,\,
\ln^2 {\mu^2  \over n \cdot p \, (\omega - \omega^{\prime}) } \,\,  \phi_B^{-}(\omega, \mu) \nonumber \\
&& =\int_{0}^{\infty} \,\, d \omega  \,\,
\left [ {2 \, \theta(\omega^{\prime}-\omega)\over \omega-\omega^{\prime}} \,
\ln {\mu^2  \over n \cdot p \, (\omega^{\prime} - \omega) } \right ]_{\oplus} \phi_{B}^{-}(\omega, \mu)
+ \left [\ln^2 {\mu^2  \over n \cdot p \, \omega^{\prime} }  -{\pi^2 \over 3}  \right ] \,,  \hspace{0.4 cm}  \\
&& {1 \over \pi} \,\, {\rm Im_{\omega^{\prime}}}  \, \int_0^{\infty}
\, {d \omega \over \omega  - \omega^{\prime} - i 0} \,\,
\ln^2 {\omega^{\prime}  - \omega \over \omega^{\prime} } \,\,  \phi_B^{-}(\omega, \mu) \nonumber \\
&& =- \int_{\omega^{\prime}}^{\infty} \,\, d \omega  \,\,
\left [ \ln^2 {\omega - \omega^{\prime}  \over \omega^{\prime} } -{\pi^2 \over 3}  \right ] \,
{d \over d \omega}\phi_{B}^{-}(\omega, \mu) \,,  \\
&& {1 \over \pi} \,\, {\rm Im_{\omega^{\prime}}}  \, \int_0^{\infty}
\, {d \omega \over \omega  - \omega^{\prime} - i 0} \,\,
\ln {\omega^{\prime}  - \omega \over \omega^{\prime} } \,
\ln {\mu^2  \over n \cdot p \, (\omega - \omega^{\prime})} \,  \phi_B^{-}(\omega, \mu) \, \nonumber \\
&& =\int_{0}^{\infty} \,\, d \omega  \,\,
\left [ {\theta(\omega^{\prime}-\omega)\over \omega-\omega^{\prime}} \,
\ln {\omega^{\prime}-\omega  \over \omega^{\prime}} \right ]_{\oplus} \phi_{B}^{-}(\omega, \mu) \nonumber \\
&& \hspace{0.4 cm} + {1 \over 2} \, \int_{\omega^{\prime}}^{\infty} \,\, d \omega  \,\,
\left [ \ln^2 { \mu^2  \over n \cdot p \, (\omega - \omega^{\prime}) }
- \ln^2 { \mu^2  \over n \cdot p \, \omega^{\prime} } + {\pi^2 \over 3}  \right ] \,
{d \over d \omega}\phi_{B}^{-}(\omega, \mu) \,, \\
&& {1 \over \pi} \,\, {\rm Im_{\omega^{\prime}}}  \, \int_0^{\infty}
\, {d \omega \over \omega  - \omega^{\prime} - i 0} \,\,
\ln {\omega^{\prime}  - \omega \over \omega^{\prime} } \,\,  \phi_B^{-}(\omega, \mu) \nonumber \\
&& =- \int_{\omega^{\prime}}^{\infty} \,\, d \omega  \,\,
\ln {\omega - \omega^{\prime}  \over \omega^{\prime} } \,
{d \over d \omega}\phi_{B}^{-}(\omega, \mu) \,.
\end{eqnarray}

\section {Two-point QCD sum rules for $f_B$ and $f_{\pi}$}
\label{Appendix C}

For completeness, we collect the two-point sum rules for the $B$-meson decay constant
$f_B$ in QCD \cite{Jamin:2001fw,Gelhausen:2013wia}  including NLO corrections to
the  perturbative term and to the $D=3$ quark condensate part:
\begin{eqnarray}
f_B^2&=&\frac{e^{m_B^2/\overline{M}^2}\,  m_b^2}{m_B^4} \,
\bigg \{ \int_{m_b^2}^{\bar{s}_0} \, ds \, e^{-s/\overline{M}^2} \,
{3 \over 8 \, \pi^2} \left [  {(s-m_b^2)^2 \over s} + {\alpha_s \, C_F  \over \pi } \,
\rho_{pert}^{(1)} (s, m_b^2) \right ]  \nonumber \\
&& + e^{-m_b^2/\overline{M}^2} \, \bigg [ -m_b \, \langle \bar q q \rangle  \,
\left ( 1 + {\alpha_s \, C_F  \over \pi } \, \rho_{q \bar q}^{(1)} (s, m_b^2)  \right )
- {m_b \langle \bar q G q \rangle  \over 2 \, \overline{M}^2} \, \left ( 1- {m_b^2 \over 2\overline{M}^2 } \right ) \nonumber \\
&& + {1 \over 12} \, \big \langle {\alpha_s \over \pi}  GG \big \rangle
- {16 \, \pi \over 27} \, { \alpha_s \langle \bar q q  \rangle^2 \over \overline{M}^2 }\,
\left ( 1- {m_b^2 \over 4 \, \overline{M}^2} - {m_b^4 \over 12 \, \overline{M}^4} \right ) \bigg ] \bigg \} \,.
\end{eqnarray}
where $\overline{M}^2$ and $\bar{s}_0$ are the Borel parameter and the effective threshold,
and the relevant NLO spectral functions are given by
\begin{eqnarray}
\rho_{pert}^{(1)} (s, m_b^2) &=& {\bar x \, s \over 2} \,  \bigg \{ \bar x \left [ 4\, {\rm Li_2}(x)
+ 2\, \ln x \, \ln \bar x - (5 - 2 x)\, \ln \bar x  \right ] \, \nonumber \\
&&  + (1- 2 x) \, (3-x)  \, \ln x + 3 \, (1-3x) \, \ln {\mu^2 \over m_b^2}  + {17 -33 x \over 2}\bigg \} \,,  \\
\rho_{q \bar q}^{(1)} (s, m_b^2) &=& - {3 \over 2} \, \left [  \Gamma \left (0, {m_b^2 \over \overline{M}^2} \right )
\, e^{m_b^2 / \overline{M}^2}  - \left ( 1- {m_b^2 \over\overline{M}^2 }\,
\left ( \ln {\mu^2 \over m_b^2} + {4\over 3}\right ) \right ) -1 \right ] \,, \,\,
\end{eqnarray}
with $x=m_b^2/s$,  $\bar x =1-x$ and the incomplete $\Gamma$ function defined as
\begin{eqnarray}
\Gamma(n, x)= \int_x^{\infty} \, d t \,\,\,  t^{n-1}\, e^{-t} \,.
\end{eqnarray}

The two-point QCD sum rules of the pion decay constant $f_{\pi}$ including the perturbative term
at ${\cal O} (\alpha_s)$ reads \cite{Colangelo:2000dp}
\begin{eqnarray}
f_{\pi}^2&=& M^2 \, \bigg [ {1 \over 4 \, \pi^2} \, \left (1-e^{-s_0/M^2} \right ) \,
\left (1 + {\alpha_s(M) \over \pi } \right )
+ {1 \over 12 \, M^4} \, \big \langle {\alpha_s \over \pi}  GG  \big \rangle  \nonumber \\
&& \hspace{0.8 cm} + {176 \, \pi \over 81} \, { \alpha_s \langle \bar q q  \rangle^2 \over M^6 }\bigg ]\,\,.
\end{eqnarray}


\end{document}